\documentclass[fleqn,usenatbib]{mnras}

\usepackage{newtxtext,newtxmath}

\usepackage[T1]{fontenc}

\DeclareRobustCommand{\VAN}[3]{#2}
\let\VANthebibliography\thebibliography
\def\thebibliography{\DeclareRobustCommand{\VAN}[3]{##3}\VANthebibliography}

\usepackage{graphicx}   
\usepackage{amsmath}    
\usepackage{subcaption}
\usepackage[dvipsnames]{xcolor}
\usepackage{bm}
\let\vec\bm

\usepackage{xpatch}
\usepackage{siunitx}
\usepackage[normalem]{ulem}
\usepackage{upgreek}
\usepackage{orcidlink}
\usepackage{hyperref}

\makeatletter
\xpatchcmd\NAT@citex
{%
	\@citea\NAT@hyper@{%
		\NAT@nmfmt{\NAT@nm}%
		\hyper@natlinkbreak{\NAT@aysep\NAT@spacechar}{\@citeb\@extra@b@citeb}%
		\NAT@date
	}%
}
{%
	\@citea
	\NAT@nmfmt{\NAT@nm}%
	\NAT@aysep\NAT@spacechar
	\NAT@hyper@{\NAT@date}%
}
{}{}
\xpatchcmd\NAT@citex
{%
	\@citea\NAT@hyper@{%
		\NAT@nmfmt{\NAT@nm}%
		\hyper@natlinkbreak{\NAT@spacechar\NAT@@open\if*#1*\else#1\NAT@spacechar\fi}%
		{\@citeb\@extra@b@citeb}%
		\NAT@date
	}%
}
{
	\@citea
	\NAT@nmfmt{\NAT@nm}%
	\NAT@spacechar\NAT@@open\if*#1*\else#1\NAT@spacechar\fi
	\NAT@hyper@{\NAT@date}%
}
{}{}
\makeatother

\newcommand{\bnabla}{\bm{\nabla}}
\newcommand{\bcdot}{\bm{\cdot}}

\newcommand{\Mvir}{M_{200}}
\newcommand{\Rvir}{R_{200}}
\newcommand{\Msun}{\mathrm{M}_\odot}
\newcommand{\muG}{\upmu\mathrm{G}}
\newcommand{\kb}{k_{\textup{B}}}
\newcommand{\para}{\parallel}
\newcommand{\mH}{m_{\textup{p}}}
\newcommand\bb[1]{\mbox{\boldmath{$#1$}}}
\newcommand\del{\nabla}
\newcommand\btimes{\bm\times}

\renewcommand{\b}{\bb{b}}
\newcommand{\be}{\begin{equation}}
\newcommand{\en}{\end{equation}}
\newcommand{\f}{\frac}

\title[The PICO-Cluster Project]{The PICO-Cluster Project: presenting the galaxy cluster sample and studying magnetic field growth, Faraday rotation and Braginskii heating}

\author[T. Berlok et al.]{%
\parbox{0.95\textwidth}{%
Thomas Berlok$^{1,2}$\thanks{E-mail: tberlok@nbi.ku.dk, cpfrommer@aip.de, epuchwein@aip.de, rosie@mpa-garching.mpg.de}\orcidlink{0000-0003-0466-603X}, 
Christoph  Pfrommer$^{2}$\orcidlink{0000-0002-7275-3998}, 
Ewald Puchwein$^{2}$\orcidlink{0000-0001-8778-7587}, 
Rosie Talbot$^{3}$\orcidlink{0000-0001-9393-7879}, 
Rüdiger Pakmor$^{3}$\orcidlink{0000-0003-3308-2420}, 
Lorenzo Maria Perrone$^{2}$\orcidlink{0000-0002-8380-9643}, 
Rainer Weinberger$^{2}$\orcidlink{0000-0001-6260-9709}, 
Volker Springel$^{3}$\orcidlink{0000-0001-5976-4599}, 
Joseph Whittingham$^{2}$\orcidlink{0000-0002-0198-8490}, 
Larissa Tevlin$^{2}$\orcidlink{0000-0002-6497-4467}, 
Niklas Dusch$^{2}$\orcidlink{0009-0006-4199-818X}, 
Martin Sparre$^{2}$\orcidlink{0000-0002-9735-3851}
}
\\%
\\%
$^{1}$Niels Bohr Institute, University of Copenhagen, Blegdamsvej 17, 2100 Copenhagen, Denmark\\%
$^{2}$Leibniz-Institut f{\"u}r Astrophysik Potsdam (AIP), An der Sternwarte 16, D-14482
Potsdam, Germany\\%
$^{3}$Max-Planck Institute for Astrophysics, Karl-Schwarzschild-Str. 1,
D-85741 Garching, Germany
}

\date{Accepted XXX. Received YYY; in original form ZZZ}

\pubyear{\the\year{}}

\begin{document}
\label{firstpage}
\pagerange{\pageref{firstpage}--\pageref{lastpage}}
\maketitle

\begin{abstract}
Galaxy clusters constitute a microcosm of the Universe and offer a unique laboratory for studying plasma astrophysics, encompassing processes such as cosmic-ray acceleration and non-thermal radio emission, turbulence, weakly collisional plasma physics, and transformative mechanisms in galaxy evolution. To investigate these phenomena, we introduce the PICO-Cluster project, studying ``Plasmas In COsmological Clusters'' using a suite of high-resolution cosmological zoom-in simulations of massive galaxy clusters with masses ${\gtrsim}10^{15}\,\mathrm{M}_\odot$ selected from a parent simulation box with a comoving side length of $1~h^{-1}\mathrm{Gpc}$. In this work, we present 24 baseline simulations performed with the moving-mesh \textsc{Arepo} code and the IllustrisTNG galaxy formation model, achieving a baryonic mass resolution of up to $1.4\times10^{6}\,\mathrm{M}_\odot$. The initial conditions are carefully designed to exclude low-resolution particle contamination within the high-resolution region; as a result, all clusters remain free of such contamination out to at least $2.7\,R_{200}$ at all times. Our galaxy and cluster properties agree with recent simulations and many observational constraints, including scaling relations and thermodynamic profiles. The magnetic energy within the cluster is numerically converged once the small-scale dynamo has saturated, yielding a remarkably tight volume-averaged plasma-beta of $\beta\approx100$ inside $R_{200}$ across our sample after redshift $z\sim1.2$. Faraday rotation measure profiles, which trace the line-of-sight magnetic field and electron density, decline with cylindrical radius; notably, the mean decreases more rapidly than the root-mean-square due to the increasing relative contribution of galaxies at larger radii. Finally, viscous heating rates in Braginskii theory are highly intermittent and, on average, approach radiative cooling rates in the cluster outskirts. 
\end{abstract}




\section{Introduction}
\label{sec:introduction}

Galaxy clusters are the largest collapsed objects in cosmic history and sit atop the mass hierarchy \citep{Voit2005}. In contrast to galaxies such as the Milky Way, in which several components (thermal plasma, turbulence, cosmic rays, magnetic fields) share similar energy densities -- complicating the theoretical understanding of interactions between the various energy forms -- in galaxy clusters, the intracluster medium (ICM) exhibits a well-defined hierarchy of energy densities: for relaxed clusters, the thermal energy of the plasma exceeds that of the bulk kinetic by a factor of about 5--10 \citep{Lau2009}, which itself is about 5--10 times larger than the energy of turbulence \citep{Perrone2026a}, magnetic fields and (potentially) cosmic rays \citep{Ruszkowski2023}. Yet despite this apparent simplicity, the theoretical description of the ICM is far from trivial and there are many puzzles associated with observations of the thermal and non-thermal energy components and their theoretical modelling. Focusing on the most massive clusters (with virial masses $M_{200}\gtrsim10^{15}\,\rmn{M}_\odot$), we identify eleven outstanding key problems: 

\begin{enumerate}
\item Why is there a bimodal distribution of central ICM cooling times? X-ray observations of a heterogeneous archival sample of low-redshift clusters indicate that approximately half of the population is characterized by short central cooling times (defining cool-core systems) and correspondingly low entropies, while the remainder exhibits longer central cooling times, defining non-cool-core clusters \citep{Cavagnolo2009,Hudson2010}. Yet, simulations cannot reproduce this dichotomy for massive galaxy clusters and worse, even have problems simulating massive cool-core clusters at all \citep{Pakmor2023}. Is this failure caused by incomplete modelling of active galactic nucleus (AGN) jet feedback \citep{Weinberger2023}, overly massive central black holes relative to galaxy--supermassive black hole (SMBH) scaling relations \citep{Weinberger2026}, or a scenario in which cosmic ray-induced inverse Compton emission masquerades as thermal bremsstrahlung \citep{Hopkins2025}?

\item Does heating by AGN jets solve the cooling flow problem and if so, which heating process is responsible? In the absence of heating, the hot gaseous atmospheres of clusters are expected to cool and form stars at rates of up to several hundred $\rmn{M}_\odot~\rmn{yr}^{-1}$  \citep{Peterson2006}. Instead, the observed cooling and star formation rates (SFRs) are suppressed to levels well below those predicted by unimpeded cooling flows. Evidently, radiative cooling is balanced by a heating mechanism, which is likely linked to AGN jet-inflated radio lobes that coincide spatially with the X-ray cavities, though its precise nature remains uncertain. The interaction between gas cooling, ensuing star formation, and nuclear activity thus appears to be governed by a tightly coupled, self-regulated feedback cycle \citep{McNamara2007,McNamara2012}. It is an important challenge to identify the dominant heating process \citep{Meenakshi2026}. Among the many suggested mechanisms are dissipation of mechanical heating by lobes, sound waves, or turbulence \citep{Churazov2001,Brueggen2002,Kunz2011a,Zhuravleva2014b}, turbulent mixing of plasma with different temperatures \citep{Kim2003}, cosmic ray heating \citep{Guo2008,Pfrommer2013,Jacob2017b}, and shock heating \citep{Nulsen2005}. Major AGN outflows move baryons beyond the virial regions of groups and low-mass clusters, causing a back-reaction on the cold dark matter (DM) distribution, thereby modifying the matter power spectrum \citep{vanDaalen2011}. This poses a challenge for deriving cosmological parameters from weak-lensing surveys \citep{Semboloni2011} such as EUCLID and ROMAN, which seek to measure the matter power spectrum with unprecedented precision, and underscores the need for a more detailed understanding of how AGN energy couples to the plasma.

\item Why is the turbulent energy density in the bulk of the ICM only a few per cent of that in the thermal plasma in quiescent clusters? Previous cosmological cluster simulations predicted turbulent energies at the level of 20--40 per cent \citep{Vazza2011a,Miniati2015b,Schmidt2016,Groth2025}, but recent observations of turbulent line broadening in the cluster centres with the XRISM X-ray micro-calorimeter indicate a turbulent pressure support of $2{-}4$ per cent \citep{XRISMCollaboration2025,XRISMCollaboration2025a}. Is this because of an erroneous attribution of bulk flows to turbulence \citep{Perrone2026a}, because of observational biases \citep{Vazza2026}, or because of numerical limitations of previous simulations?

\item The plasma in galaxy clusters is weakly collisional, so heat and momentum should be transported along magnetic field lines as described by Braginskii magnetohydrodynamics \citep[MHD;][]{Braginskii1965,Schekochihin2006}. Are there observable consequences of this plasma property? Anisotropic heat conduction in combination with magnetic draping at cold fronts \citep{Dursi2008} implies sharper contact discontinuities in agreement with X-ray observations \citep{ZuHone2013a}. Anisotropic viscosity, meanwhile, should suppress the growth of the Kelvin--Helmholtz instability at tangential discontinuities \citep{Berlok2019a}, which is in qualitative agreement with X-ray observations \citep{ZuHone2015a}. Finally, Whistler--electron interactions are expected to suppress thermal conduction along magnetic field lines relative to the classical Spitzer value \citep{Roberg-Clark2016,Drake2021}. Cosmological simulations will be required to understand the overall impact of these plasma processes on the (thermo-)dynamics and observable structures \citep{Talbot2025}.

\item What is the origin of cluster magnetic fields and how do they scale with redshift and cluster mass? Because of the weakly collisional property of clusters today, this theoretically necessitates the inclusion of kinetic plasma effects \citep{Schekochihin2009}. In the resulting plasma dynamo scenario, magnetic seed fields in the ICM would arise from plasma instabilities: turbulence in a weakly collisional plasma drives pressure anisotropies that trigger the Weibel instability \citep{Zhou2022,Zhou2024}, while kinetic firehose and mirror instabilities transfer magnetic energy to larger scales and provide effective viscosity at Larmor scales, enabling subsequent amplification by the small-scale dynamo \citep{StOnge2018,Squire2019,StOnge2020}. However, this picture assumes that clusters are weakly collisional throughout their entire cosmic history which is not true during proto-cluster formation at redshifts around six. During this time, magnetic seed fields can be amplified via a small-scale fluctuating dynamo \citep{Kazantsev1968} when the plasma is still collisional. Galactic feedback grows the magnetic coherence length to scales of several tens of kpc, so that it always exceeds the particle mean free path, justifying the use of MHD for studying the cluster dynamo \citep{Tevlin2025}. Such scenarios are best tested by comparing synthetic Faraday rotation measure (RM) observations from cosmological cluster simulations against real observational data.

\item What is the origin of radio halos? Early studies identified a correlation between radio halo power and X-ray luminosity, along with an apparent bimodal distribution separating radio halos from radio-quiet clusters at fixed X-ray luminosity \citep{Brunetti2009}. However, analyses based on early Planck cluster catalogue data revealed a global scaling between radio halo power and the integrated Sunyaev--Zel'dovich (SZ) signal, without clear evidence for such a bimodality \citep{Basu2012}. This discrepancy likely reflects an X-ray selection bias toward lower-mass, cool-core systems, which are typically radio-halo quiet. Hybrid models reproduce this behaviour \citep{Zandanel2014}: in these scenarios, central radio mini-halos arise from electrons produced via hadronic cosmic-ray interactions with ICM protons \citep{Pfrommer2004}, while the more extended halo component requires additional energization, possibly through turbulent re-acceleration \citep{Brunetti2014}. Radio observations of mass-selected cluster samples offer a more nuanced picture \citep{Cuciti2021,Cuciti2023}: radio halos are predominantly found in merging clusters, with their power correlated with cluster mass. In contrast, clusters lacking radio halos are generally relaxed, and upper limits on their diffuse emission fall below the established correlation. The fraction of clusters hosting radio halos declines from $\sim$70 per cent at high masses to $\sim$35 per cent at masses $M_{500}<6\times10^{14}\,\rmn{M}_\odot$, where a bimodality in radio emissivity becomes apparent. Recent observations have revealed extended radio halos surrounding mini-halos \citep{vanWeeren2024}, as well as halos reaching out to the cluster periphery \citep{Cuciti2022}. These structures appear as smooth extensions of the inner halo component, provided that embedded compact sources are carefully subtracted \citep{Rajpurohit2025}. Identifying the plasma physical cosmic-ray transport processes powering radio halos \citep{Ensslin2011} and how this relates to the cosmological cluster growth remains a challenge \citep{Pinzke2017}.

\item Do radio relics indeed trace merger-driven shock waves, and how can they be used to probe the physics of cluster peripheries? In particular, how are electrons accelerated at these low-Mach number shocks relative to the high-Mach shocks observed at supernova remnants? The association between relics and merger shocks, as originally proposed by \citet{Ensslin1998}, now forms the basis of the standard model for radio relics \citep[see e.g.][and references therein]{Brunetti2014, vanWeeren2019}. However, recent high-resolution radio observations have exposed several tensions, including the fact that (1) Mach numbers inferred from radio data are generally higher than the X-ray-derived equivalents \citep{Wittor2021,Lee2025}; (2) cooling-length arguments imply magnetic fields an order of magnitude stronger than those in the surrounding ICM \citep{vanWeeren2010}; and (3) downstream spectral index evolution appears inconsistent with standard cooling models \citep{Rajpurohit2020}. These discrepancies can be reconciled if the interaction between merger and cluster accretion shocks is taken into account, as well as the subsequent propagation of the shock-compressed sheet through density fluctuations \citep{Whittingham2026a}. In this scenario, radio emission preferentially traces the high-Mach number tail of the distribution of shock strengths, while X-rays probe the mean (addressing point 1). Meanwhile, shear motions, induced partially by hydrodynamic instabilities, are able to amplify the peak magnetic field strength to $\upmu$G levels. These are, in turn, preferentially probed by synchrotron emission (addressing point 2). Finally, the same motions redistribute electrons of different ages, producing the observed spectral variations (point 3). Whilst this scenario has been successful thus far, key questions remain: for example, radio polarization appears to suggest a quasi-perpendicular shock geometry \citep[where the shock front is aligned with the magnetic field;][]{vanWeeren2010} but supernova remnant studies indicate that efficient electron acceleration to radio-emitting energies takes place primarily at quasi-parallel shocks \citep{Winner2020}. Furthermore, relic luminosities cannot be explained by diffusive shock acceleration of thermal electrons alone, but instead require a pre-existing population of fossil electrons \citep{Markevitch2005,Kang2011,Pinzke2013}. It remains unclear whether such electrons -- potentially injected by structure formation shocks or AGN activity \citep[see citations given in section 1 of][]{Whittingham2026a} -- are sufficiently abundant, and what role AGN jets play in establishing the fossil electron reservoir in galaxy clusters. Finally, relics exhibit a diverse range of features and morphologies. Modelling them  in cosmological \citep{2023Boess} and high-resolution idealised simulations \citep{Whittingham2026b} is required to understand the cause of this.

\item Why have galaxy clusters not yet been detected at gamma-ray energies? Early work using idealized ICM models \citep{Colafrancesco1998,Pfrommer2004} and later cosmological simulations including cosmic-ray acceleration at structure formation shocks and supernovae with advective cosmic-ray transport \citep{Pfrommer2007,Pfrommer2008,Pinzke2010} predicted observable emission. However, extensive observational efforts have yielded no detections from the bulk of clusters \citep{Ackermann2010,Ackermann2014,Arlen2012,Aleksic2012b,Ahnen2016}. Reducing acceleration efficiencies in more physically complete models -- incorporating cosmic-ray streaming and diffusion from dense cores to cluster outskirts -- can suppress gamma-ray emission below current observational limits \citep{Ensslin2011,Wiener2013,Wiener2019}. However, it remains to be seen whether the proposed solutions hold in cosmological MHD simulations with modern treatments of cosmic-ray transport \citep[e.g.,][]{Thomas2019,Diesing2026}.

\item How can we improve our understanding of transformational processes of galaxies in cluster environments? In particular, how do jellyfish galaxies form and how does star formation proceed in the extreme conditions of jellyfish tails? The seminal work of \citet{Gunn1972} established the foundation for understanding galaxy transformation in clusters, identifying ram-pressure stripping of the interstellar medium (ISM) -- driven by the headwind experienced during infall -- as a key mechanism behind the excess of passive ellipticals and lenticulars. High-resolution wind-tunnel simulations are well suited for isolating the roles of individual physical processes. In multiphase ISM, ram pressure efficiently removes low-density gas, truncating the outer disc while compressing the denser inner regions \citep{Tonnesen2009}. The draping of cluster magnetic fields over galaxies and their stripped tails suppresses Kelvin--Helmholtz instabilities and promotes star formation \citep{Pfrommer2010, Ruszkowski2014, Sparre2024a, Lee2026}. While dense molecular clumps in the near wake primarily originate from stripped ISM, filaments in the distant tail arise from mixing between the circumgalactic and intracluster media with cold ISM, followed by radiative cooling \citep{Tonnesen2021, Lee2022, Sparre2024b}. In contrast, cosmological simulations are indispensable for population-level studies \citep{Yun2019, Rohr2023}, while recent observations indicate that ICM shocks may be key to explaining the longest known tails of ram-pressure-stripped galaxies \citep[e.g.,][]{Edler2026}, highlighting the importance of modelling galaxies within realistic cosmological environments.

\item What fraction of stars resides in cluster galaxies versus the intracluster light? Simulations to date tend to predict brighter intracluster light than observed. In cosmological runs, stars are preferentially stripped as satellites traverse dense cluster regions, affecting both surviving and disrupted systems \citep{Jeon2026}. The amount of intracluster light produced is rather insensitive to resolution and integration accuracy once the galaxies that dominate the stellar mass budget are reasonably well resolved, and its mass fraction is nearly independent of halo mass \citep{Puchwein2010}. The tension with observations may partly reflect difficulties in detecting faint, extended intracluster light at large radii: its diffuse nature often necessitates stacking and is easily lost in sky background noise \citep{Zibetti2005}, while even deep imaging leaves its boundary ill-defined \citep{Montes2022,Brough2024}. Using consistent techniques in both observations and simulations may partly alleviate these tensions \citep{Montenegro-Taborda2025}. Conversely, tidal stripping of low-mass galaxies is sensitive to numerical resolution \citep{Lovell2025}, likely affecting the satellite population and detailed properties of the intracluster light. In simulations, these trends are often superimposed on a resolution-drift of the star-formation efficiency, complicating the analysis. In addition, some intracluster stars may form in situ within the ICM, e.g., in cold gas clouds of multiphase jellyfish tails, a channel still only partially captured in current cosmological simulations.

\item Are galaxy groups and clusters effectively baryonically closed, as was long assumed, such that their baryon-to–DM ratio matches the cosmic mean? The tension between early-Universe measurements of the baryon fraction from the cosmic microwave background on large scales \citep{Komatsu2011,Planck2018_Cosmology} and late-time constraints from X-ray and SZ observations on group and small cluster scales \citep[e.g.,][]{Gonzalez2015,Eckert2021} constitutes one of the most persistent challenges to the cold DM paradigm, commonly known as the missing baryons problem. While previous SZ observations and simulations concluded that massive clusters with virial masses $M_{200}\gtrsim10^{15}\,\mathrm{M}_\odot$ are effectively closed systems with closure radii comparable to their virial radii \citep{Chiu2018,Ayromlou2023}, the baryon fraction at the group mass scale $M_{200}\sim(10^{13}$--$10^{14})\,\mathrm{M}_\odot$ remains very uncertain. \citet{Popesso2026}, who stack eROSITA data of a nearly complete sample of optically selected groups, find lower average gas fractions in the group regime in comparison to earlier studies of bright X-ray groups \citep{Sun2009,Lovisari2015}. This finding is confirmed by studies of the kinetic SZ effect \citep{Bigwood2024,Hadzhiyska2025,Hadzhiyska2026,Qu2026}. Interestingly, galaxy simulation projects that differ in how AGN feedback is implemented arrive at different gas fractions at the group scale (\citealt{Ayromlou2023} using data of EAGLE: \citealt{Crain2015,Schaye2015}; IllustrisTNG: \citealt{Weinberger2017a,Pillepich2018a}; SIMBA: \citealt{Dave2019}). It remains to be seen whether simulations with improved baryonic physics can illuminate this problem.
\end{enumerate}

To address these questions, we present the PICO-Cluster project: a large sample of zoom-in simulations of massive cosmological galaxy clusters ($M_{200}\gtrsim10^{15}\,\mathrm{M}_\odot$). The overarching objective of this project is to explore, in a controlled and systematic manner, how the inclusion of additional plasma-physical processes alters the thermodynamic and dynamical properties of galaxy clusters relative to our base model, the IllustrisTNG galaxy formation framework \citep{Weinberger2017a,Pillepich2018a}, and how this depends on the formation history and cosmological environment. In particular, we aim to explore how synthetic observations derived from our various simulation models can be leveraged to gain deeper insight into the underlying physical processes. These processes include anisotropic viscosity \citep{Berlok2020}, anisotropic heat conduction \citep{Talbot2025}, cosmic-ray populations of both protons \citep{Pakmor2016CRs,Pfrommer2017a} and electrons \citep{Winner2019,Whittingham2026a}, active galactic nucleus feedback implemented through low-density, momentum-driven jets \citep{Weinberger2017b,Weinberger2023,Ehlert2023,Jlassi2026}, as well as an accurate measurement of turbulence driven by cluster mergers and AGN feedback using real-space filtering with the \textsc{Turbocluster} library \citep{Perrone2026a}.

For the 24 baseline PICO-Cluster simulations we adopt the IllustrisTNG galaxy formation model. Our simulations are thus broadly comparable to the TNG300 and TNG-Cluster simulations \citep{Nelson2019a,Nelson2024}. They differ, however, in three key aspects: (i) improved initial conditions that ensure a contamination-free high-resolution region out to $2.7\,R_{200}$; (ii) the use of the \textsc{Arepo-2} code version, an evolved version of \textsc{Arepo} \citep{Springel2010a} originating from the MillenniumTNG project \citep[MTNG,][see Section~\ref{sec:Arepo} for further technical details]{Pakmor2023}, thereby enabling unprecedented resolution in simulations of massive clusters; and (iii) a distinct scientific focus on plasma processes in the ICM. In contrast to many previous galaxy cluster simulation projects \citep{Hahn2017,Barnes2017b,Henden2018,Bassini2020,Pellissier2023}, the IllustrisTNG model employs a kinetic AGN feedback injection at low accretion rates, which was found to work well at the scale of massive galaxies (\citealt{Weinberger2017a}; see also \citealt{Han2026}, for a zoom-in cluster simulation with kinetic AGN feedback). Within the PICO-Cluster project, we use this as a baseline model but explore more detailed jet models in a companion paper \citep{Weinberger2026b}. PICO-Cluster is complementary to The-Three-Hundred project, which examined how simulations compare to observations in terms of fundamental galaxy cluster properties and scaling relations \citep{Cui2018} as well as the DARKSKIES project, a suite of 100 galaxy clusters simulated with self-interacting DM \citep{Harvey2025}.

The outline is as follows. In Section~\ref{sec:methodology}, we present our computational code and galaxy formation model. In Section~\ref{sec:ICs}, we detail the construction of our initial conditions that prevent the problem of low-resolution particles contaminating the zoom-in regions. In Section~\ref{sec:verification}, we verify our simulations by comparing to previous successful simulation projects, discuss numerical convergence, and validate the simulated clusters against observational scaling relations and thermodynamic profiles. We discuss magnetic field growth in our cluster sample and correlations of magnetic, thermal and kinetic energies in Section~\ref{sec:magnetic_fields}, and study the emerging Faraday rotation measure signal in Section~\ref{sec:FRM}. In Section~\ref{sec:Braginskii}, we study viscous heating estimates using Braginskii viscosity theory, and we conclude in Section~\ref{sec:conclusions}.

\section{Methodology}
\label{sec:methodology} 

\subsection{General considerations}
\label{sec:general_considerations}

PICO-Cluster is a suite of cosmological zoom-in simulations targeting massive galaxy clusters. The project is based on a large parent simulation of a periodic cosmological volume with a comoving side length of $1~h^{-1}\mathrm{Gpc}$, within which 543 clusters with masses $\Mvir > 10^{14.9}\,\mathrm{M}_\odot$ are identified at redshift $z=0$. From this sample, 24 systems are re-simulated at higher resolution using the zoom-in technique (see Section~\ref{sec:ICs}).

In this paper, we focus exclusively on the baseline simulations adopting the IllustrisTNG galaxy formation framework (see details in Section~\ref{sec:illustris-tng}), which serves as the reference model for this simulation suite, while investigations of alternative physics variants are deferred to future work. We note, however, that the first successes in answering our scientific questions, as presented in Section~\ref{sec:introduction}, have already been achieved using some clusters of our PICO-Cluster suite: (i) four clusters have been re-simulated with non-radiative MHD to study the impact of galaxy formation physics on the cluster magnetic dynamo \citep{Tevlin2025}, (ii) one cluster has been re-simulated with two different variants of anisotropic thermal conduction \citep{Talbot2025}, and one late merging cluster has been examined to (iii) investigate the interplay between merger and accretion shocks in the context of radio relic formation \citep{Whittingham2026a}, and (iv) to characterise intracluster turbulence \citep{Perrone2026a}.

Throughout this work, all cluster masses are defined as $M_{200} = \tfrac{4}{3} \uppi R_{200}^3 \times 200 \rho_\mathrm{crit}$, where $R_{200}$ is the radius of a sphere centred on the cluster within which the average density is 200 times the critical density of the Universe, $\rho_\rmn{crit}$ (and an analogous definition for $M_{500}$). We assume that plasma effects equilibrate the electron and ion temperatures on small enough scales so that we always have an ion-electron thermal equilibrium: $T_\rmn{e}=T_\rmn{i}=T$ on our resolved scales. We have adopted a $\Lambda$ cold dark matter ($\Lambda$CDM) cosmology with parameters inferred by the \citet{Planck2018_Cosmology}. We use the following cosmological parameters: $\Omega_\mathrm{m}=0.316$, $\Omega_\mathrm{b}=0.049$, $\Omega_\Lambda=0.684$, $\sigma_8=0.81$, $n_\mathrm{s}=0.97$, and the Hubble constant is $H_0 = 100\, h\, {\rm km \, s^{-1} \, Mpc^{-1}}$ with $h = 0.673$.

\subsection{Simulation code \textsc{Arepo} and MHD}
\label{sec:Arepo}

The PICO-Cluster simulations were carried out with the moving-mesh code \textsc{Arepo} \citep{Springel2010a, Pakmor2016a, Weinberger2020}. Collisionless components, including DM, stars, and SMBHs, are modelled as discrete particles. Gravitational forces are calculated using a TreePM approach \citep{Springel2005_Gadget}, which separates long- and short-range contributions. Long-range forces are obtained via Fourier transforms on two meshes, one spanning the entire box and one covering the high-resolution region, while short-range forces are computed using an oct-tree algorithm \citep{Barnes1986}.

In \textsc{Arepo}, MHD quantities are represented on an unstructured mesh defined by the Voronoi tessellation of a set of discrete, freely movable points. When these mesh-generating points follow the gas motion, the code exhibits the benefits of Lagrangian schemes. The Voronoi mesh can be adaptively refined or de-refined and we chose the refinement strategy of maintaining cell masses within a factor of two of a specified target (for specific values adopted in each zoom level, see Section~\ref{sec:ICs}). Consequently, spatial resolution naturally adapts to structural complexity, concentrating computational effort in dynamically active regions. We use an updated code version, \textsc{Arepo}-2, which evolved from the MillenniumTNG project \citep[MTNG,][]{Pakmor2023}. \textsc{Arepo}-2 features advances in memory management, load-balancing efficiency, and incorporates the \textsc{Subfind}-HBT algorithm \citep{Springel2021}, which enables the storage of group and subhalo properties at a higher cadence than full simulation snapshots and facilitates the efficient construction of merger trees.\footnote{The original \textsc{Subfind} algorithm \citep{Springel_SubfindOriginal_2001} in \textsc{Arepo} is prohibitively expensive at zoom factor 24 (hereafter Z24; see resolution overview in Table~\ref{tab:resolution_levels}), which corresponds to a 24 times better spatial resolution and a $24^3$ better mass resolution. Completing these simulations was only possible thanks to the more efficient \textsc{Gadget4} implementation \citep{Springel2021} now ported to \textsc{Arepo}-2.} In this work, we employ on-the-fly merger trees to track the centres of galaxy clusters \citep{Springel2022}.

\textsc{Arepo} evolves the MHD equations in comoving coordinates using a second-order finite-volume Godunov scheme \citep{Pakmor2011, Pakmor2013} with an HLLD Riemann solver \citep{Miyoshi2005}. To control divergence errors in the magnetic field, we employ the Powell 8-wave scheme \citep{Powell1999}, which advects magnetic divergence with the flow via non-conservative source terms, improving numerical accuracy and maintaining stability. The comoving implementation has been verified against analytic solutions for the comoving Alfvén and magnetosonic waves \citep{Berlok2022}.
Simulations using this MHD approach successfully reproduce analytic growth rates and magnetic curvature statistics expected for a fluctuating dynamo in a forming galaxy \citep{Pfrommer2022}, as well as a variety of observational diagnostics such as Faraday rotation measures (FRMs) in spiral galaxies \citep{Pakmor2018} and the Milky Way \citep{Reissl2023}. The Auriga simulations run with \textsc{Arepo} reproduce observed equipartition magnetic field strengths\footnote{Equipartition magnetic field strengths seem to be robust estimates of true field values \citep{Chiu2025}.} and radial profiles across cosmologically forming disk galaxies \citep{Pakmor2014,Pakmor2017}, dwarf and massive galaxies, galaxy groups \citep{Pakmor2024}, and the circumgalactic medium \citep{Pakmor2020}. Adopting steady-state modelling of cosmic ray spectra \citep{Werhahn2021a,Werhahn2026} provides another test. Here, the model successfully reproduces observed radio synchrotron spectra and correlations with the far-infra-red luminosity \citep{Werhahn2021c,Pfrommer2022} and the observed intensity and polarization maps of edge-on galaxies \citep{Chiu2024}.

In the PICO-Cluster simulations, we initialize a uniform magnetic seed field with a comoving strength of $10^{-14}~\mathrm{G}$ at $z=127$, sufficient to reach saturation in Milky-Way-mass galaxies by $z\sim1.5$ at our numerical resolution \citep{Pakmor2014} while remaining too weak to artificially suppress star or galaxy formation \citep{Marinacci2016}. This choice, though larger than expected from cosmological Biermann battery processes \citep{Kulsrud2008}, is justified by our finite numerical resolution. Across a wide range in strengths, the initial seed field has minimal impact on subsequent evolution, as dynamo amplification rapidly erases memory of the initial field and leads to similar saturation values \citep{Pakmor2013,Pakmor2014,Liu2022}. 

\subsection{The IllustrisTNG galaxy formation model}
\label{sec:illustris-tng}

We model radiative physics using the IllustrisTNG galaxy formation model \citep{Weinberger2017a,Pillepich2018a}, which incorporates star formation and associated feedback \citep{Vogelsberger2013,Pillepich2018a}, seeding and growth of SMBHs, AGN feedback \citep{Springel2005, Springel2005_BHs,Weinberger2017a}, and gas cooling and heating \citep{Vogelsberger2013}. Simulations employing this model successfully reproduce observed stellar mass functions in galaxy groups and clusters \citep{Pillepich2018a}, as well as galaxy clustering statistics, including two-point correlation functions of low-redshift galaxies consistent with Sloan Digital Sky Survey measurements \citep{Springel2018}.

Star formation and the ISM are modelled with an effective subgrid approach \citep{Springel2003}. Star particles form stochastically once the hydrogen density exceeds $n_\mathrm{H} > 0.106~\mathrm{cm}^{-3}$, calibrated to match the observed Kennicutt-Schmidt relation \citep{Schmidt1959,Kennicutt1989} and assuming a Chabrier initial mass function \citep{Chabrier2003}. Mass loss and metal enrichment\footnote{Metallicity is defined as the ratio of metal mass-to-total cell mass, $Z = M_Z/ M_\mathrm{total}$. We adopt a solar metallicity of $\mathrm{Z}_\odot = 0.0127$, corresponding to a metal mass fraction of 1.27 per cent in the Sun.} are calculated at each time step according to the initial mass function \citep{Pillepich2018a}. Stellar feedback is implemented via a kinetic wind model: wind particles are launched isotropically from star-forming cells with velocities correlated to the local DM velocity dispersion and adjusted for redshift. They  decouple from hydrodynamics, interacting only gravitationally, and recouple when encountering gas below a density threshold ($n_\mathrm{H} < 0.0053~\mathrm{cm}^{-3}$, corresponding to 5 per cent of the star formation threshold) or after a travel time of $0.025/H(z)$, where $H(z)^{-1}$ is the Hubble time. At high redshift, the shorter Hubble time and higher critical densities make the time criterion dominant. Upon recoupling, wind particles deposit mass, metals, thermal energy, and momentum into the surrounding gas.

Gas cooling and heating follow the implementation of \citet{Vogelsberger2013}. Primordial cooling and heating are treated via ionization equations accounting for recombination, collisional ionization, and cooling rates \citep{Cen1992,Katz1996}, along with Compton cooling off the cosmic microwave background \citep{Ikeuchi1986}. Metal-line cooling is computed using CLOUDY tables \citep{Wiersma2009}. A spatially uniform, redshift-dependent ionizing ultra-violet background is included \citep{FaucherGiguere2009}, with self-shielding corrections applied in dense regions \citep{Rahmati2013}.

AGN physics follows the IllustrisTNG implementation \citep{Springel2005,Vogelsberger2013,Weinberger2017a}. Each halo exceeding a friends-of-friends mass of $5  \times 10^{10}~h^{-1}\mathrm{M}_\odot$ is seeded with a SMBH of mass $8  \times 10^{5}~h^{-1}\mathrm{M}_\odot$ at the location of its densest cell. SMBHs accrete gas from their surroundings assuming the Bondi--Hoyle formalism \citep{Bondi1944,Bondi1952}, growing in mass while converting a fraction of the accreted rest-mass energy into feedback. The feedback operates in a two-mode scheme depending on the accretion rate relative to the Eddington rate ($\dot{M}_{\mathrm{BH}}/\dot{M}_{\mathrm{Edd}}$), with a BH mass-dependent threshold of $\chi = \min\{0.002 \times [M_\mathrm{BH}/(10^8 \mathrm{M}_\odot)]^2, 0.1\}$. Below this threshold, kinetic AGN feedback drives SMBH winds \citep{Weinberger2017a}; above it, quasar-mode feedback releases thermal energy isotropically to neighbouring gas cells \citep{Springel2005,Springel2005_BHs}, which is most prominent at high redshift.

\begin{figure*}
        \includegraphics{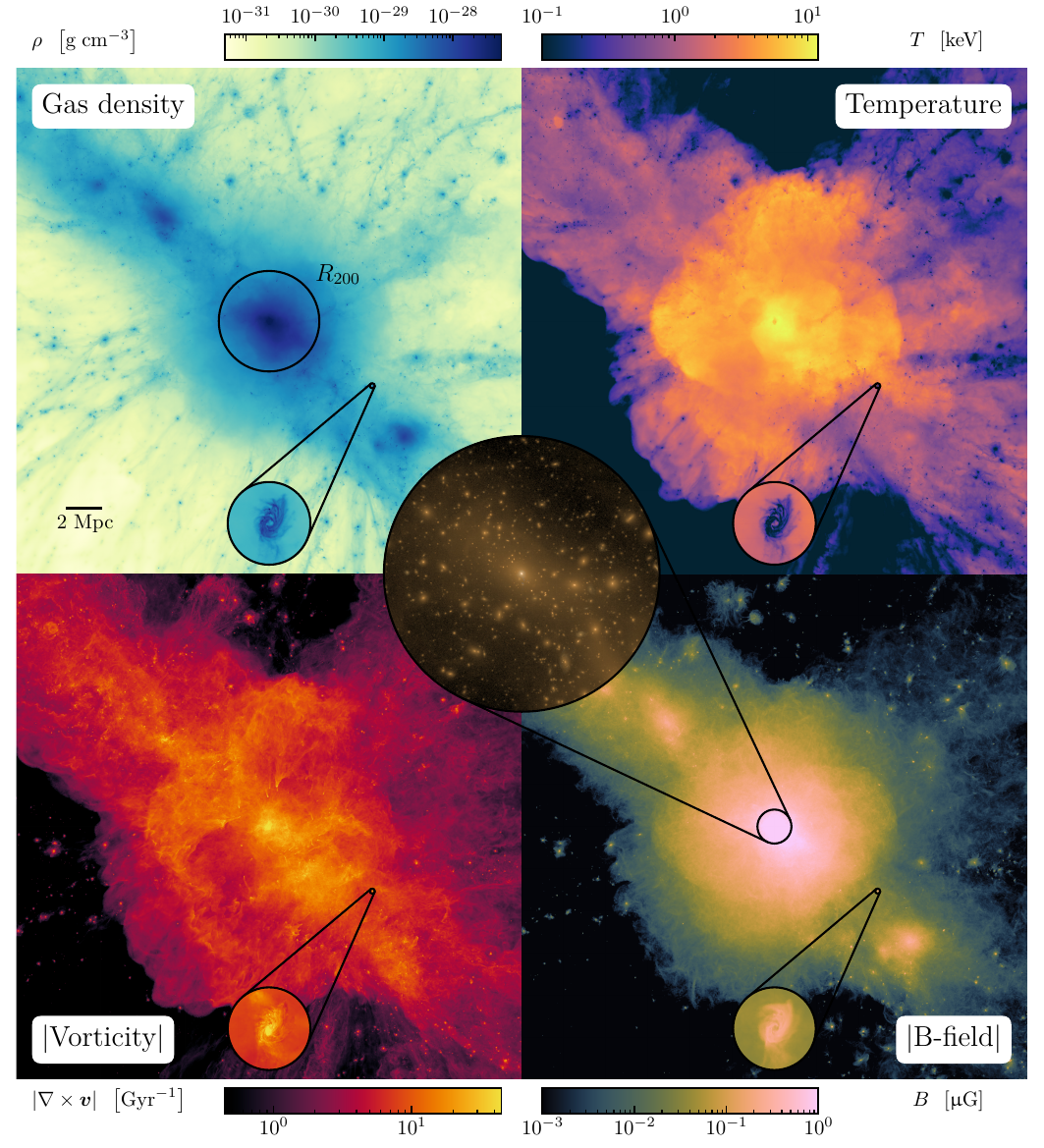}
        \caption{A large-scale view of a very massive ($M_{200} = 2.7\times10^{15}\,\mathrm{M}_\odot$) galaxy cluster (halo 4) simulated at a so far unprecedented resolution for this mass scale (baryonic mass resolution $1.4\times10^{6}\,\mathrm{M}_\odot$) using the IllustrisTNG galaxy formation model. We show deep projections (15 Mpc) of gas density, temperature, vorticity and magnetic field at $z=0$. In the centre, a synthetic SDSS $gri$ composite image of the stellar light shows a collection of mainly red, elliptical galaxies, as expected for a massive galaxy cluster, as well as a diffuse intracluster light component. Magnetic fields, turbulence and temperature are some of the key quantities that are affected by plasma scale processes, and thus of prime interest for the PICO-Cluster project. The zoom-in view shows an individual spiral galaxy external to the cluster. The black circle in the density panel shows the virial radius $R_{200}$. Similar images for all PICO-Clusters can be accessed via the \href{https://pico-cluster.aip.de}{PICO-Cluster website} and the Supplementary Material to this paper.}
        \label{fig:zoom24_halo4}
\end{figure*}

Figure~\ref{fig:zoom24_halo4} shows the environment of a very massive galaxy cluster ($M_{200} = 2.7\times10^{15}\,\mathrm{M}_\odot$) at a resolution previously unattained for such systems when employing a realistic galaxy formation model. The vorticity projection shows tails of jellyfish galaxies as well as turbulence generated within the accretion shocks that are clearly visible as discontinuities in the temperature map at cluster-centric radii ${\sim}2\,R_{200}$. The magnetic field exhibits a smooth distribution that correlates well with the gas density on these scales and shows similar discontinuities as seen in temperature and vorticity, suggesting its origin in a small-scale fluctuating dynamo that is driven by turbulence generated by processes at the accretion shocks \citep{Tevlin2025}. 

In the centre of Figure~\ref{fig:zoom24_halo4}, we show a synthetic SDSS $gri$ composite image of the stellar light, which indicates a population of mostly quenched galaxies, as expected for massive galaxy clusters. To make this, we follow the method outlined in \citet{Whittingham2021}, where photometric properties of the star particles are mapped to the RGB channels, scaled according to algorithms in \citet{Lupton2004}, and given an additional transparency factor set proportional to the maximum binned $V$-band luminosity.

\section{Initial conditions for the PICO-Cluster simulations}
\label{sec:ICs}

\subsection{Parent simulation and halo selection}

\begin{figure*}
    \includegraphics[width=\textwidth]{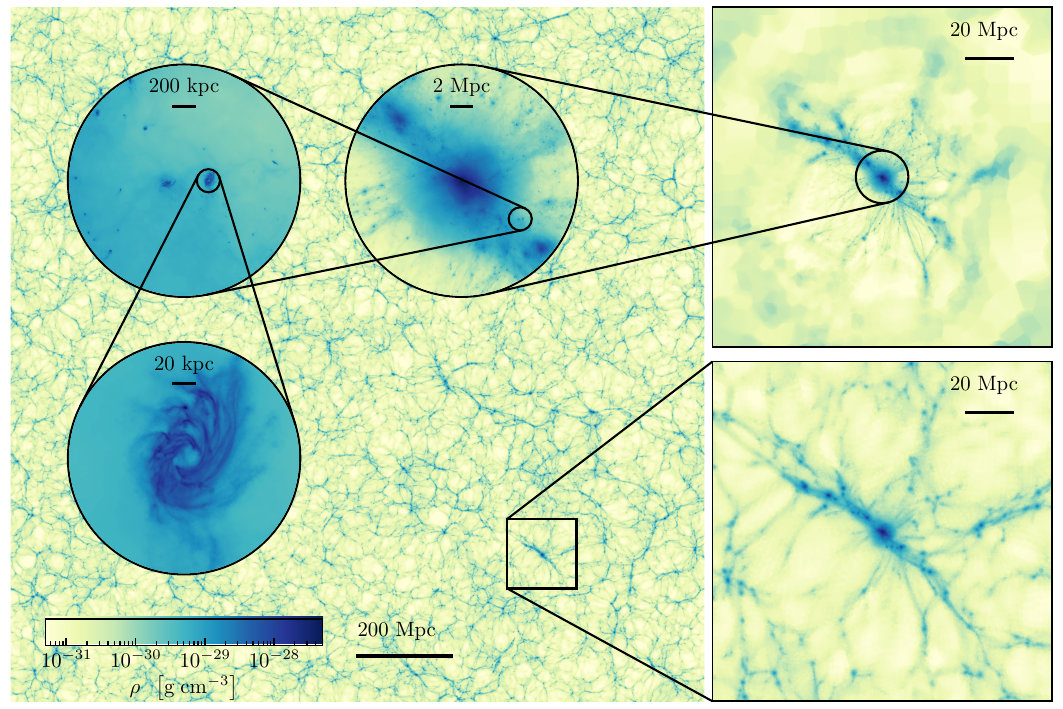}
        \caption{Illustration of the zoom-in technique. In the background, we show a thin gas projection (10~$h^{-1}$Mpc) of the full width (1~$h^{-1}$Gpc) corresponding to 1 per cent of the volume of the parent simulation at $z=0$. The cosmic web is seen to contain a myriad of knots at which massive galaxy clusters reside. The lower right panel zooms in on a string of clusters found in the parent simulation. On the top right, we show the IllustrisTNG re-simulation using the zoom-in technique of halo~4, which allows the galaxy cluster to be better resolved by a zoom factor 24. The corresponding zooms into the cluster and a disc galaxy outside of the cluster demonstrate the superior resolutions achieved with this technique.}
        \label{fig:parent_box}
\end{figure*}

Our PICO-Cluster halos were selected from a dedicated full-volume parent simulation box with a comoving side length of $1~h^{-1}\mathrm{Gpc}$. We have simulated this parent box with the \textsc{Arepo} code using $1024^3$ DM particles and the same initial number of gas cells, but following only self-gravity and non-radiative MHD. A cosmology consistent with that provided by the \citet{Planck2018_Cosmology} was adopted (cosmological parameters stated in Section~\ref{sec:general_considerations}). The initial conditions for the parent simulation were generated using the \textsc{N-GenIC} code \citep{Springel_N-GenicASCL_2015}, based on a matter power spectrum computed with the \textsc{CAMB} code \citep{Lewis_CambCode_2000}.

Figure~\ref{fig:parent_box} shows a projection of the gas density of the cosmic web using 1 per cent of the volume of the parent simulation at $z=0$ (image dimensions are $1~h^{-1}\mathrm{Gpc}\times1~h^{-1}\mathrm{Gpc}$ with a depth of $10~h^{-1}\mathrm{Mpc}$). The projection is centred along the line-of-sight on halo 4 (the fifth most massive cluster in the cosmological volume).\footnote{This cluster was selected because it was the subject of a resolution study.} The inset on the lower right shows a projection of the vicinity of the cluster in the parent simulation. The mass resolution of the parent simulation is $M_\mathrm{DM} = 1\times10^{11}\,\mathrm{M}_\odot$, and while the halo-4 cluster contains ${\approx}2.2\times10^4$ DM particles inside $\Rvir$, the Voronoi cells are so large outside $\Rvir$ that their outlines are visible. The zoom-in technique, explained in detail in Section~\ref{sec:particle_load}, concentrates the resolution inside a high-resolution region and reduces the resolution everywhere outside. Our highest resolution zoom-in simulation uses almost as many resolution elements as the parent simulation to simulate a tiny fraction of the cosmological volume but at $24^3 = 13{,}824$ times better mass resolution. The upper and lower panels on the right-hand side of Fig.~\ref{fig:parent_box} show the same region but in a zoom-in re-simulation (with IllustrisTNG) and the parent simulation (with non-radiative MHD), respectively. The upper right panel clearly displays more small-scale structure inside the high-resolution region, while retaining the correct local cosmological environment and a transition to very low resolution far away from the cluster. We use DM particles with different particle masses to create this transition with low-mass particles inside the cluster, high-mass particles outside of it, and a buffer region in-between consisting of particles with intermediate masses (see Section~\ref{sec:particle_load} for details).

We present in this paper re-simulations at spatial zoom factors 4, 8, 12 and 24, for which the high resolution DM particles have masses that are smaller by factors $4^3$, $8^3$, $12^3$ and $24^3$ than the DM particles used in the parent simulation. The high-resolution DM particle mass and target gas mass can be found in Table~\ref{tab:resolution_levels}. The high-resolution DM, star, and black hole particles share a fixed comoving softening length of twice the values for $\varepsilon_\mathrm{DM}$ displayed in Table~\ref{tab:resolution_levels} at $z > 1$, transitioning to a fixed physical softening of $\varepsilon_\mathrm{DM}$ at $z \leq 1$. Gas softening is adaptive, tied to the local cell size, with a minimum comoving value of a quarter of $\varepsilon_\mathrm{DM}$.

\begin{table}
 \caption{\label{tab:resolution_levels}
 An overview of the resolutions employed in the simulations. The columns show (i) the zoom factor relative to the parent simulation, (ii) the DM particle mass and (iii) baryonic cell mass in the zoom region, (iv) fixed physical gravitational softening length $\varepsilon_\mathrm{DM}$ at $z<1$, (v) corresponding resolutions of cosmological box simulations.}
\small{
\begin{center}
\setlength{\tabcolsep}{2pt}
\renewcommand{\arraystretch}{1.}
\begin{tabular}{cllcl} \hline
~~Zoom~~  & DM mass [$\Msun$] & Gas mass [$\Msun$] & $\varepsilon_\mathrm{DM}$ [$h^{-1}\mathrm{kpc}$] & equivalent \\
factor & & & & resolution \\
1      &  $1.0\times 10^{11}$  &      $1.9\times 10^{10}$ & 19.53 & \\
4      &  $1.6\times 10^{9}$   &     $3.0\times 10^{8}$ & 4.88 & $\sim$TNG300-2.5  \\
8      &  $2.0\times 10^{8}$   &     $3.7\times 10^{7}$ & 2.44 & $\sim$MTNG  \\
12     &   $5.9\times 10^{7}$  &      $1.1\times 10^{7}$ & 1.63 & $\sim$TNG300  \\
24     &   $7.4\times 10^6$      &  $1.4 \times 10^6$ & 0.81 & $\sim$TNG100 \\
\hline
\end{tabular}
\end{center}
}
\end{table}

We select our sample of 24 clusters to have approximately a log-uniform distribution in cluster mass, $M_{200}$, at redshift zero, with clusters chosen solely based on their halo mass in the parent simulation. We sorted all 543 halos with $M_{200} > 10^{14.9} M_\odot$ by mass and assigned each of them a zero-indexed halo number such that halo 0 is the most massive halo. First, we select the five most massive halos, which have $M_{200} > 10^{15.4} \Msun$ in the parent simulation. We then randomly selected four halos in each of the five 0.1-dex mass bins between $10^{14.9} \Msun$ and $10^{15.4} \Msun$ (i.e. 20 halos), but then excluded one burdensome object\footnote{Halo 20 is undergoing a major merger at $z=0$ and has the largest friends-of-friends mass in the parent simulation. As a result, the required high-resolution region is exceptionally large (similar to halos 56 and 542; see Section~\ref{sec:avoiding_contamination}), making the simulation too expensive for this initial study.}.  An overview of the 24 selected clusters is given in Table~\ref{tab:cluster_overview}. We present in this paper re-simulations of these halos at zoom factor 12 (Z12) and a convergence study of halo 4 at zoom factors up to 24 (Z24). In comparison to the IllustrisTNG cosmological volume simulations, Z12 approximately corresponds to TNG300 while our Z24 simulation has similar resolution to TNG100 \citep{Nelson2019a}. The highest mass clusters in TNG100 and TNG50 were $3.8\times10^{14}\, \Msun$ and $1.8\times10^{14}\, \Msun$, respectively, and the resolution used in TNG-Cluster corresponds roughly to our Z12 resolution. Our Z24 simulation, which contains a cluster which is ${\approx} 7$ times more massive than the most massive cluster found in TNG100, is thus the highest resolution IllustrisTNG simulation of a galaxy cluster with $M \gtrsim 10^{15} \Msun$.

\begin{table*}
        \centering
        \caption{This Table shows the characteristics of each cluster that has been simulated in PICO-Cluster with IllustrisTNG physics. The columns show (i) sample number, (ii) spherical overdensity halo number, defined by rank-ordering $M_{200}$ at $z=0$ (or, alternatively, by rank-ordering friends-of-friends (FoF) halo masses at $z=0$), (iii) zoom factor, (iv) $R_{500}$ and (v) $R_{200}$ in the zoom-in IllustrisTNG simulation at $z=0$, (vi) $M_{500}$ and (vii) $M_{200}$ in the zoom-in IllustrisTNG simulation at $z=0$, (viii) $M_{200}$ in the parent non-radiative simulation at $z=0$, (ix) minimum radius of the closest contamination particle in the zoom-in simulation across all output times, (x) number of snapshots/catalogue snapshots in the output, and (xi) the number of high-resolution DM particles in the simulation. }
        \label{tab:cluster_overview}
        \begin{tabular}{llcccccccrr}
                \hline
                \noalign{\vskip 2pt}
                \# & Halo (FoF) & Zoom & $R_{500}$ & $R_{200}$ &
                 $M_{500} $
                & $M_{200} \left[10^{15}\mathrm{M}_\odot\right]$ & $M_{200} \left[10^{15}\mathrm{M}_\odot\right]$ & $\mathrm{min}(R_\mathrm{min}/R_{200})$ & \# full / cat. & $N_\mathrm{dm}$ \\
                \noalign{\vskip 2pt}
                 & & factor & $\left[\mathrm{Mpc}\right]$ & $\left[\mathrm{Mpc}\right]$ & $\left[10^{15}\mathrm{M}_\odot\right]$ & 
                 Zoom-in TNG & Parent non-rad. & & snapshots & $\left[10^{6}\right]$ \\
                 \noalign{\vskip 2pt}
                \hline
                \noalign{\vskip 2pt}
1 	& 0 (8)      	& 12 	& 1.968 	& 3.153 	& 2.009 	& 3.304 	& 3.026      	& 4.57	& 76/306 	& 117.6\\
2 	& 1 (15)     	& 12 	& 1.990 	& 3.025 	& 2.076 	& 2.916 	& 2.882      	& 2.74	& 75/305 	& 137.9\\
3 	& 2 (14)     	& 12 	& 1.812 	& 3.002 	& 1.567 	& 2.851 	& 2.849      	& 3.72	& 75/305 	& 110.7\\
4 	& 3 (11)     	& 12 	& 1.956 	& 2.955 	& 1.972 	& 2.720 	& 2.669      	& 3.12	& 262/307 	& 125.8\\
5 	& 4 (6)      	& 4 	& 1.877 	& 2.960 	& 1.743 	& 2.732 	& 2.664      	& 5.00	& 261/306 	& 5.5\\
 	&            	& 8 	& 1.921 	& 2.957 	& 1.867 	& 2.725 	&            	& 4.42	& 261/306 	& 39.3\\
 	&            	& 12 	& 1.921 	& 2.957 	& 1.867 	& 2.725 	&            	& 4.20	& 261/306 	& 127.5\\
 	&            	& 24 	& 1.925 	& 2.956 	& 1.879 	& 2.721 	&            	& 3.31	& 304/349 	& 976.1\\
6 	& 8 (19)     	& 12 	& 2.026 	& 2.994 	& 2.192 	& 2.830 	& 2.421      	& 4.21	& 75/305 	& 106.0\\
7 	& 9 (7)      	& 12 	& 1.862 	& 2.873 	& 1.699 	& 2.500 	& 2.411      	& 3.93	& 75/305 	& 114.7\\
8 	& 19 (79)    	& 12 	& 1.824 	& 2.715 	& 1.599 	& 2.109 	& 2.062      	& 4.60	& 75/305 	& 78.1\\
9 	& 22 (35)    	& 12 	& 1.723 	& 2.699 	& 1.347 	& 2.072 	& 1.970      	& 3.97	& 75/305 	& 83.8\\
10 	& 33 (128)   	& 12 	& 1.725 	& 2.584 	& 1.353 	& 1.818 	& 1.775      	& 4.84	& 75/305 	& 67.0\\
11 	& 41 (45)    	& 12 	& 1.641 	& 2.541 	& 1.164 	& 1.729 	& 1.709      	& 3.37	& 75/305 	& 84.8\\
12 	& 49 (16)    	& 12 	& 1.572 	& 2.485 	& 1.024 	& 1.617 	& 1.630      	& 3.91	& 75/305 	& 109.6\\
13 	& 56 (65)    	& 12 	& 1.536 	& 2.464 	& 0.955 	& 1.577 	& 1.538      	& 4.75	& 75/305 	& 99.6\\
14 	& 98 (169)   	& 12 	& 1.559 	& 2.348 	& 0.997 	& 1.364 	& 1.293      	& 4.02	& 75/305 	& 59.8\\
15 	& 104 (145)  	& 12 	& 1.526 	& 2.342 	& 0.936 	& 1.354 	& 1.288      	& 3.62	& 75/305 	& 66.4\\
16 	& 112 (155)  	& 12 	& 1.527 	& 2.328 	& 0.939 	& 1.329 	& 1.268      	& 2.91	& 75/305 	& 63.7\\
17 	& 194 (374)  	& 12 	& 1.316 	& 2.141 	& 0.601 	& 1.034 	& 1.105      	& 4.40	& 260/305 	& 47.3\\
18 	& 239 (368)  	& 12 	& 1.404 	& 2.166 	& 0.728 	& 1.071 	& 1.049      	& 4.50	& 75/305 	& 62.6\\
19 	& 260 (321)  	& 12 	& 1.372 	& 2.217 	& 0.681 	& 1.149 	& 1.021      	& 3.79	& 260/305 	& 53.9\\
20 	& 261 (653)  	& 12 	& 1.327 	& 2.110 	& 0.615 	& 0.989 	& 1.019      	& 4.21	& 75/305 	& 41.9\\
21 	& 344 (210)  	& 12 	& 1.361 	& 2.068 	& 0.664 	& 0.932 	& 0.929      	& 4.53	& 75/305 	& 60.3\\
22 	& 417 (776)  	& 12 	& 1.276 	& 2.012 	& 0.547 	& 0.858 	& 0.865      	& 4.41	& 75/305 	& 72.1\\
23 	& 420 (1002) 	& 12 	& 1.262 	& 2.033 	& 0.530 	& 0.885 	& 0.863      	& 2.87	& 75/305 	& 45.5\\
24 	& 542 (127)  	& 12 	& 1.186 	& 1.924 	& 0.440 	& 0.751 	& 0.795      	& 3.80	& 75/305 	& 72.3\\
\hline
        \end{tabular}
\end{table*}

\subsection{Creating the unperturbed particle load for the zoom-in initial conditions}
\label{sec:particle_load}

For each selected halo, we created zoom-in initial conditions using our new \textsc{CosmoZoomIC} code (full details of the code will be provided in Puchwein et al., in prep.). We search for all DM particles within a radius of $5 \, R_{200}$ from the halo centre, as defined by the minimum of the gravitational potential found by the \textsc{Subfind} (sub)halo-finder code \citep{Springel_SubfindOriginal_2001}. This is significantly larger than the splashback radius ($\lesssim 2 R_{200}$, \citealt{More_SplashbackRadius_2015}) and, thus, ensures that all particles that have already crossed the cluster boundary are included. Not including them when finding the Lagrangian region of the cluster would result in contamination at earlier times. Also, such a large selection radius allows us to study the outskirts and environments of the clusters. Using the particle ID numbers of the selected particles, they are then traced back from the selection redshift, 0, to the initial redshift, 127, of the parent simulation to obtain their initial positions. Next, we subtract their initial displacement in the initial conditions, as given by the Zel'dovich approximation, to get their true Lagrangian positions. These are then used to flag the region that will be simulated at high resolution in the zoom-in simulation (see Section~\ref{sec:avoiding_contamination} for some additional modifications that are used to avoid contaminating particles).

To this end, a hierarchical tree structure is created. Its leaf node centres will define the Lagrangian particle positions of the zoom-in simulation. For each level of the tree, a different value for the number of daughter nodes can be set, that is then used for all nodes at that level that need to be refined. This allows more flexibility in the resolution, e.g., compared to using a simple oct-tree. The tree is constructed by starting with a root node that has the same comoving side length as the parent simulation box, $1~h^{-1}\mathrm{Gpc}$. We then refine the tree level-by-level until we reach the minimum spatial resolution that we would like to guarantee for the boundary particles in the zoom-in simulation. This choice will only play a role far away from the selected halo, as the halo's environment will be further refined as described below. For our adopted choice, this results in a tree with $128^3$ leaf nodes. We then refine the tree further in the Lagrangian region of the selected halo (and its environment out to $5 \, R_{200}$) down to the resolution corresponding to the parent simulation. This is done by sequentially inserting the Lagrangian positions of the selected parent simulation particles. We split each node that already contains such a particle into the number of daughter nodes that was chosen for the node's level. The particle originally occupying the node and the newly added particle are then attached to the appropriate daughter nodes based on their positions. As our parent simulation grid dimension is 1024, this results in a tree with leaf node side lengths equal to 1/1024 of the (parent) simulation box side length within the Lagrangian region of the selected halo.

Depending on the resolution that we would like to achieve in the high-resolution region of the zoom-in simulation, we choose a zoom factor (see Table~\ref{tab:resolution_levels}). To achieve the desired resolution, we have to further refine our tree in the selected Lagrangian region. For example, for a zoom factor of 12, we split each node in that region into $12^3$ leaf nodes, which accordingly results in a particle mass in the high-resolution region of the zoom simulation that is smaller by a factor of $12^3$ than the parent simulation particle mass. We perform this splitting using intermediate tree levels with the number of daughter nodes given by the cubes of the prime factors of the zoom factor, e.g., for our example of a zoom factor of 12, we split the relevant nodes into $3^3$, then $2^3$, and then again $2^3$ daughter nodes. We use the smallest prime factor last, as this will result in the lowest resolution difference between the high-resolution region and the nearest layer of boundary particles. Using this approach, we obtain a high-resolution region that can have an arbitrary shape that follows the Lagrangian region of the cluster. This allows us to concentrate the numerical effort in the region where it is actually needed.

We complete setting up the unperturbed particle distribution of the zoom-in simulation by making the transition between the high-resolution region and the coarse resolution grid covering the rest of the box volume in a more gradual manner. This is achieved by refining the tree around the high-resolution region. More precisely, for each leaf node that is not part of the high-resolution region, it is checked whether its side length is larger than the distance of its centre from the nearest edge of the high-resolution region multiplied by a tree-opening-angle of 0.3. If this is the case, the node is refined to the next finer tree level. This procedure is repeated iteratively until no more nodes are found that require refinement. The leaf node centres of this tree define the unperturbed particle positions for the zoom-in run. The final resolution corresponds to the desired zoom factor in the high-resolution region, which is surrounded by several layers of progressively coarser resolution until it reaches the base grid resolution that covers the remainder of the box volume. 
Particles in the innermost boundary layer have a mass that is 8 times larger than that of the high-resolution particles (as the smallest prime factor of all zoom factors we use is 2, see above). 

\subsection{Adding perturbations}

Finally, we need to add perturbations consistent with the starting redshift of the zoom-in simulation to the Lagrangian distribution of particles. We elected to start the zoom-in runs at the same initial redshift, 127, as the parent box. Displacements and velocity perturbations are computed using the Zel'dovich approximation. We start by adding those perturbation modes that were present in the parent simulation by drawing mode amplitudes that are consistent with the assumed matter power spectrum and phases in Fourier space. Importantly, we draw them in the same order as in the \textsc{N-GenIC} code and use the same random number seed as was used to generate the initial conditions of the parent simulation. This ensures that we obtain the same realization and form the same structures as in the parent simulation, just with a locally adjusted resolution. For each tree level of the zoomed particle distribution that is coarser or equal to that in the parent simulation, modes within a sphere in wave number ($k$)-space are included in this manner up to the Nyquist frequency corresponding to the particle spacing on the considered tree level. For tree levels with a smaller spacing, including that covering our high-resolution region, we include modes up to the Nyquist frequency of the parent simulation's initial particle grid in this first step. In both cases, the velocity perturbations are then transformed back to position space and linearly interpolated to the particle positions (if needed). The spatial displacements are then computed from the velocity perturbations, assuming the Zel'dovich approximation.

\begin{figure*}
        \includegraphics[trim = 0 15 0 0]{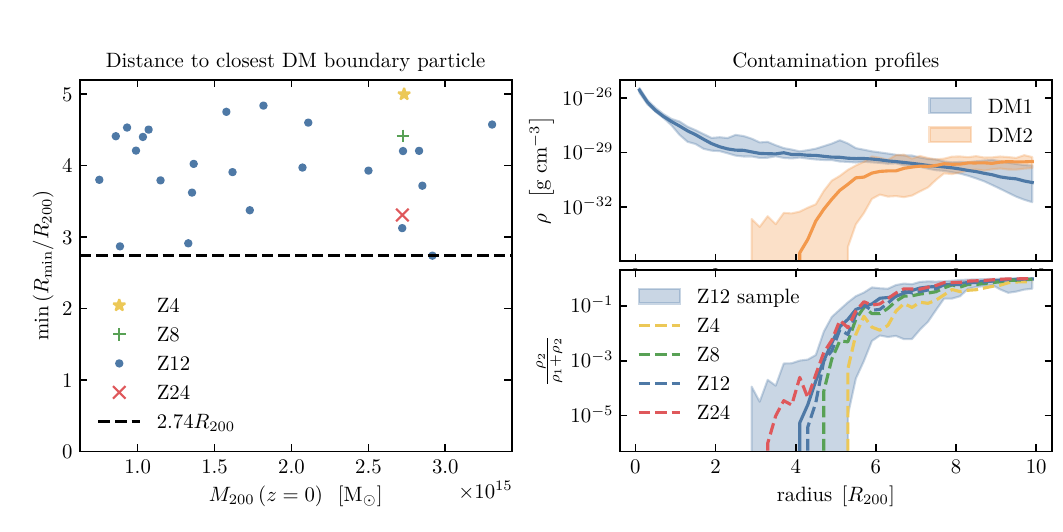}
        \caption{Left panel: minimum cluster-centric radius, $R_\mathrm{min}$, in units of $R_{200}$, of the closest low-resolution DM boundary particle \emph{across all output times} versus the $M_{200}$ mass at $z=0$ in each zoom-in simulation analysed here. These are far outside the splashback zone at ${\sim}2 R_{200}$. Top right panel: median and 16th--84th percentiles of the DM mass density distributions of high-resolution (DM1) and low-resolution DM boundary particles (DM2) for the full sample at $z=0$. Bottom right panel: mass density fraction of low-resolution boundary particles for the full sample and the halo-4 simulations at different resolution. This demonstrates that the contamination radius decreases with increasing resolution. The Z24 simulation is nevertheless contamination-free inside $r/R_{200} = 3.3$ at all times.}
        \label{fig:contamination_radius}
\end{figure*}

Because we simulate our selected clusters in zoom-in simulations at higher resolution, we still need to appropriately add additional small-scale power that was not present in the parent simulation for all tree levels that are finer than the parent simulation resolution. Again, for each tree level, we would like to include modes up to its Nyquist frequency. Doing this with a grid that covers the entire box would, however, require very large Fourier transforms, e.g., for a zoom factor of 24 we would need to transform a $(24 \times 1024)^3 = 24,576^3$ grid, which would be computationally very expensive. We avoid this by computing the small-scale contribution to the perturbations using a grid with a shorter side length, but the same number of grid points, 1024, along each dimension (with the exception of the zoom factor 24 run for which we use 2048). Specifically, we choose an integer fraction of the full box's side length and tile the box with periodic replications of the small-scale perturbations from this grid. Of course, this results in a small-scale perturbation field that is correlated at the scale of this grid's side length. We ensure this side length exceeds the diameter of the high-resolution region, so that it does not overlap with multiple periodic copies of any feature in the small-scale perturbation field. Additionally, when these perturbations are combined with the large scale perturbations, similar features in the small-scale perturbation field would translate into different cosmic structures. We set the side length of the small grid at $1/13$ of the full box, equating to roughly $77~h^{-1}\mathrm{Mpc}$. Modes with $k$-values ranging between the Nyquist frequency of the parent simulation and that of the current (higher-resolution) tree level are then added to the particle displacements and velocities. We note that the wavelength of the longest mode used from the small-scale grid is ${\sim} 1/40$ of the small-scale field side length. This scale separation further limits residual artificial correlations of the small scale perturbation field. At the initiation of the zoom-in simulation, total matter particles in the zoom-in initial conditions are split into their DM and baryonic components, and then spatially displaced diagonally by half a diagonal grid spacing to prevent particle clustering. This approach effectively uses the same mass-weighted transfer function for both baryons and dark matter. Using separate transfer functions is, however, not expected to make a notable difference for massive halos.

\subsection{Avoiding contamination by low-resolution particles}
\label{sec:avoiding_contamination}

In the previous subsections, we have described the basic technique for creating zoom-in initial conditions for the PICO-Cluster simulations. We found, however, that some additional tweaks are needed to ensure that all clusters in the zoom-in simulations, as well as their close environment, are free of contamination with low-resolution particles (see Fig.~\ref{fig:contamination_radius}). Such contamination would reduce the effective resolution in regions traversed by these low-resolution particles and might cause numerical heating of nearby high-resolution particles.

The orbits of individual particles can be stochastic in $N$-body simulations. This can result in particles in the parent simulation not ending up within the selection radius, i.e., within $r<5\, R_{200}$ (see Section~\ref{sec:particle_load}), despite all its initial neighbours doing so. In this case, flagging the Lagrangian region by tracing back the particles found within the selection radius results in a `hole' in the Lagrangian region. Due to small changes in the particle orbit in the zoom-in run compared to the parent simulation, this often results in contaminating low-resolution particles originating from the `hole' region to end up  within the selection radius in the zoom-in simulation.

\begin{figure*}
        \includegraphics[trim= 0 10 0 0, width=\textwidth]{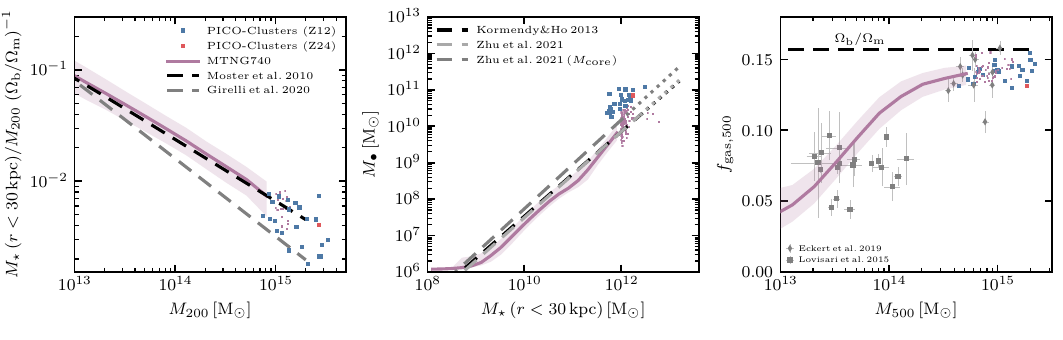}
        \caption{The figure compares integrated quantities of PICO-Clusters to the MTNG simulation and observations at $z=0$. From left to right, we show the high-mass end of the brightest cluster galaxy stellar mass--halo mass relation, the mass of the central SMBH as a function of host galaxy stellar mass (within 30 kpc), and the halo gas-to-total mass fraction within $R_{500}$. Our PICO-Cluster simulations are largely consistent with MTNG; the remaining minor discrepancies can be attributed to the inclusion of MHD relative to the MTNG simulation (see text for details).}
        \label{fig:validation_MTNG}
\end{figure*}

We avoid this by filling such holes as follows. First, we note that a regular grid was used for the unperturbed initial particle positions in the parent simulation. Thus, flagging the Lagrangian region amounts to selecting which of the grid points should become part of the high-resolution region. After this initial flagging, holes are filled by identifying unselected grid points for which at least 14 of their 26 direct neighbours are flagged. If so, we add the grid point in question to the high-resolution region as well. We iteratively repeat this check until no more grid points can be added. This typically results in `holes' and low-resolution `peninsulas' being added to the high-resolution region, while changing the total number of selected particles by only a few per cent. In a second step, we flag one additional layer of grid points to become high-resolution, i.e.\ we add all grid points that are direct neighbours of previously selected points. This increases the selected number of grid points in our case more substantially by ${\sim}20$--$30$ per cent. This updated flagged region is then used for refining the tree as discussed in Section~\ref{sec:particle_load}. This typically results in zoom-in simulations that are free of contamination out to several virial radii. We have tested this with collisionless (DM only) simulations before running the significantly more expensive IllustrisTNG-physics simulations.

We encountered, however, three somewhat pathological cases in which two or three clusters were in an early stage of a major merger at $z=0$. The infalling secondary or tertiary cluster can then redirect particles into the selection (high-resolution) region in a manner that is highly sensitive to small differences between the parent and zoom-in simulations. To avoid contamination in these cases, we decided to include all particles that are either within the selection radius around the centre of the main cluster or the infalling cluster(s). This also allows us to make the infalling cluster(s) free of contamination, so that it can be analysed as well. For halo 542, the three clusters are already in the same friends-of-friends group at $z=0$ and we use a selection radius of 6 times the group's (main halo's) $R_{200}$ around the three centres. For halo 56, using a selection radius of $6 \, R_{200}$ around the main halo centre was sufficient.

Finally, we want to point out that the changes in the included physics, e.g., the feedback in the IllustrisTNG model, results in differences in the gravitational potentials that are large enough to divert some particles to smaller cluster-centric radii compared to the non-radiative parent simulation. As a consequence, the minimum cluster-centric radius of the closest low-resolution boundary particle in the zoom-in simulation is typically somewhat smaller than the selection radius (of $5 \, R_{200}$ in most cases), as is illustrated in Fig.~\ref{fig:contamination_radius}. We find that this contamination radius typically moves in with increasing resolution. This is likely either because of the increased structure of the Lagrangian regions in the initial conditions that better resolve the gravitational force distribution or because of random trajectories resulting from particle-particle scattering despite our conservative choice of the DM softening length. We defer the study of this effect to future work and state that all PICO-Clusters are completely free of contamination within $2.74 \, R_{200}$ across all output times.

\section{Verification and validation of PICO-Clusters}
\label{sec:verification}

In this section, we will verify and validate the PICO-Cluster simulation suite. In this context, verification means that the code correctly solves the mathematical model and thus we focus on internal consistency and precision. To this end, we compare global scaling relations between the masses of different components of our simulations to a re-analysis of the MTNG simulation (Section~\ref{sec: previous sims}) and perform convergence studies (Section~\ref{sec: convergence}). By contrast, validation implies that the model itself accurately represents reality, which is challenging to prove rigorously but we support our model by comparing our results to observational data (Sections~\ref{sec: observational data} and \ref{sec: cluster profiles}).

\subsection{Global scaling relations between halos, stars, SMBHs and the ICM}
\label{sec: previous sims}

In Fig.~\ref{fig:validation_MTNG}, we assess how galaxy, SMBH, and cluster properties of our PICO-Cluster simulations compare to the MTNG simulation and recent observations. The plots represent a subsample of the main calibration plots of the IllustrisTNG galaxy formation model \citep{Weinberger2017a,Pillepich2018a}. However, this model was calibrated in cosmological volumes of side length 25~$h^{-1}$Mpc, which did not allow for the formation of massive galaxy clusters. Thus, these plots provide a test of the IllustrisTNG model in extreme cluster environments, especially during the time of early cluster assembly at high redshifts and in the high-pressure ICM environments at low redshifts. Following \citet{Pakmor2023}, we also updated the observational data in each panel, while we refer to \citet{Pillepich2018a} for the original data used in the model calibration. All shown quantities are inferred from observations and do not represent direct observables. A more rigorous comparison would require forward-modelling PICO-Clusters to generate synthetic observational data. This is beyond the scope of this work but will be addressed in future studies.

The first panel of Fig.~\ref{fig:validation_MTNG} shows the stellar mass--halo mass relation for central galaxies and their halos at the high-mass end ($M_\mathrm{200}>10^{13}\,\mathrm{M_\odot}$), normalized by the cosmic baryon fraction. Stellar masses are computed within a fixed physical radius of 30 kpc centred on the main subhalo. The central galaxies of PICO-Clusters extend the MTNG results to higher masses and are consistent with abundance matching \citep{Moster2010} and other observationally motivated reconstructions of the stellar--halo mass relation \citep[e.g.][]{Girelli2020}. On average, PICO-Cluster galaxies have slightly lower stellar masses than MTNG, which itself is in excellent agreement with abundance matching results.

\begin{figure*}
        \includegraphics{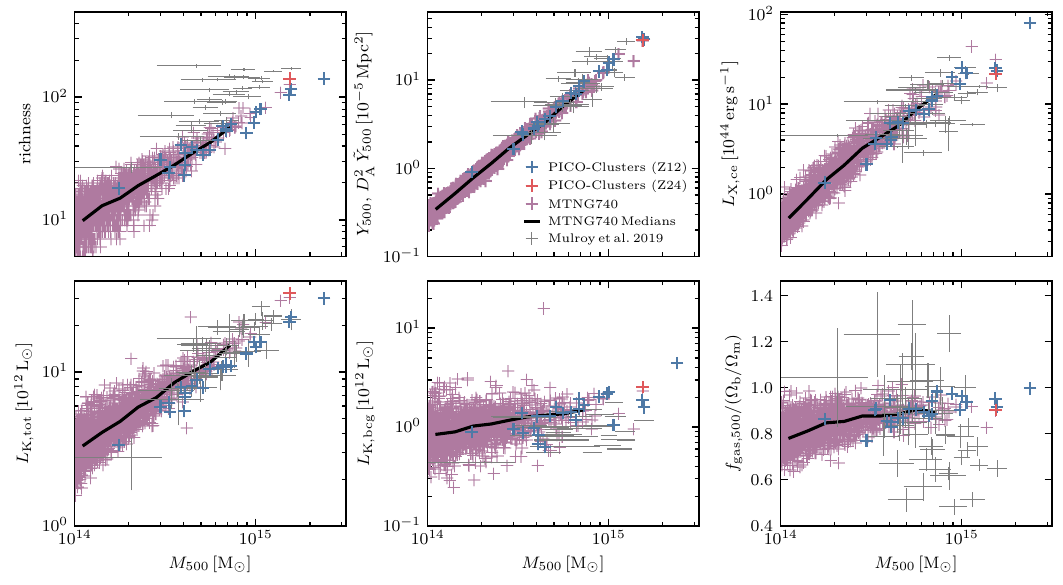}
        \caption{Cluster scaling relations of our PICO-Clusters and all MTNG galaxy clusters with $M_{500} >10^{14}\mathrm{M}_\odot$ at $z=0.25$. The panels display, as a function of $M_{500}$, the cluster richness (top left), the integrated Compton-$y$ parameter (top centre), the core-excised $[0.7{-}10]~\mathrm{keV}$ X-ray luminosity (using the radial range $0.15\,R_{500}<r<R_{500}$, top right), the total K-band luminosity, $L_\rmn{K,tot}$ (bottom left), the K-band luminosity of the brightest cluster galaxy, $L_\rmn{K,bcg}$ (measured via the stellar mass within 30 kpc, bottom centre), and the gas fraction (bottom right). The black line denotes the binned median of the MTNG clusters, while the grey points represent observational data for 41 galaxy clusters at $0.15<z<0.3$ \citep{Mulroy2019}. Overall, the PICO-Cluster results show good agreement with both the MTNG simulations and the observational measurements.}
        \label{fig:cluster_scaling_relations}
\end{figure*}

The second panel of Fig.~\ref{fig:validation_MTNG} presents the relation between the mass of the most massive SMBH and the stellar mass of the central galaxy. Unlike \citet{Pakmor2023}, our MTNG re-analysis includes only stars within a fixed physical radius of 30 kpc -- chosen to match observations -- rather than using stellar half-mass radii. We compare against the observed SMBH mass--bulge stellar mass relation of \citet{Kormendy2013}, which was used to calibrate the IllustrisTNG model for galaxy halos forming abundantly in cosmological volumes of $(25~h^{-1}\,\rmn{Mpc})^3$. We further include more recent relations linking SMBH mass to both stellar bulge and core masses \citep{Zhu2021}, the latter exhibiting a comparable slope yet a slightly elevated normalization. At fixed stellar mass, PICO-Clusters host SMBHs that are $\sim$0.5 dex more massive than in MTNG and exceed the core-mass relation of \citet{Zhu2021} by a factor of $\sim$2--3. Part of this offset from observations may stem from unintended behaviour of the SMBH repositioning scheme in IllustrisTNG \citep{Borrow2023}. In this numerical model, massive galaxies entering a cluster can lose their central SMBHs, which then rapidly merge with the cluster's central SMBH.

The smaller stellar and larger SMBH masses in PICO-Clusters compared to MTNG are a consequence of including magnetic fields in the simulation, which increases the efficiency of mass accretion onto SMBHs and hence, their feedback strength \citep{Pillepich2018a,Pakmor2023}. This allows our simulations to grow larger SMBHs in comparison to the MTNG simulation, which was run without magnetic fields. The larger SMBH masses cause increased AGN feedback, which better suppresses star formation, thus causing smaller stellar masses and lower K-band luminosities in central and satellite galaxies (Fig.~\ref{fig:cluster_scaling_relations}, bottom left and middle panels).

The third panel of Fig.~\ref{fig:validation_MTNG} shows the gas fraction within $R_\mathrm{500}$ as a function of $M_\mathrm{500}$, compared to observations of galaxy groups \citep{Lovisari2015} and clusters \citep{Eckert2016,Eckert2019}. The PICO-Cluster gas fractions are consistent with MTNG. Notably, the observational data exhibit substantially larger scatter at fixed halo mass than the simulations. Further work on forward modelling of mock X-ray and gravitational lensing observables is required to determine whether the remaining discrepancies in clusters and in the associated scatter are driven primarily by observational biases or reflect limitations of the galaxy formation model.

\subsection{Mass--observable scaling relations}
\label{sec: observational data}

Figure~\ref{fig:cluster_scaling_relations} shows six popular mass--observable scaling relations for galaxy clusters at $z=0.25$. We chose this redshift because the larger cosmic volume yields more extensive observational samples, while the relatively low redshift still permits well-controlled selection functions. We compare our PICO-Clusters and all MTNG clusters with $M_\mathrm{500} > 10^{14}\,\mathrm{M_\odot}$ to the LoCuSS sample of 41 well-studied clusters \citep{Mulroy2019} at redshifts between $z=0.15$ and $z=0.3$. While the virial masses $M_\mathrm{500}$ of the simulated galaxy clusters are computed using a spherical overdensity estimate around the potential minimum of the cluster, the observed scaling relations adopt weak lensing-based mass estimates. Weak-lensing masses are known to be biased tracers of cluster virial masses due to contributions from the triaxial halo shapes, uncorrelated large-scale matter projections along the line of sight, and cluster mergers \citep{Becker2011,Lee2023,Giocoli2025}, but we do not attempt to model these uncertainties here.

The cluster scaling relations of Figure~\ref{fig:cluster_scaling_relations} show as a function of $M_{500}$ the cluster richness (top left), the integrated Compton-$y$ parameter, $Y_{500}$ (top centre), the core-excised $[0.7{-}10]~\mathrm{keV}$ X-ray luminosity, $L_\mathrm{X,ce}$ (using the radial range $0.15\,R_{500}<r<R_{500}$, top right), the total K-band luminosity, $L_\rmn{K,tot}$ (bottom left), the K-band luminosity of the brightest cluster galaxy, $L_\rmn{K,bcg}$ (measured via the stellar mass within 30 kpc, bottom centre), and the gas fraction (bottom right). We refer the reader to \citet{Pakmor2023} for a detailed description of the modelling of the cluster richness and the K-band luminosities of cluster galaxies. 

First, we discuss the stellar content of PICO-Clusters. While the richness of PICO-Clusters agrees well with MTNG, both the total K-band luminosity and, to a lesser extent, that of the brightest cluster galaxy fall below the median of the MTNG scaling relations, possibly owing to the stronger AGN feedback in PICO-Clusters. The interpretation of the discrepancy of the observed and simulated richness is not straightforward, as it may stem from differences in the cluster richness of the PICO-Cluster simulations or the observational measurements, or alternatively from a bias in the weak lensing mass estimates of the observed clusters. In addition, the brightest cluster galaxies in PICO-Clusters appear roughly twice as bright in the K-band as their observed counterparts, while the total K-band luminosity is about half that of observations. Nevertheless, the scatter and slope of the simulated scaling relations remain broadly consistent with observations. For a detailed discussion of the discrepancies in cluster richness, $L_\rmn{K,tot}$, and $L_\rmn{K,bcg}$ between observations and the MTNG simulations, and by extension PICO-Clusters, including the enhanced scatter in the observed scaling relations relative to the simulated ones, we refer the reader to \citet{Pakmor2023}.

We now turn our discussion to ICM observables, namely the SZ effect, X-ray emission and gas fractions. We compute the integrated Compton-$y$ parameter within a sphere of radius $R_{500}$ as \citep{Battaglia2012a}
\begin{equation}
\label{eq:Y500}
Y_{500} = \frac{\sigma_\mathrm{T}}{m_\mathrm{e}c^2}\int_{V_{500}} \mathrm{d}V\, P_\mathrm{e}
= \frac{(\gamma-1)\sigma_\mathrm{T}}{m_\mathrm{e}c^2} X_\mathrm{e} X_\mathrm{H} \mu E_\mathrm{th},
\end{equation}
where $\sigma_\mathrm{T}$ denotes the Thomson cross-section, $m_\mathrm{e}$ the electron mass, and $c$ the speed of light. Here $P_\mathrm{e}$ is the electron pressure, $\gamma=5/3$ is the adiabatic index, $X_\mathrm{e}=n_\mathrm{e}/n_\mathrm{H}=1.158$ is the electron-to-hydrogen number density ratio for a fully ionised gas with a hydrogen mass fraction of $X_\mathrm{H}=0.76$, and $\mu=0.588$ is the mean molecular weight of a fully ionised primordial plasma. The quantity $E_\mathrm{th}$ represents the total thermal energy of the gas within $R_{500}$. For comparison with observations, we compute
\begin{equation}
\label{eq:Y500_tilde}
Y_{500}=D_\mathrm{A}^2 \tilde{Y}_{500},
\quad\mbox{where}\quad 
\tilde{Y}_{500}=\frac{\sigma_\mathrm{T}}{m_\mathrm{e}c^2}\int_{\delta \Omega}\mathrm{d}\Omega \int_D\mathrm{d}l P_\mathrm{e}
\end{equation}
is the Compton-$y$ parameter integrated over the solid angle $\delta \Omega$ subtended by the cluster. The line-of-sight integral extends from the observer to the cluster at a distance $D$ (while only receiving relevant contributions from the cluster's virial region) and $D_\mathrm{A}$ is the angular diameter distance at $z=0.25$ in the MTNG cosmology.

We compute the core-excised X-ray luminosity, $L_\rmn{X}$, in the $[0.7{-}10]~\mathrm{keV}$ energy band by integrating the X-ray emissivity, $\varepsilon$, over the cluster volume  ($V_\rmn{ce}$), restricted to the radial range $0.15 R_{500}<r<R_{500}$:
\be
	L_\rmn{X,ce} = \int_{V_\rmn{ce}} \rmn{d}V \varepsilon.
    \label{eq:L_X}
\en
Here, we only consider the bremsstrahlung emissivity, which is justified for our very massive clusters. We calculate the thermal X-ray emissivity by generalizing equation (5.14a) in \citet{Rybicki1979} to a fully ionised multi-component plasma. The X-ray emissivity becomes (see details in Appendix~\ref{sec:x_ray_emissivity})
\be
	\label{eq:x_ray_emissivity}
	\varepsilon = n_\mathrm{e} n_\mathrm{H} \Lambda \ ,
\en
where $n_\mathrm{e}$ is the electron number density, $n_\mathrm{H}$ is the hydrogen number density, and the cooling function is given by
\begin{align}
	\label{eq:Lambda_xray}
	 \Lambda = \f{2^5\uppi e^6}{3 m_\mathrm{e} c^3} \left(\f{2\uppi}{3 k_\mathrm{B} m_\mathrm{e}}\right)^{1/2} \f{k_\mathrm{B}T^{1/2}}{h} \left(\mathrm{e}^{-E_\mathrm{low}/k_\mathrm{B}T}-\mathrm{e}^{-E_\mathrm{high}/k_\mathrm{B}T}\right) \nonumber \\
	 \times  \sum_s \bar{g}_{\mathrm{B},\,s} Z_s^2 \f{n_\mathrm{s}}{n_\mathrm{H}} \ .
\end{align}
Here $e$ is the electron charge, $\kb$ Boltzmann's constant, $h$ Planck's constant, $T$ the temperature and $E_\mathrm{low}$ ($E_\mathrm{high}$) the lower (upper) cut in X-ray photon energy. The contribution from each ion species, denoted with subscript $s$, depends on ion number density $n_s$, the charge number $Z_s$ and the frequency-and-temperature averaged Gaunt factor, $\bar{g}_{\mathrm{B},\,s}$. We assume here a fully ionised hydrogen-helium plasma with primordial composition, i.e., $n_\mathrm{He}/n_\mathrm{H} = 0.079$ and assume for simplicity a species-independent Gaunt factor, $\bar{g}_{\mathrm{B},\,s}=1.3$. Finally, we note that we have $n_\mathrm{e}= \sum_s Z_s n_\mathrm{s} = 1.158 \, n_\mathrm{H}$ in this approximation.

As before, the ICM scaling relations of the PICO-Cluster sample shown in Fig.~\ref{fig:cluster_scaling_relations} represent a natural extension of MTNG to higher-mass systems, with the additional inclusion of magnetic fields. The PICO-Cluster gas fractions are exactly on the median relation of MTNG. In contrast, the integrated Compton-$y$ parameters are slightly above the MTNG relation. This likely results from enhanced AGN feedback relative to MTNG, driven by the more massive SMBHs in the PICO-Cluster simulations due to the inclusion of MHD (see Section~\ref{sec: previous sims}). This increases the thermal pressure and the integrated Compton-$y$ parameter, which is proportional to the volume-integrated thermal pressure (see equation~\ref{eq:Y500}). By contrast, the X-ray luminosity is proportional to the volume-integrated square of the gas density (equations~\ref{eq:L_X} and \ref{eq:x_ray_emissivity}) so that it is more sensitive to the core density profile. The influence of AGN feedback on the intracluster gas extends beyond $0.1 R_{500}$, irrespective of whether the cluster hosts a cool core. Hence, despite excising the core,  the median X-ray luminosity of our sample is slightly lower relative to MTNG. 

The simulated cluster mass-ICM observable scaling relations for $Y_{500}$, $L_\mathrm{X,ce}$, and $f_\mathrm{gas,500}$ are well within the observational scatter,\footnote{The core-excised bolometric X-ray luminosity reported by \citet{Mulroy2019} is derived by fitting a single-temperature model to the $[0.7{-}10]~\mathrm{keV}$ spectrum and extrapolating the surface brightness profile to $R_{500}$; reproducing this procedure in detail lies beyond the scope of the present work.} which is in all cases larger than the scatter in the simulated scaling relations \citep[see][for a more detailed discussion]{Pakmor2023}. Future work on careful mock observations is needed to establish whether this is due to observational biases or a shortcoming of the IllustrisTNG galaxy formation model.

\begin{figure*}
        \includegraphics[trim = 0 10 0 0]{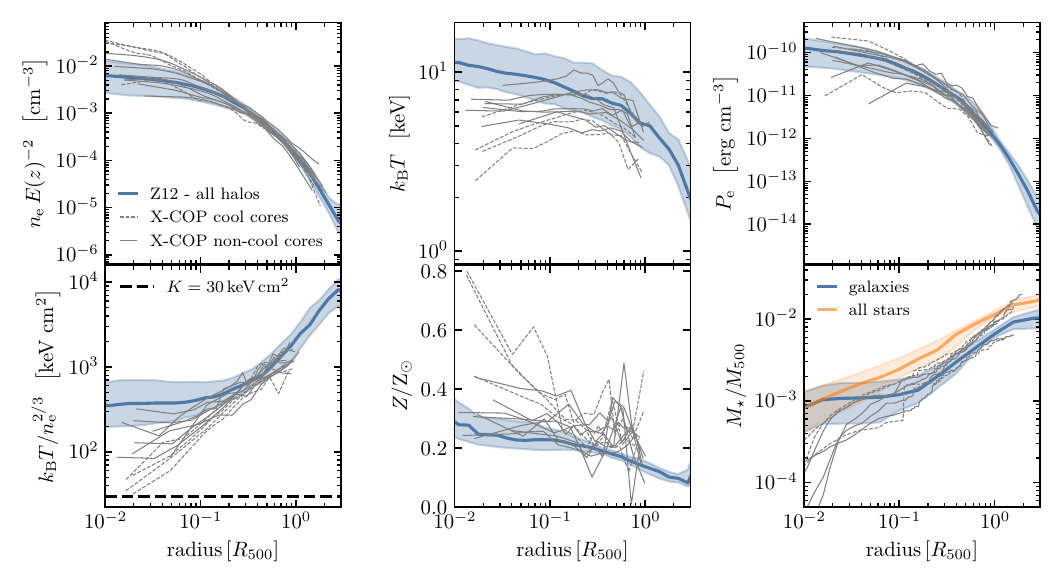}
        \caption{The figure compares the median radial profiles of several thermodynamic quantities at $z=0$, with the coloured bands indicating the 16th--84th percentile range, to profiles from X-COP observations shown in grey \citep{Ghirardini2019,Ghizzardi2021}. Cool core clusters with central entropy $K < 30 \;\mathrm{keV\,cm^{2}}$ are shown with dashed lines. The panels show profiles for the electron density as derived from the volume-weighted X-ray emissivity (equation~\ref{eq:<ne>}), the X-ray emission-weighted temperature (equation~\ref{eq:<T>}), the electron pressure and cluster entropy (computed from the individual density and temperature profiles), X-ray emission-weighted metallicity, and cumulative stellar mass profile scaled by $M_{500}$. We show the total stellar mass including the intracluster light (orange) as well as the stellar mass profile that only accounts for the brightest cluster galaxy (i.e., stars within 30~kpc) and all galaxies with $M_\star>10^9\mathrm{M}_\odot$ (blue), similar to the observed profiles. Overall, the shapes of the simulated thermodynamic profiles are in good agreement with observational results, except in the cluster cores, where the temperature and entropy profiles reflect the well-known underrepresentation of massive cool-core clusters in the simulations. The simulated metallicity and stellar mass profiles (that only account for galaxies) are at the lower envelope of the observed profiles due to observational biases and/or incomplete numerical convergence.}
        \label{fig:radial_profiles_obs}
\end{figure*}

\subsection{Galaxy cluster profiles}
\label{sec: cluster profiles}

Having discussed integrated properties of the various galaxy cluster components, we now turn to radial profiles of ICM thermodynamic quantities and gas metallicity. Figure~\ref{fig:radial_profiles_obs} compares the median PICO-Cluster profiles to observations of nearby galaxy clusters from the X-COP project \citep{Eckert2019,Ettori2019,Ghirardini2019}. X-COP provides thermodynamic and metallicity profiles \citep{Ghizzardi2021} for 12 nearby clusters at redshifts $0.04 < z < 0.1$ and masses in the range $4\times 10^{14}\,\mathrm{M_\odot} \lesssim M_\mathrm{500} \lesssim 10^{15}\,\mathrm{M_\odot}$, on average a factor of two lighter than the PICO-Cluster sample. These X-COP results are based on deep XMM-Newton X-ray and millimetre observations, enabling high-precision reconstructions of the intracluster gas properties. The X-COP sample is selected from the Planck all-sky SZ catalogue and comprises the most significant detections within this redshift range.

To remove the (weak) mass dependence of both the simulated and observed clusters, we scale all radii by $R_{500}$. By contrast, we do not rescale the thermodynamic quantities, as their mass dependence is much smaller than the intrinsic scatter associated with the cooling state in the cluster core, i.e.\ whether or not a cluster hosts a cool core. To focus our discussion on the PICO-Cluster population rather than on individual outliers, we show median profiles and coloured bands indicating the 16th--84th percentile range. All cluster profiles compare three-dimensional, spherically averaged radial profiles of the PICO-Clusters to deprojected radial profiles inferred from two-dimensional observational maps.

For PICO-Clusters, we calculate radial profiles using the following observationally motivated method: using equation~\eqref{eq:x_ray_emissivity} for the bremsstrahlung emissivity, $\varepsilon$, we calculate the emission-weighted radial temperature profile as
\be
    \langle T \rangle (r) = \frac{\int_{\Delta V(r)} T \,\varepsilon\,\mathrm{d}V}{\int_{\Delta V(r)} \varepsilon\,\mathrm{d}V}
    \label{eq:<T>}
\en
inside a set of radial shells with volumes $\Delta V(r)$. We then obtain the electron density profile by solving equation~\eqref{eq:x_ray_emissivity} for $n_\mathrm{e}$ (using that $n_\mathrm{e} = X_\mathrm{e} n_\mathrm{H}$) and find
 \be
     \langle n_\mathrm{e}\rangle(r) = \sqrt{\frac{X_\mathrm{e} \langle\varepsilon\rangle}{\Lambda(\langle T\rangle)}} \,,
     \label{eq:<ne>}
 \en
where $\langle\varepsilon\rangle=\langle\varepsilon\rangle(r)$ is the volume-weighted X-ray emissivity profile and we have highlighted how $\Lambda$ depends on the temperature profile $\langle T\rangle(r)$. The entropy and pressure profiles were constructed from the $\langle n_\mathrm{e}\rangle$ and $\langle T\rangle$ profiles (equations~\ref{eq:<T>} and \ref{eq:<ne>}), i.e., $K_\mathrm{e} = \langle \kb T\rangle \langle n_\mathrm{e}\rangle^{-2/3}$ and $P_\mathrm{e} = \langle n_\mathrm{e}\rangle \langle \kb T\rangle$. The metallicity profile was computed using emission-weighting in the same manner as for the temperature profile. While this method is better observationally motivated than the volume-weighted density and mass-weighted temperature profiles, we have tried both and note that the two methods give essentially indistinguishable results.

Overall and in particular at larger radii, we find good agreement between the PICO-Cluster thermodynamic profiles of density, temperature, pressure, and entropy and those from X-COP. Interestingly, the scatter among the simulated clusters seems to match that of the non-cool core clusters. However, the limited sample size precludes a thorough assessment of this property.

The central entropy, density, and temperature profiles, however, reveal a systematic discrepancy between the simulations and the observations (top panels of Fig.~\ref{fig:radial_profiles_obs}). The broad spread in those quantities in the X-COP sample reflects the presence of both cool-core (dashed lines) and non–cool-core systems (solid lines), with the former exhibiting a central temperature drop, enhanced core densities \citep{Vikhlinin2006}, and low central entropies $K_\mathrm{e}=k_\mathrm{B}T/n_\mathrm{e}^{2/3}<30~\mbox{keV cm}^2$ \citep{Cavagnolo2009,Hudson2010}.\footnote{The X-COP sample relies on \textit{XMM-Newton} data, whose angular resolution is insufficient to resolve the cooling regions of these clusters. Higher-resolution \textit{Chandra} observations would reveal the characteristically low entropy values of cool cores,  $K_\mathrm{e}<30~\mbox{keV cm}^2$ \citep{Cavagnolo2009}.} Such cool-core clusters are entirely absent in the baseline PICO-Cluster simulations, which instead display a central temperature plateau and a comparatively elevated entropy floor of 200--700~$\mbox{keV cm}^2$ (bottom left panel of Fig.~\ref{fig:radial_profiles_obs}).

We argue in our companion paper \citep{Weinberger2026b} that this is likely a direct consequence of two effects, (i) the different energy deposition of AGN jets and the kinetic feedback mode in IllustrisTNG and (ii) the overly massive SMBH in PICO-Clusters. In \citet{Weinberger2026b}, we demonstrate that PICO-Clusters run with AGN jet feedback \citep{Weinberger2023,Ehlert2023} self-regulate to cool-core states, whereas the kinetic AGN feedback mode in IllustrisTNG produces non-cool-core clusters from identical initial conditions. This contrast reflects a fundamental difference in how energy couples to the ICM: the kinetic mode drives shocks that rapidly dissipate the injected energy \citep[][their figures 1–3]{Weinberger2017a}, while AGN jets instead inflate buoyantly rising low-density lobes, driving subsonic turbulence and mixing gas across a range of temperatures \citep[][their figure 10]{Meenakshi2026}. This gentler coupling preserves the cool-core state, keeping the entropy below the threshold characteristic of non-cool-core clusters.

Secondly, the central entropy profile in those simulations reflects a time-averaged equilibrium between kinetic AGN feedback heating of the radiatively cooling ICM as well as Bondi accretion, which implies an equilibrium entropy that scales with the SMBH and cluster mass as $K_\mathrm{e}\propto M_\bullet^{4/3}M_{500}^{-1}$ \citep{Weinberger2026}. Adopting typical SMBH masses for our PICO-Clusters, we obtain central equilibrium entropies of
\begin{align}
    K_\mathrm{e} &\approx 300 \,\mathrm{keV}\,\mathrm{cm}^2 \, \left(\frac{\epsilon_\rmn{BH}}{0.2}\right)^{2/3} \, \left(\frac{M_\bullet}{8\times10^{10}\,\mathrm{M}_\odot}\right)^{4/3} \, \left(\frac{M_{500}}{10^{15}\,\mathrm{M}_\odot}\right)^{-1},
\end{align}
where we adopted the IllustrisTNG kinetic AGN feedback efficiency of $\epsilon_\rmn{BH}=0.2$ \citep{Weinberger2017a}. This argues for two improvements in future cosmological cluster simulations: they should model (i) SMBH growth more realistically and (ii) account for feedback from AGN jets rather than kinetic AGN feedback as low-momentum density AGN jet feedback is able to maintain the cool core state while self-regulating the cooling ICM \citep{Ehlert2023,Weinberger2023,Weinberger2026}.

\begin{figure*}
        \includegraphics{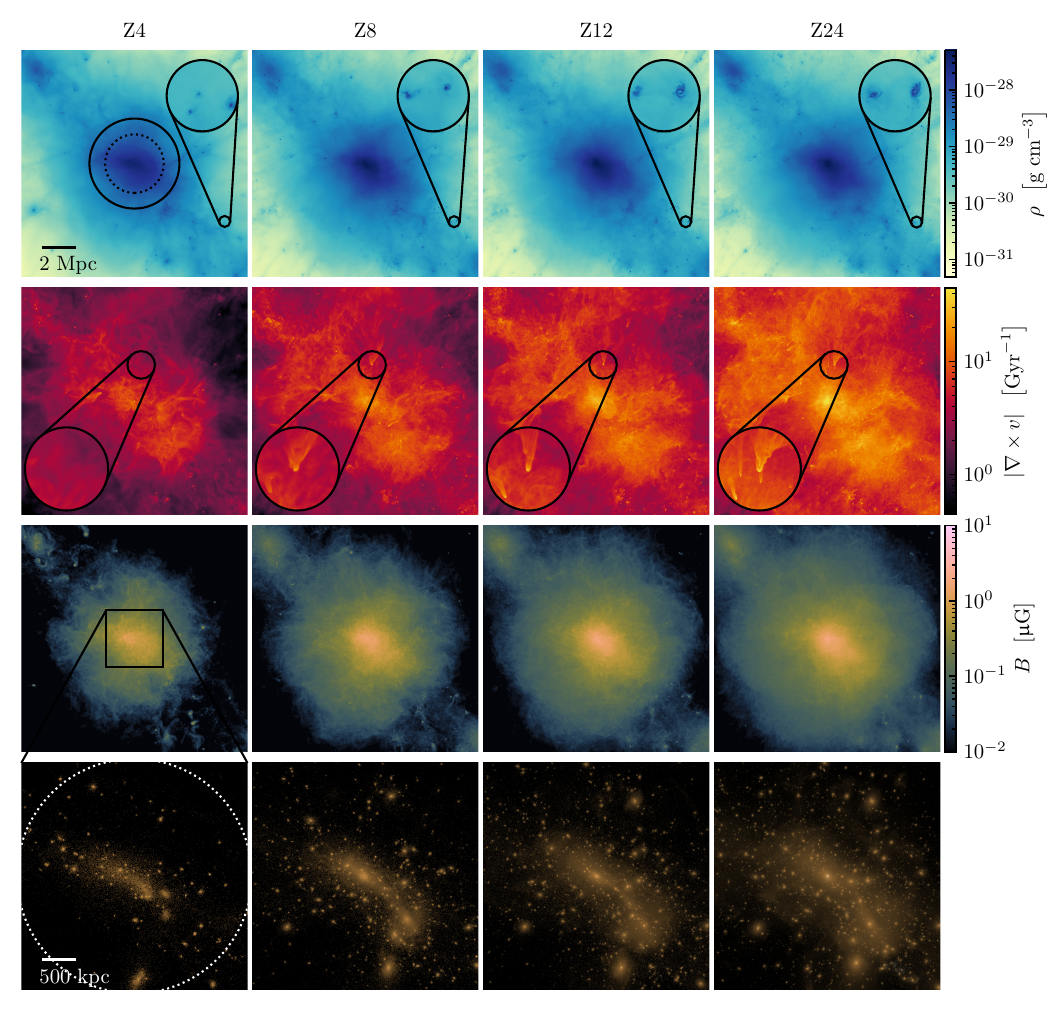}
        \caption{Convergence properties of the PICO-Cluster simulations, which employ the IllustrisTNG galaxy formation model and \textsc{Arepo-2}. We use four different numerical resolutions (Z4, Z8, Z12 and Z24) and show projections with a depth of 10~$h^{-1}$ Mpc of gas density (top row), vorticity (second row), magnetic field strength (third row) as well as synthetic SDSS $gri$ composite images of the stellar particles (bottom row) at $z=0$. The dotted (solid) circle in the top left panel indicates $R_{500}$ ($R_{200})$. The consistency of the density distribution and the presence and locations of resolved subhalos across resolutions attest to \textsc{Arepo}'s integration accuracy and the quality of the initial conditions, with higher resolutions revealing additional detail and low-mass subhalos. This trend is also evident in the vorticity maps, where turbulent wakes form behind galaxies moving through the ICM. The circular insets in this row illustrate two such subhalos at Z8 and Z12, and several additional ones at Z24. The magnetic field is seen to converge first at the centre of the cluster, and then progressively at larger radii. The exception is the very central region where the field strength keeps increasing with resolution due to the slow convergence of the galaxy formation model (the small size of the affected region makes it not very evident here, but see Fig.~\ref{fig:convergence_radial_profiles}). Similarly, the intracluster light toward the cluster core becomes brighter with increasing resolution.}
        \label{fig:convergence_panels}
\end{figure*}

\begin{figure*}
        \includegraphics[trim = 0 10 0 0]{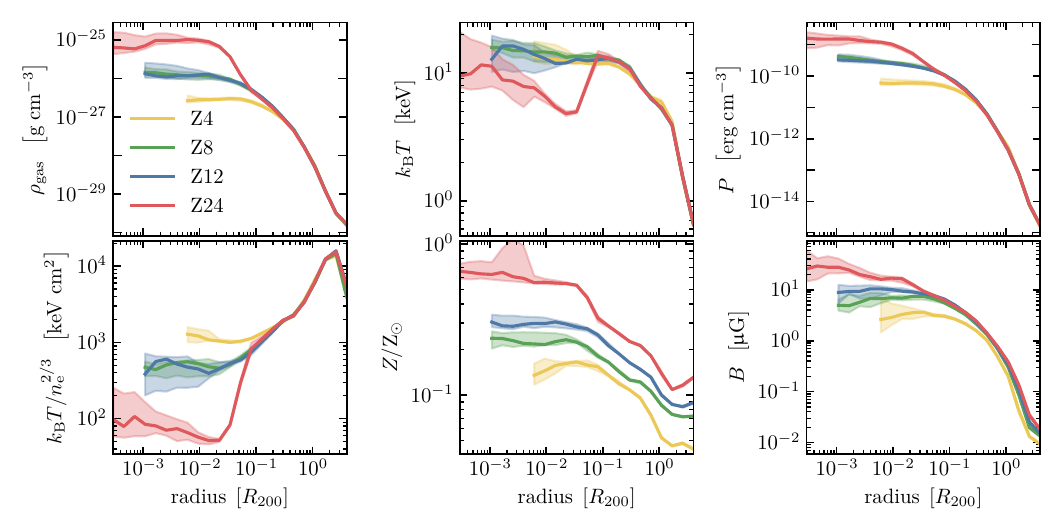 }
        \caption{Convergence study of median radial profiles of halo 4 between 13.4 and 13.8 Gyr with the bands indicating the 16th-84th percentile range of this time average. The panels show (volume-weighted) gas density, (mass-weighted) temperature, (volume-weighted) pressure, entropy (derived from the density and temperature profiles), (mass-weighted) metallicity, and magnetic field strength (as derived from the volume-averaged magnetic energy density). With the exception of metallicity, the outer profiles are well converged (but note the slight resolution-dependence of magnetic field strength outside $R_{200}$). By contrast, the profiles in the cluster core region are not converged because star formation in the IllustrisTNG model increases with increasing resolution. This increases the gas metallicity at all radii (and predominantly at the centre), as well as the gas density and pressure in the deeper central potential. The increased cooling leads to a significant drop in the central temperature and entropy at the highest resolution level (Z24).}
        \label{fig:convergence_radial_profiles}
\end{figure*}

The PICO-Cluster mean metallicity profile of our Z12 sample is smoothly declining while the observed profiles of most individual X-COP clusters show bumps at large radii. Temporarily enhanced metallicity profiles in the cluster outskirts can arise in PICO-Clusters during ongoing mergers, though this elevation does not persist. Moreover, the simulated mean metallicity profile is at the lower end of the observed metallicities (bottom middle panel of Fig.~\ref{fig:radial_profiles_obs}) with a better agreement with non-cool cores than with cool cores. The discrepancy could be either due to observational biases, incomplete numerical convergence (see Section~\ref{sec: convergence}) or the inability of the kinetic AGN feedback to maintain cool cores: high-resolution and deep \textit{Chandra} abundance maps indicate a highly inhomogeneous distribution of metals in the ICM \citep{Sanders2004,Sanders2007,Sanders2016}, which may bias the observationally inferred abundance profiles (or the bremsstrahlung-emission weighted profiles) toward higher values, as locally enhanced metallicity increases cooling rates, leading to gas compression and, in turn, further enhancing the cooling. Modelling this effect would require dedicated mock observations of our simulations, which we postpone to future work. On the other side, our simulated metallicity profile increases with resolution because of the slow convergence of the IllustrisTNG galaxy formation model at the mass scale of groups and clusters \citep{Pakmor2023} as we will discuss in Section~\ref{sec: convergence}.

At the bottom right panel of Fig.~\ref{fig:radial_profiles_obs}, we show two cumulative stellar mass profiles, each normalized by $M_{500}$ to account for the factor-of-two difference in cluster mass between the simulated and observed samples: one profile encompassing all stars including the intracluster light (orange), and one constructed to mirror observational selections, retaining only the brightest cluster galaxy (stars within 30 kpc) and satellite galaxies with $M_\star>10^9\mathrm{M}_\odot$ (blue). The plateau in the galaxy stellar mass profile at small radii arises from our simplified definition for the brightest central galaxy and the problems of the \textsc{Subfind(HBT)} algorithm to accurately recover the subhalo when it is close to the centre of the halo \citep{Muldrew2011,Knebe2011}. At small radii, the substantial variance in the scaled cluster profiles can be attributed to the diversity of cluster merger histories, whereas the scatter narrows considerably as one approaches the virial radius, which dominates the stellar mass of all galaxies as well as the total stellar mass including the intracluster light. This suggests that the stellar mass is a good proxy of cluster mass for massive clusters.

However, the simulated stellar mass in the brightest central galaxies of our clusters exceeds the observations \citep{Ghizzardi2021}, mirroring our results on the K-band luminosity of the brightest cluster galaxy, $L_\rmn{K,bcg}$ (bottom central panel of Fig.~\ref{fig:cluster_scaling_relations}). By contrast, the total stellar mass profiles of stars in galaxies, which reflect observational selection, are at the lower envelope of most of the observed profiles of \citet{Ghizzardi2021}, again reflecting our results on the total K-band luminosity, $L_\rmn{K,tot}$ (bottom left panel of Fig.~\ref{fig:cluster_scaling_relations}). As galaxies with stellar masses of ${\sim}10^{11}\,\rmn{M}_\odot$ contribute most to the total stellar mass budget, the IllustrisTNG galaxy formation model at PICO-Cluster Z12 resolution (comparable to TNG300) falls short of the observed stellar mass by approximately 30 per cent at fixed galaxy abundance \citep[see figure 2 of][]{Pakmor2023}. Moreover, the simulations contain nearly as much stellar mass in intracluster light as in galaxies themselves, suggesting that this overabundance relative to observations \citep{Puchwein2010} may further contribute to the discrepancy in stellar mass profiles. On the other side, the large variance of the observed scaled stellar mass profiles at large radii may signal unaccounted systematics in converting stellar light to mass. We defer a more detailed study of this puzzle to future work.

\begin{figure}
        \includegraphics[trim = 0 10 0 0]{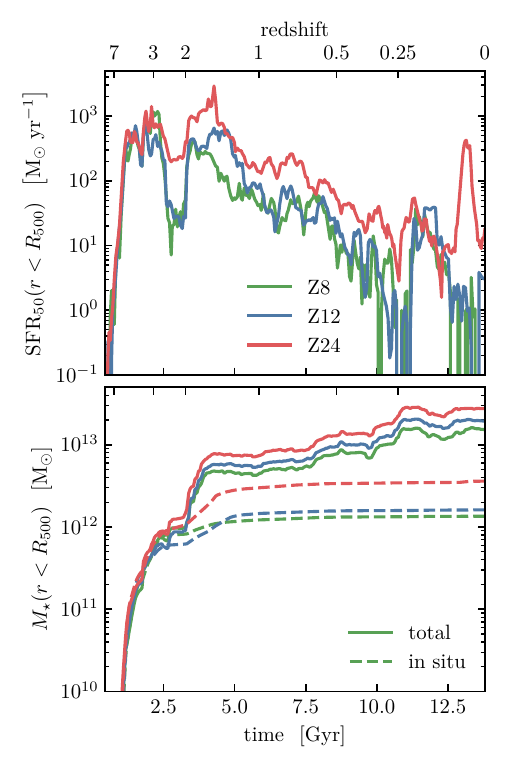}
        \caption{Top panel: SFR of halo 4 within $R_{500}$ tracked back in time. The SFR increases with resolution, with the recent starburst in the Z24 simulation not appearing at lower resolutions. Bottom panel: time evolution of the stellar mass within $R_{500}$. The solid lines show the total stellar mass while the dashed lines represent the stars formed in situ so that the difference indicates accreted stars.  SFR$_{50}$ is calculated by averaging the initial mass of all stellar particles inside $R_{500}$ that have an age less than $\tau = 50$ Myr. The `in situ' stellar mass is calculated as the time integral of this star formation rate. As expected, most of the stars belonging to cluster galaxies formed outside the cluster, owing to quenching processes associated with the cluster environment. }
        \label{fig:SFR}
\end{figure}

\subsection{Convergence properties of the simulations}
\label{sec: convergence}

To address the convergence properties of our simulation suite, we ran halo 4 at different numerical resolutions. Figure~\ref{fig:convergence_panels} shows projections of the density, vorticity, magnetic field strength and synthetic SDSS $gri$ composite images of the cluster (see Section~\ref{sec:illustris-tng}) for those different numerical resolutions. The sequence of panels at increasing resolution demonstrates \textsc{Arepo}'s integration accuracy and the quality of the initial conditions, with the existence and locations of resolved subhalos remaining consistent across resolutions, while higher resolutions reveal additional low-mass subhalos and substructures. This is exemplified in the density field that has the same overall appearance at all resolutions but with more galaxies at higher resolutions as is shown in the zoom-in insets. 

This is also clearly seen in the vorticity maps, where turbulent wakes appear behind galaxies as the ram-pressure-stripped ISM mixes with the ambient ICM as a result of the Kelvin--Helmholtz instability. The circular insets contain two such subhalos at resolutions Z8 and Z12, but several additional ones at Z24. Because the simulations are performed with ideal MHD (i.e., without explicit dissipation), it is expected that additional structure and higher vorticity values would appear at even higher resolution. In Kolmogorov subsonic turbulence, as expected in galaxy clusters, the velocity at scale $\lambda$ scales as $\varv_\lambda \propto \varv (\lambda/L)^{1/3}$, where $\varv$ is the velocity at the injection scale $L$ (e.g. \citealt{Schekochihin2022}). Hence,  vorticity scales with length as $|\bm\omega|=|\bnabla\btimes\bm{\varv}|\approx\varv_\lambda/\lambda\propto \lambda^{-2/3}$, implying that it is inherently not numerically converged in ideal (magneto-)hydrodynamics. The bulk kinetic energy in the cluster is however converged (see Section~\ref{sec:B_evolution}). The magnetic dynamo saturates first at the centre of the cluster, and then progressively at larger radii. This is because resolution decreases with radius, which in turn lowers the growth rate of the magnetic dynamo \citep{Tevlin2025}. The most significant difference between the different resolution levels is seen in the synthetic SDSS $gri$ composite images which show increasingly more dwarf galaxies and intracluster light at larger resolution. 

Figure~\ref{fig:convergence_radial_profiles} addresses numerical convergence of our radial profiles of gas density, temperature, pressure, entropy, metallicity, and magnetic field strength. At large radii, all thermodynamic profiles are well converged as expected from the dominant DM mass component that primarily shapes the cluster potential on large scales and hence, shapes the thermodynamic profiles there.

The profiles in the cluster core region are not converged because star formation in the IllustrisTNG model somewhat increases with increasing resolution (there is an increase by 30 per cent in stellar mass for every factor eight improvement in mass resolution as shown in the plot of cumulative stellar mass as a function of time\footnote{Throughout the paper, time refers to the cosmic time.} in the bottom panel of Fig.~\ref{fig:SFR}, and Appendix~A of \citealt{Pillepich2018b}). This leads to an increase in metal production in the central and satellite galaxies, with metals subsequently distributed to the surrounding circumgalactic medium via stellar winds and transported into the ICM through galactic outflows and AGN-driven winds (bottom middle panel in Fig.~\ref{fig:convergence_radial_profiles}).

The increased stellar mass with resolution\footnote{Note that we do not re-calibrate between resolution levels to test  convergence of the IllustrisTNG model in massive galaxy clusters.} is enhanced in the cluster centre because IllustrisTNG feedback has not been calibrated to quench these extremely dense proto-cluster regions at high redshift, which experience large mass accretion rates, and within the high-pressure environment we observe in the cluster centres today. This increased stellar mass forms preferentially between redshifts $z\approx2$--3 (bottom panel of Fig.~\ref{fig:SFR}) and deepens the central potential, which compresses the gas further, and increases the central metallicity via feedback-driven outflows. Both effects lead to stronger cooling, which enhances the central gas density, pressure, and metallicity. The increased gas cooling lowers the central temperature and consequently, lowers the central entropy, albeit not to the low levels observed in massive cool-core clusters \citep{Cavagnolo2009,Hudson2010}. 

The magnetic field strength has already converged at the Z8 resolution, although the region outside $R_{200}$ and the central field strength continues to increase with improved resolution. The latter increase is a direct consequence of the flux-freezing condition of ideal MHD and the increased central gas density in the highest resolution simulation (we will come back to the relation between $B$ and $\rho$ in the next section).

We confirmed that the increase of the central gas density in the Z24 simulation is not just the result of the last major starburst at $z\sim0.05$ in Z24 but in fact was built up earlier via the elevated SFR in comparison to the low-resolution runs (see top panel of Fig.~\ref{fig:SFR}). This elevated star formation already produces more stars during cluster assembly at around $z\sim3$ (see bottom panel of Fig.~\ref{fig:SFR}). A comparison between the total amount of stars and the stars formed inside the cluster ('in situ') demonstrates that most of the stars contained in clusters formed outside of the galaxy cluster. This is a consequence of cluster-related galaxy quenching processes such as ram-pressure stripping of the ISM and the circumgalactic medium and increased tidal interactions in proto-cluster group environments.

\section{Magnetic field growth}
\label{sec:magnetic_fields}

\begin{figure*}
        \includegraphics{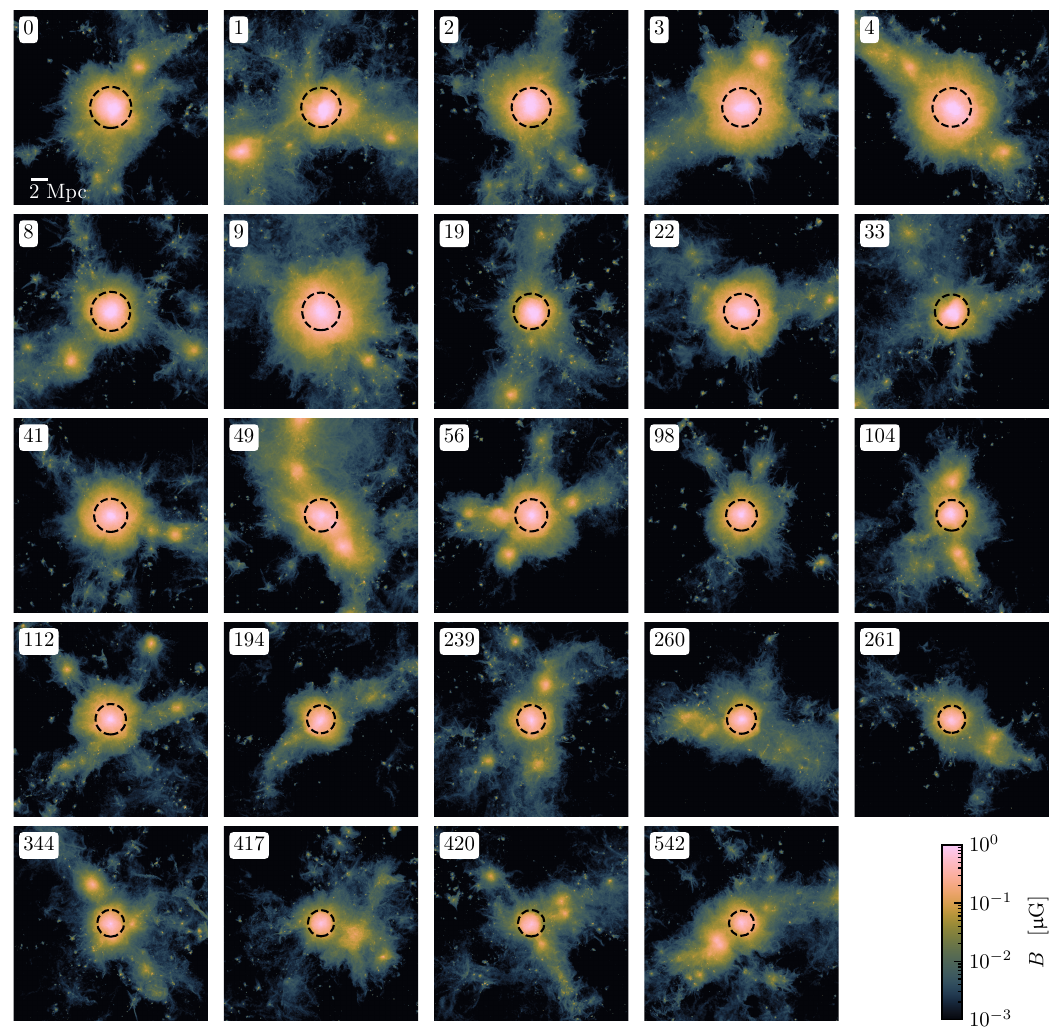}
        \caption{Projections of magnetic field strength at $z=0$ for the 24 galaxy clusters re-simulated at Z12 resolution in PICO-Clusters (depth 10~$h^{-1}$ Mpc). Similar images have been produced for gas density, vorticity, temperature, metallicity, pressure, dissipated shock energy and stellar content and can be accessed via the \href{https://pico-cluster.aip.de}{PICO-Cluster website} and the Supplementary Material to this paper. The circles indicate $R_{200}$.}
        \label{fig:B_panels}
\end{figure*}

Given the importance of magnetic fields for the plasma physical processes that we intend to investigate with the PICO-Cluster project (see Section \ref{sec:introduction}), we provide a more detailed study of the magnetic field properties found in the PICO-Cluster simulations. We have already established in Fig.~\ref{fig:convergence_radial_profiles} that the magnetic field strength is converged as a function of radius at $z=0$, except for very large radii and the central regions where both gas densities and magnetic field strength increases with resolution. We have also already noted the strong correlation between density, $B$-field and vorticity seen in Fig.~\ref{fig:zoom24_halo4}.

We now proceed by taking a look at the magnetic field distribution of the full PICO-Cluster sample in Fig.~\ref{fig:B_panels},
which shows an overview of the magnetic field strengths found in our simulations using deep $10~h^{-1}\mathrm{Mpc}$ projections.\footnote{These images were obtained by calculating the volume-weighted root mean square magnetic field strength along $4096^2$ lines of sight with the line integrals performed using the GpuRayProjector in Paicos \citep{Berlok2024}.}
A comparison with similar projections of vorticity and gas density (see Fig.~\ref{fig:zoom24_halo4}) reveals that regions of high magnetic field strength are correlated with dense, high-vorticity regions. Theoretical studies have made predictions for the correlation between $B$ and $\rho$: a simple argument shows that isotropic compression of a gas parcel leads to an increase in magnetic field strength that scales as $B\propto \rho^{2/3}$ \citep{Kulsrud2005}. Some dynamo theory papers also make predictions for the resulting $B$--$\rho$ scaling. Of particular interest are the works of \citet{Xu2020}, who predict $B\propto\rho^{1/2}$ for the saturated dynamo state and \citet{MuhammedIrshad2026}, who predict that $B\propto\rho^{5/6}$ during collapse of a sphere. We compare the PICO-Cluster simulations with these correlations in the following section.

\begin{figure*}
        \includegraphics[trim = 0 10 0 0]{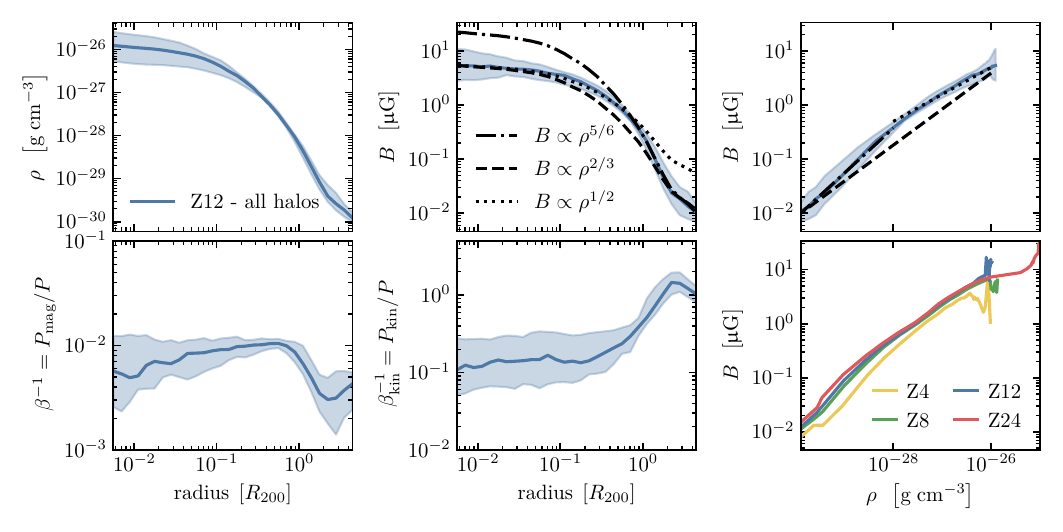}
        \caption{Radial profiles for plasma density, $\rho$ (top left), magnetic field strength, $B$ (top middle), the inverse of the plasma-$\beta$ (the ratio of magnetic-to-thermal pressure, bottom left), and the inverse of $\beta_\mathrm{kin}$ (the ratio of kinetic-to-thermal pressure, bottom middle) at $z=0$. The right column shows the correlation between $B$ and $\rho$ at each radius, as derived from the radial profiles, for the full sample (top) and for different  resolutions (bottom).}
        \label{fig:B_correlations}
\end{figure*}

\subsection{Magnetic fields at $z=0$}
\label{sub:magnetic_fields_at_z_0_}

In the ICM of galaxy clusters, the magnetic fields observed at low redshifts have been amplified and are kept at equipartition with the turbulent kinetic energy by a small-scale dynamo (see \citealt{Rincon2019} for a review). The character of the small-scale dynamo and its saturation level depends in general on the structure of the turbulent velocity flow and on microphysical properties of the plasma, such as the magnetic Prandtl number $\rmn{Pm}$, which is defined as the ratio of viscosity to magnetic resistivity. In realistic typical ICM conditions $\rmn{Pm} \gg 1$ \citep{Schekochihin2006}. However, since the PICO-Cluster simulations presented here were performed with ideal MHD, we rely on numerical dissipation on the grid scale for both viscosity and resistivity, which effectively implies $\rmn{Pm} \sim1$. In the PICO-Cluster sample, the tiny initial seed for the magnetic field is amplified at high redshifts by a combination of adiabatic compression during the protocluster formation, a gravitationally-driven dynamo, and pollution from galaxies via feedback and ram-pressure stripping \citep{Tevlin2025}.

We analyse here the magnetic field properties at $z=0$. Figure~\ref{fig:B_correlations} shows an analysis of the magnetic field and density dependence. The upper left panel shows the median of the volume-weighted plasma density for the PICO-Cluster simulations at Z12 resolution with the coloured band indicating the 16-84th percentiles. The upper middle panel shows the volume-weighted root-mean-square magnetic field strength as a function of radius. The black lines indicate the three different theoretical predictions that we discussed in the previous section ($2/3$ for isotropic collapse, \citealt{Kulsrud2005}, $1/2$ in \citealt{Xu2020} and $5/6$ in \citealt{MuhammedIrshad2026}): the central regions of the clusters appear to be well matched with the $B\propto\rho^{1/2}$ scaling predicted in \citet{Xu2020} while the outer parts appear to follow the $B\propto\rho^{5/6}$ scaling predicted in \citet{MuhammedIrshad2026}. However, this region is also where the distribution of plasma density and magnetic strength becomes highly non-uniform in space (see Fig.~\ref{fig:B_panels}), so that the angular averaging used to calculate $\rho(r)$ and $B(r)$ becomes problematic. We also show the $B\propto\rho^{2/3}$ scaling expected from isotropic collapse, which appears to match the $B$-profile at small and large radii well, but has a discrepancy at intermediate radii.

The upper right panel of Fig.~\ref{fig:B_correlations} gives a better view of the $B$--$\rho$ scaling by using logarithmic axes, so that power laws appear as straight lines. The slope is seen to match overall with a $2/3$ scaling, but with a $1/2$ scaling at high densities and a $5/6$ scaling at low densities matching the data somewhat better.
Finally, the lower right panel shows a resolution study of the $B$--$\rho$ scaling for halo 4, which we have simulated at Z4, Z8, Z12 and Z24 resolutions. This reveals that the profiles are well converged already at Z12 resolution.

We have calculated the plasma-$\beta$, i.e., the ratio of thermal-to-magnetic pressure, by using the radial profiles of $B$ and the volume-weighted thermal pressure, $P$. We show $\beta^{-1}$ in the lower left panel of Fig.~\ref{fig:B_correlations}. There is a large scatter in the values at small radii, presumably because processes such as AGN feedback, mergers and radiative cooling create stochastic variations in both the central magnetic field and pressure (see also Figs.~\ref{fig:radial_profiles_obs} and \ref{fig:convergence_radial_profiles}). The increase of $\beta^{-1}$ towards the virial radius is likely a direct consequence of the saturated state of the small-scale dynamo and an increasing turbulent-to-thermal pressure profile with radius. Confirming this hypothesis will require calculating the turbulent kinetic energy profile, which we leave to future work (Perrone et al., in prep). The behaviour at radii larger than the virial radius is likely influenced by incomplete numerical convergence of the magnetic field in these regions (as seen in Fig.~\ref{fig:convergence_radial_profiles}). We find that the median $\beta$ for the PICO-Cluster sample is in the range $100$ to $300$ with a relatively weak dependence on radius.

We have calculated the kinetic energy density of the gas,\footnote{The kinetic energy density is estimated as follows: We first compute the physical velocity field by adding the Hubble flow to the peculiar velocities, i.e. $\vec{\varv} = \vec{u} + H(z) \vec{r}$ where $\vec{u}$ is the peculiar velocity, $H(z)$ is the Hubble function and $\vec{r}$ is the physical coordinate in a coordinate system centred on the main subhalo. We select a spherical region of radius 10~$h^{-1}$Mpc and subtract the centre-of-mass velocity of the gas. The resulting velocity field is then mainly due to the internal dynamics of the cluster and not the movement of the cluster through the computational volume. We use the physical velocity field to calculate the kinetic energy density $e_\mathrm{kin} = \rho \vec{\varv}^2/2$ and the closely related kinetic pressure $P_\mathrm{kin} = \rho \vec{\varv}^2/3$ (see \citealt{Perrone2026a} for a discussion).} $e_\mathrm{kin}$, and compared it to the thermal energy density of the clusters, by defining $\beta_\mathrm{kin} =  P / P_\mathrm{kin}$ where the kinetic pressure is given by $P_\mathrm{kin} = 2 e_\mathrm{kin} / 3$ (see \citealt{Perrone2026a}). We show the inverse of this ratio, $\beta_\mathrm{kin}^{-1}$, in the lower middle panel of Fig.~\ref{fig:B_correlations}. The median kinetic pressure of the PICO-Cluster sample is roughly 10 per cent of the thermal pressure for $r\lessapprox 0.3\,R_{200}$ and then rises steeply so that the outermost regions are generally kinetically rather than thermally dominated. These large kinetic energies in the outskirts are mostly probing infall in the form of large-scale bulk motions and are situated outside of the cluster accretion shocks. In order to gain a handle on how much of the kinetic energy of the ICM is due to turbulence and how much is due to bulk motions, we intend to use the tools developed in \citet{Perrone2026a} to calculate the \emph{turbulent} kinetic energy fraction of the cluster sample (Perrone et al., in prep).

\begin{figure*}
        \includegraphics[trim = 0 10 0 0]{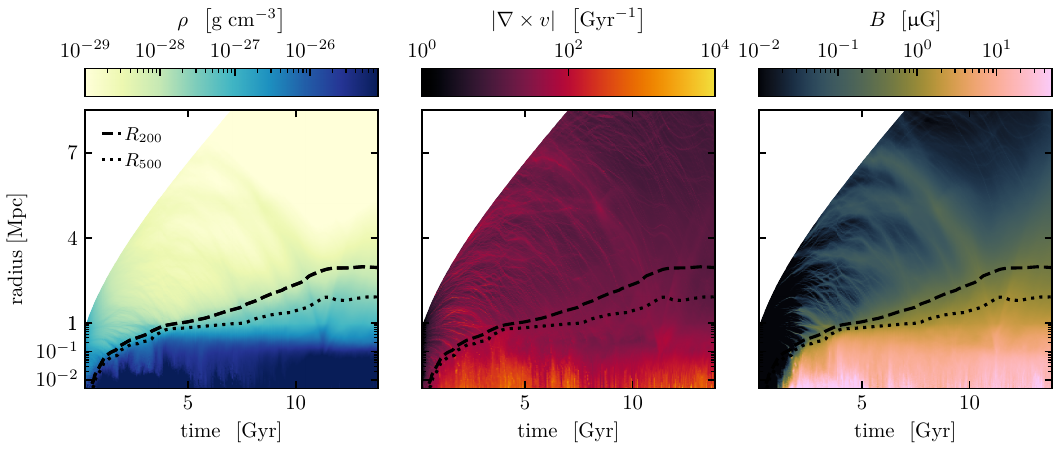 }
        \caption{Radial profiles (centred on the progenitors of the main subhalo at $z=0$, physical radii are shown on the vertical axis) have been stacked to construct images that display the time evolution of radial profiles in the Z24 simulation of halo 4. The frequent simulation output allows precise tracking of the cluster evolution in time. We show the gas density (left), vorticity (middle), and magnetic field strength (right). The dashed and dotted lines indicate the cluster radii ($R_{200}$ and $R_{500}$, respectively). Note that the $y$-axis is logarithmic (linear) for radii smaller (larger) than 1 Mpc. Trajectories of galaxies are seen as thin stripes in vorticity at large radii and AGN feedback events are seen in bright yellow at small radii.
        Wider bands that cross the virial radius are cluster mergers (seen particularly well in the $B$-field map) and the ``V-shaped'' pattern most notable around 11 Gyr indicates a major cluster merger at late times. The magnetic field grows in time and expands to larger radii; this is due to the combination of a magnetic dynamo in the ICM, accretion of pre-enriched material and $B$-field pollution by galaxy feedback and enrichment. A detailed analysis of these processes has been presented in \citet{Tevlin2025} for the Z12 simulation of halo 3. }
        \label{fig:x_t_plots}
\end{figure*}

\begin{figure}
        \includegraphics[trim = 0 10 0 0]{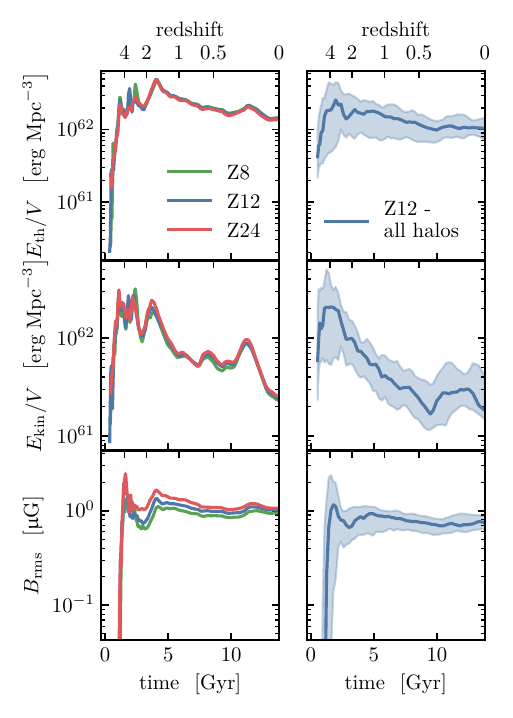}
        \caption{Evolution of energy densities (thermal and kinetic) and the mean $B$-field inside $\Rvir$. We find excellent convergence properties for the resolution study of halo 4 (left column) and the PICO-Cluster sample quickly saturates with a median $B_\mathrm{rms}$-field of ${\approx} 1\;\muG$. For halo 4 (left), the kinetic energy density shows peaks that coincide with violent events, while these peaks are washed out in the full cluster sample (right), where the solid lines display medians for the full sample and the bands are 16th-84th percentiles.}
        \label{fig:energy_evolutions}
\end{figure}
\begin{figure}
        \includegraphics[trim = 0 10 0 0]{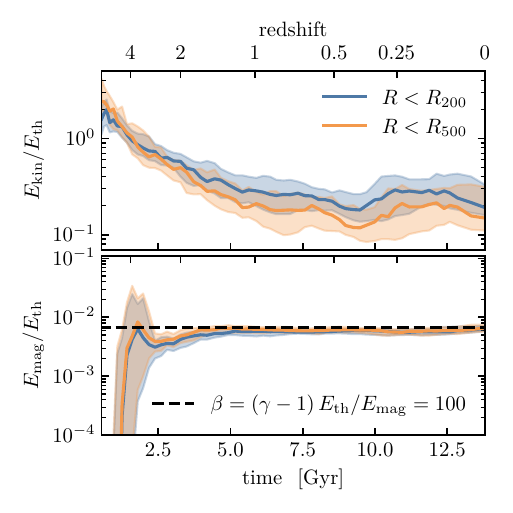}
        \caption{We show the time evolution of $E_\mathrm{kin}/E_\mathrm{th}$ (top) and $E_\mathrm{mag}/E_\mathrm{th}$ (bottom) inside $R_{200}$ and $R_{500}$. The solid lines display medians for the full sample and the bands are 16th-84th percentiles. The spread in the ratio of magnetic-to-thermal energies inside $R_{200}$ and $R_{500}$ is extremely small and constant in time after the initial growth phase. As indicated with a dashed black line, the value of this ratio corresponds to $\beta\sim 100$.}
       \label{fig:evolution_of_energy_ratios}
\end{figure}

\subsection{Time evolution of magnetic fields}
\label{sec:B_evolution}

The previous subsection focused on the magnetic fields as they appear at $z=0$. We now study the time evolution of the magnetic fields. Figure~\ref{fig:x_t_plots} shows the time evolution of radial profiles of gas density, vorticity and magnetic field strength. These profiles were generated by tracking the main subhalo back in time using the on-the-fly generated merger trees produced by \textsc{Arepo-2}. We used logarithmically spaced bins out to a maximum co-moving radius of 10 $h^{-1}\mathrm{cMpc}$ but display all quantities in physical coordinates.\footnote{The high redshift white regions in the plot are thus not empty, they have just not been computed as they are far outside the region of interest.}
The vertical axis is logarithmic for $r<1$~Mpc and linear for $r>1$~Mpc. This allows us to simultaneously view the internal cluster dynamics and the tracks of minor and major mergers.

The density is seen to consist of a high-density core, which at $t \lessapprox 4$ Gyr roughly tracks the growth of $\Rvir$. At later times, the evolution of the core radius is decoupled from the evolution of $\Rvir$, with the core radius slowly decreasing while $\Rvir$ continues to grow. This behaviour has been observed for relaxed clusters by \citet{Mostoghiu2019} (see their figure 2), and occurs because late-time mass accretion preferentially accumulates in the outskirts of the cluster.

Minor and major mergers occur when the tracks visible in density, vorticity and magnetic field strength cross the dashed line indicating $\Rvir$. AGN feedback events are visible at low radii as bright yellow stripes in vorticity.
The magnetic field strength is seen to grow quickly inside the central galaxy \citep{Tevlin2025}, saturating with a mean value of ${\sim}10\,\muG$ inside $r<100$~kpc (${\sim}15\,\muG$ inside $r<30$~kpc).  The magnetic field grows with the cluster; that is, as $\Rvir$ increases, so too does the extent of the magnetized region.

Figure~\ref{fig:energy_evolutions} shows the time evolution of the mean thermal energy density (top), kinetic energy density (middle) and root mean square magnetic field strength\footnote{Note that the magnetic energy density is $E_\mathrm{mag}/V = B_{\mathrm{rms}}^2/8\uppi$, where $E_\mathrm{mag}$ is the total magnetic energy inside $\Rvir$. We show $B_{\mathrm{rms}}$ for convenience.} (bottom) inside $\Rvir$. In the left column, we show these quantities for halo 4 at three different resolutions, where peaks in kinetic energy density are due to galaxy cluster mergers. The thermal and kinetic energy densities are seen to be very well converged, while the magnetic field energy densities systematically increase with resolution. 
Since the magnetic energy is dominated by contributions from radii $>200$~kpc and is thus insensitive to the resolution-dependent amplification of the central magnetic field, we are therefore justified to discuss the magnetic field properties at Z12 resolution.

The root-mean-square magnetic field is exponentially amplified at early times, and saturates with a roughly constant mean value that converges towards ${\approx} 1 \muG$. The thermal and kinetic energy densities similarly undergo rapid growth at early times. In the right column of Fig.~\ref{fig:energy_evolutions}, we show the thermal and kinetic energy densities and the median magnetic field inside $\Rvir$ for the Z12 PICO-Cluster simulations. The individual galaxy cluster merger events are here smeared out due to ensemble-averaging, but we nevertheless see the same overall trend as in halo 4. The median field increases rapidly at early times and saturates with a value of roughly $1 \muG$.

\subsection{Scaling relation for magnetic field strength}

\begin{figure*}
        \includegraphics{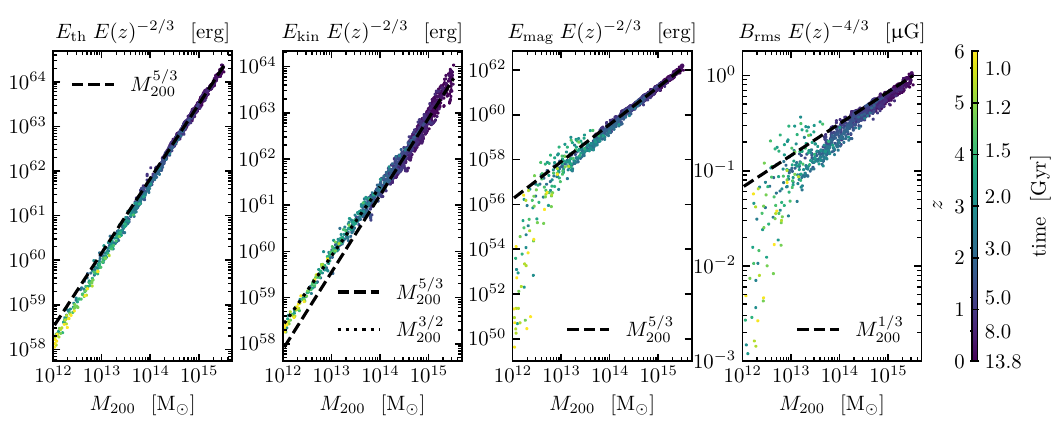}
        \caption{Cluster scaling relations for thermal (first panel), kinetic (second panel), and magnetic (third panel) energies as well as the root-mean-square magnetic field strength  inside $\Rvir$. The points are from the 24 different clusters at Z12 resolution and their colours indicate the redshift. We have added power scaling laws to guide the eye. We observe that the magnetic energy (root-mean-square field strength) scales with $M_{200}^{5/3}$ ($M_{200}^{1/3}$) at low redshift.}
        \label{fig:cluster-energy-scalings}
\end{figure*}

Magnetic dynamos use turbulent motions to amplify magnetic fields.
In turbulent box studies, the so-called dynamo efficiency is defined as the ratio of magnetic energy density to (turbulent) kinetic energy density, i.e., $\epsilon = E_\mathrm{mag} / E_\mathrm{turb}$. The dynamo efficiency depends on whether the driving is subsonic or supersonic, compressive or solenoidal, and on microphysical properties of the plasma \citep{Federrath2011b,Kriel2022,Beattie2023}. The PICO-Cluster simulations have a strong radial gradient in all relevant quantities, and so we do not expect a universal and global $\epsilon$ to describe the growth of the magnetic field.
In addition, it is unfortunately not straightforward to estimate $E_\mathrm{turb}$ (see \citealt{Perrone2026a}), and so we will here use $E_\mathrm{kin}$, which includes both bulk and turbulent motions, instead.

Figure~\ref{fig:evolution_of_energy_ratios}
shows the time evolution of the ratio of kinetic-to-thermal energy (top) and the ratio of magnetic-to-thermal energy (bottom), where the energies are obtained by integrating the energy densities over the volume enclosed within $\Rvir$. The kinetic-to-thermal energy ratio is seen to exceed one at high redshift because of (i) vigorous AGN and stellar feedback, (ii) higher merger rates at high redshift, and (iii) shorter cooling times in lower-mass progenitors at high redshift, implying that the hydrostatic atmosphere cannot be fully established and resulting in lower thermal energy content. By contrast, at lower redshift, infalling gas is heated and decelerated at the accretion shock outside the clusters, establishing a hydrostatic cluster atmosphere and causing a lower kinetic-to-thermal energy ratio of around 20 per cent. From the radial profile of $\beta_\mathrm{kin}^{-1}$ shown in the middle lower panel of Fig.~\ref{fig:B_correlations}, we infer that this large kinetic energy fraction is due to in-falling material at $r\gtrapprox 0.3\,R_{200}$. A detailed analysis using a filtering approach reveals that the turbulent-to-thermal energy densities on scales $\lesssim 50 ~ \si{kpc}$ are much lower at the level of 2 per cent \citep{Perrone2026a}.

The lower panel of Fig.~\ref{fig:evolution_of_energy_ratios} reveals that the ratio of magnetic-to-thermal energy grows but quickly saturates and then keeps an essentially constant value from $t=5$~Gyr ($z=1.2$) and onwards. From turbulent dynamo simulations (e.g.\ \citealt{Beattie2023}), we might have expected a constant magnetic-to-turbulent energy ratio but here we observe a constant magnetic-to-thermal ratio. The magnetic-to-thermal ratio is closely related to the plasma-$\beta$, and so we show the value of $E_\mathrm{mag}/E_\mathrm{th}$ that corresponds to $\beta=100$ with a dashed black line in the lower panel of Fig.~\ref{fig:evolution_of_energy_ratios}. The PICO-Cluster simulations are seen to follow this $\beta\sim 100$ line with very little scatter around the median. This universal trend is in line with previous results that argued that in massive clusters the equipartition plasma $\beta$ is a constant that does not depend on the specifics of the cluster nor on its mass, but only on the conversion efficiency between kinetic and magnetic energy \citep{Miniati2015b}. We note that while the volume-averaged $\beta$ follows the $\beta\sim 100$ line with little scatter in time and across the cluster sample, the underlying spatial distributions of $\beta$ are turbulent and thus highly non-uniform. This has important implications for viscous heating, as we will discuss in Section~\ref{sec:Braginskii}.

Given that the PICO-Cluster simulations exhibit an approximately constant ratio of $E_\mathrm{mag}/E_\mathrm{th}$ after 5~Gyr, it follows that the scaling relation for $E_\mathrm{mag}$ should be identical to that for $E_\mathrm{th}$. Scaling relations for $E_\mathrm{th}$ as a function of $M_{500}$ and $M_{200}$ have been widely studied in the literature, and the self-similar scaling relation is $E_\mathrm{th} \propto M_{200}^{5/3} E(z)^{2/3}$ where $E(z) = H(z) /H_0$ \citep{Kaiser1986,Kravtsov2006,Lovisari2022}. Figure~\ref{fig:cluster-energy-scalings} shows the scaling relations for $E_\mathrm{th}$, $E_\mathrm{kin}$ and $E_\mathrm{mag}$ for the Z12 simulations using a wide range in redshift (with each data point obtained from a snapshot from one of the Z12 simulations). The thermal energy roughly follows a power law with slope $5/3$, which is established already at high redshift.\footnote{Please see section 3.2.3 in \citet{Henden2018} for a more detailed discussion of the comparison between simulations, observations and the self-similar expectations.} The magnetic field energy similarly follows a power law with $E_\mathrm{mag}\propto M_{200}^{5/3}$, which is established at lower redshifts where the magnetic field amplification has reached saturation (see also Figs.~\ref{fig:energy_evolutions} and \ref{fig:evolution_of_energy_ratios}). 

We can predict the scaling for the volume-weighted root-mean-square magnetic field strength inside $R_{200}$, $B_\mathrm{rms}$, by assuming a universal volume-averaged $\beta=100$ inside $R_{200}$ (as motivated by Fig.~\ref{fig:evolution_of_energy_ratios}) and 
a self-similar scaling of thermal energy \citep{Kaiser1986} anchored such that a $10^{15}\,\mathrm{M}_\odot$ cluster has $E_\mathrm{th}= 3.12\times10^{63}$ erg at $z=0$. This yields
\be
    B_{\mathrm{rms}} = 0.67 \upmu\mathrm{G} \;\left(\f{M_{200}}{10^{15}\,\mathrm{M}_\odot}\right)^{1/3} E(z)^{4/3} \ .
    \label{eq:B_rms_200_scaling}
\en
We show the scaling predicted by equation~\eqref{eq:B_rms_200_scaling}
in the fourth panel of Fig.~\ref{fig:cluster-energy-scalings}, and find that it appears to work well for masses $M_{200} \gtrsim 10^{14} \,\mathrm{M}_\odot$. Downward deviations for smaller masses at higher redshift indicate an ongoing magnetic dynamo in the kinematic phase.
Upward deviations at lower masses indicate that these systems are not well-described by the universal $\beta$ prediction.

\begin{figure*}
        \includegraphics{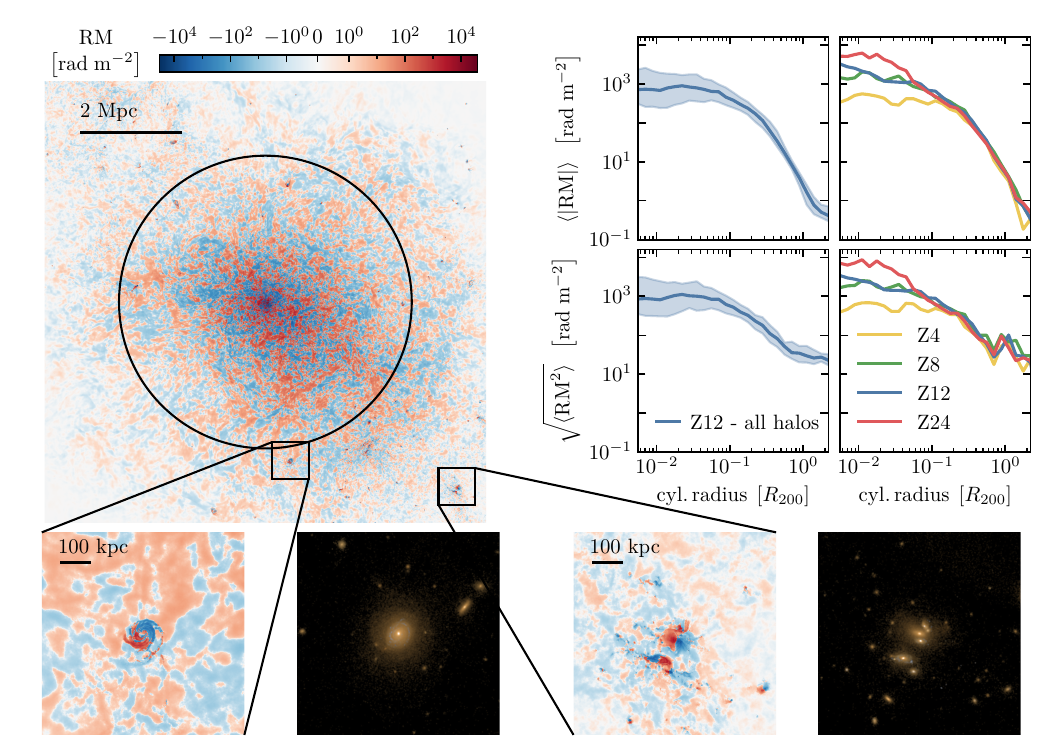}
        \caption{Faraday RM map of the Z24 simulation of halo 4 at $z=0$. The contribution of galaxies to the RM signal is highlighted with zoom-ins on selected regions and corresponding synthetic SDSS $gri$ composite images. Comparing profiles of mean pixel values in cylindrical radius, namely of $\langle |\mathrm{RM}|\rangle$ and $\sqrt{\langle \mathrm{RM}^2\rangle}$, show that $\langle |\mathrm{RM}|\rangle$ decreases more rapidly with radius. This is shown for the cluster sample (the solid blue lines display the medians for the full sample and the bands are 16th-84th percentiles) in the left two panels of the profile inset in the upper right corner, and for different resolutions of halo 4 in its right two panels. This is because the contribution of galaxies becomes relatively more important at larger radii. These strong but non-volume filling sources result in $\sqrt{\langle \mathrm{RM}^2\rangle}\gg \langle |\mathrm{RM}|\rangle$ (see Fig.~\ref{fig:FRM_histograms}).  Regarding convergence, the central signal is seen to be significantly larger at Z24 than at lower resolutions because of the increased electron density and magnetic field strength in the cluster core region.}
        \label{fig:FRM}
\end{figure*}

\begin{figure}
        \includegraphics{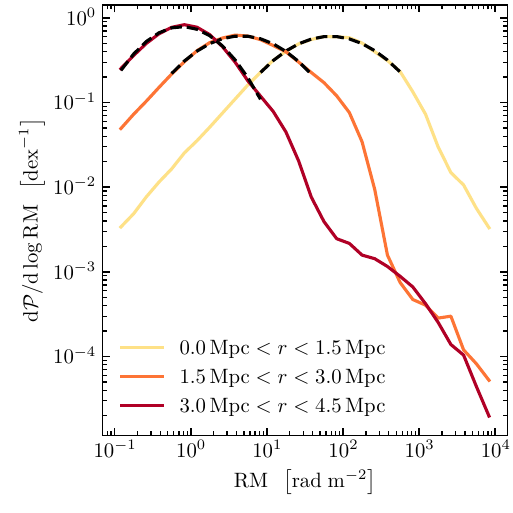}
        \caption{Probability distribution of Faraday RM of the Z24 simulation of halo 4 in three different cylindrical bins as indicated in the legend. At each radius, there is an approximate log-normal distribution in RM (black dashed) which arises from the ICM and a tail toward large RM values that mostly originates from the galaxies. The galaxies become more dominant in the outskirts where the ICM contribution declines, moving the mean RM contribution towards lower values. The $\sqrt{\langle \mathrm{RM}^2\rangle}$ becomes much larger than $\langle |\mathrm{RM}|\rangle$ for these tailed distributions, i.e.\ a factor of ${\sim} 10$ larger at $R_{200}$.}
        \label{fig:FRM_histograms}
\end{figure}

At late times, the kinetic energy roughly follows a power law with slope $5/3$, as expected from the virial equilibrium scaling: $E_\mathrm{kin}\propto E_\mathrm{th}\propto M_{200}^{5/3} E(z)^{2/3}$. We observe a flattening towards larger redshift (and smaller halo masses), roughly approaching $E_\mathrm{kin}\propto M_{200}^{3/2} E(z)^{2/3}$, which could indicate the additional non-gravitational energy input from AGN feedback.

\section{Faraday rotation}
\label{sec:FRM}

Faraday rotation refers to the rotation of the plane of polarization of linearly polarized radio waves as they propagate through a magnetized plasma, arising from the plasma’s birefringent nature. In such a plasma, the rotation angle is given by $\phi=\phi_0 + \mathrm{RM}\, \lambda^2$, where $\phi$ and $\phi_0$ are the measured and intrinsic polarization angle, respectively, and $\lambda$ is the wavelength of radiation. The Faraday RM determines the rate of rotation and is given by 
\begin{align}
    \mathrm{RM} = \frac{e^3}{2 \uppi m_\mathrm{e}^2 c^4} \int_0^L \frac{n_\mathrm{e} \bm B \bcdot \mathrm{d} \bm l}{(z+1)^2}.
    \label{equ:RM}
\end{align}
Here, $\bm l$ is the line-of-sight vector, and $L$ the physical distance from the radio source to the observer. The $(1+z)^{-2}$ factor arises from the cosmological stretching of wavelengths in an expanding universe.

Since the RM integral depends on the free electron density, it is necessary to account for the subgrid treatment of star-forming gas. Our model assumes an unresolved multiphase interstellar medium consisting of a volume-filling warm phase and dense cold clumps that contain most of the cell mass \citep{Springel2003}. When computing the contribution of star-forming gas to the RM, we include only the gas assigned to the volume-filling warm phase by the subgrid model. The mass fraction of this warm phase is calculated following \citet{Springel2003} and is assumed to be fully ionised.

Figure~\ref{fig:FRM} shows the Faraday RM map of our high-resolution Z24 simulation of halo 4 at $z=0$, which was computed using the GPU-parallel line integral implementation in \textsc{paicos} \citep{Berlok2024}. The small-scale alternation of the sign of the RM signal indicates the absence of coherent Mpc-scale magnetic fields of sufficient strengths. The small-scale fluctuations of RM correspond to a small-scale turbulent magnetic field with a coherence length of $\lesssim$150~kpc, as demonstrated through a fixed-scale filtering approach \citep{Perrone2026a}. These small-scale magnetic fields are a consequence of the small-scale dynamo, which generates and maintains cluster magnetic fields at late times \citep{Tevlin2025}. As expected from the radial profile of mass density\footnote{Note that the mass density and electron density are related via $n_\mathrm{e}=X_\mathrm{H} X_\mathrm{e}\rho/m_\mathrm{p} =0.88\rho/m_\mathrm{p}$ in fully ionized systems such as the volume-filling hot ICM.} and magnetic field strength (Fig.~\ref{fig:B_correlations}), the mean of the RM signal, $\langle |\mathrm{RM}|\rangle$, decreases with radius.

A resolution study of the mean RM signal (top right panel of Fig.~\ref{fig:FRM}) shows excellent convergence at larger radii, as expected from the three-dimensional radial $n_\mathrm{e}$ and $B$ profiles. The central regions, however, are not numerically converged and show increasing RM values with improved resolution because of the higher electron density and magnetic field strength in the cluster cores due to the slow convergence of the IllustrisTNG galaxy formation model (see Section~\ref{sec: convergence}). 

Two zoom-ins on individual galaxies with a magnetised ISM in the cluster exterior highlight the contribution of galaxies to the RM signal. The blue disc morphology in the corresponding synthetic SDSS $gri$ composite images demonstrates ongoing star formation. The dipolar RM structures of the gas-rich discs signal a dominating azimuthal disc magnetic field, which is in line with previous MHD simulations studying magnetic field growth in isolated \citep{Pakmor2013,Pfrommer2022} and cosmological galaxies \citep{Pakmor2017,Pakmor2018,Liu2022}. By contrast, in the lower right-hand panels, the dipolar RM signal is less pronounced, likely reflecting the more disrupted fields typical of merging galaxies \citep{Whittingham2023}.

Because the RM signal includes contributions from both the ICM and individual galaxies, we examine their relative roles by analyzing radial profiles of the mean absolute RM, $\langle |\mathrm{RM}|\rangle$, and its root-mean-square RM, $\sqrt{\langle \mathrm{RM}^2\rangle}$. We find that $\langle |\mathrm{RM}|\rangle$ declines more rapidly with radius than $\sqrt{\langle \mathrm{RM}^2\rangle}$ (see Fig.~\ref{fig:FRM}). To understand this behaviour, we show histograms of Faraday RM in three cylindrical radial bins in Fig.~\ref{fig:FRM_histograms}. In the ICM, the magnetic field is shaped by a small-scale dynamo \citep{Tevlin2025}, and hence its morphology is dominated by the turbulent velocity field. If we can additionally neglect self-gravity, the density probability distribution is expected to approximately follow a log-normal distribution \citep{Ostriker2001,Berkhuijsen2012,Berkhuijsen2015,Buck2022}. Since RM depends on the product of $n_e$ and $B_\parallel$ along the line of sight, and since the magnetic field strength and density are highly correlated (Fig.~\ref{fig:B_correlations}), the RM distribution should also follow a log-normal distribution. Indeed, at each radius, the RM distribution is approximately log-normally distributed as we will show in the following.

We have fitted log-normal distributions to the bulk of the RM distribution at each radial bin as follows: we define $s = \ln\left(\mathrm{RM}/ \mathrm{RM}_0\right)$ with $\mathrm{RM}_0 = 1$ $\mathrm{rad\,m}^{-2}$ and fit the distribution with the function $f(s) = A \exp\left(-(s - \mu)^2/2\sigma^2\right)$. Moving from the inside to the outside of the cluster, we obtain $\mu = (4.23, 1.53, -0.27)$ (corresponding to medians around 70, 5 and 1 $\mathrm{rad\,m}^{-2}$) and $\sigma = (1.51, 1.47, 1.20)$. The peaks of the RM distributions with 1$\sigma$ intervals are $68.7^{+242.4}_{-53.5}$, $4.6^{+15.4}_{-3.5}$ and $0.8^{+1.8}_{-0.5}$ $\mathrm{rad\,m}^{-2}$, respectively. 

In addition to the log-normal distribution, we observe an extended high-RM tail primarily originating from galaxies. As we move toward the cluster outskirts, the ICM contribution declines and the mean RM shifts to lower values. At the same time, the RM contribution from galaxies becomes increasingly dominant. We predict that with increasing resolution of radio observations, the RM signal should be resolved into individual galaxies (and AGN jets, which we currently do not model). For these tailed distributions, $\sqrt{\langle \mathrm{RM}^2\rangle}$ greatly exceeds $\langle |\mathrm{RM}|\rangle$, reaching a ratio of $\sim$10 at $R_{200}$.

By stacking the Galaxy-foreground-subtracted Faraday RM signal from 124 of the most massive Planck clusters at low redshift ($z < 0.35$), \citet{Osinga2025} detect a significant radial increase in root-mean-square RM, $\sigma_\mathrm{RM}$, toward cluster centres, with an average standard deviation within $R_{500}$ of $\sigma_\mathrm{RM} = (209 \pm 37)~\mathrm{rad~m}^{-2}$. The standard deviation outside of $R_{500}$ was found to be $\sigma_\mathrm{RM} = (28 \pm 5)~\mathrm{rad~m}^{-2}$.
The simulated values in halo 4 of PICO-Cluster are considerably larger and we obtain $\sigma_\mathrm{RM} (<R_{500}) = 433~\mathrm{rad~m}^{-2}$ and $\sigma_\mathrm{RM} (R_{500}<r<2R_{500}) = 62~\mathrm{rad~m}^{-2}$. Interestingly, excising the RM values from the central core region inside $0.1 R_{500} = 192~\mathrm{kpc}$ yields values consistent with the observations, $\sigma_\mathrm{RM} (0.1R_{500}<r<R_{500}) = 209~\mathrm{rad~m}^{-2}$. We postpone a detailed treatment of observational effects -- such as beam smoothing, which is expected to mitigate extreme RM values -- to future work and note that the PICO-Cluster RM profiles as well as that of other cosmologically simulated galaxy clusters generally obtain larger values in the central cluster regions \citep{Alonso-Lopez2026,Khadir2026}.

\section{Viscous heating estimates using Braginskii viscosity theory}
\label{sec:Braginskii}

\begin{figure*}
        \includegraphics[trim = 0 20 0 0]{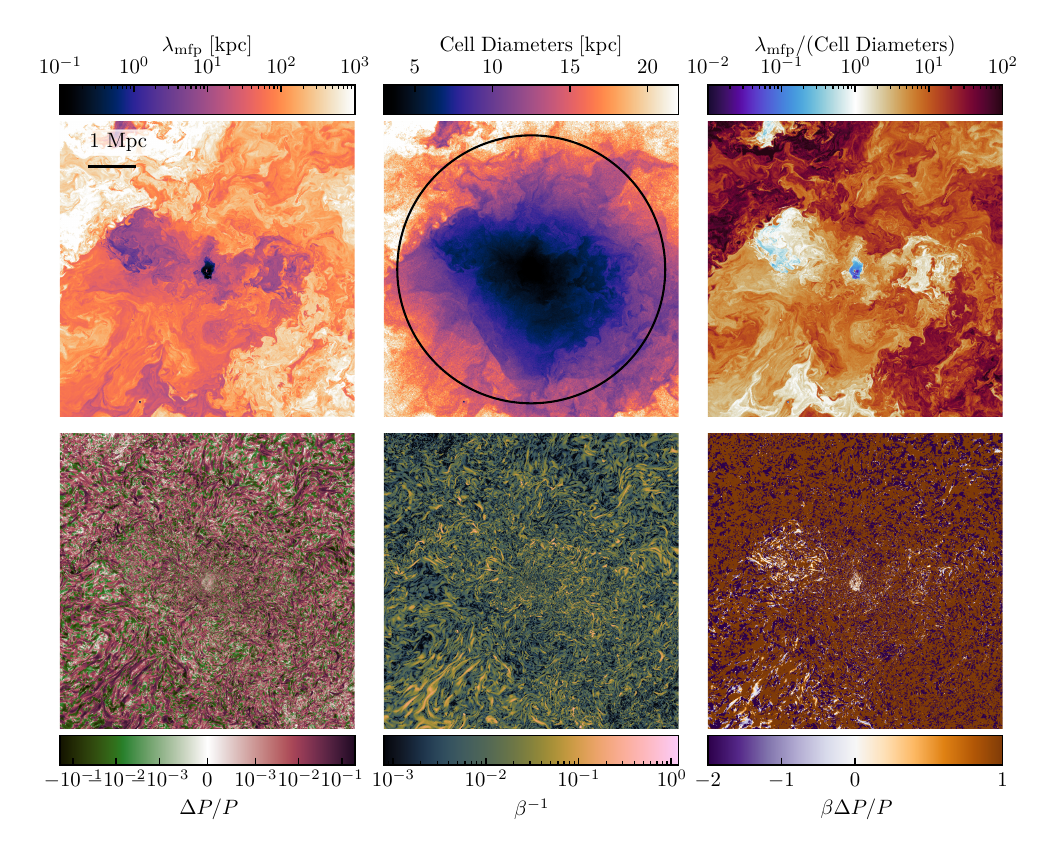}
        \caption{Slices through the Z24 simulation of halo 4 at $z=0$. Top row: the mean-free-path for ion-ion collisions (left), the cell diameters (middle), and their ratio (right). Due to the high numerical resolution, high temperature and low density, the intracluster medium in the simulation is in the regime where $\lambda_{\mathrm{mfp}}$ is larger than the cell sizes. The central region hosts the brightest cluster galaxy and is cold and dense with $\lambda_{\mathrm{mfp}}$ smaller than the cell sizes. Bottom row: We show the (unlimited) pressure anisotropy, $\Delta P / P$ (left), the inverse plasma beta, $\beta^{-1}$, (middle), and the instability parameter, $\beta \Delta P / P$ (right). There is a turbulent and intermittent structure in the pressure anisotropy and the magnetic strength. Mirror and firehose instabilities are unstable in regions where the $B$-field is weak and the pressure anisotropy is large. The colour scale limits in the lower right panel correspond to the threshold for mirror ($\beta \Delta P / P > 1$) and firehose instability ($\beta \Delta P / P<-2$), see equation~\eqref{eq:fire_and_mirror}. The driving of motions by accretion, mergers, and AGN is thus sufficient to drive the pressure anisotropy to these limits.}
        \label{fig:brag_6_panels}
\end{figure*}

Several heating processes operate in galaxy clusters \citep{Meenakshi2026}, including shock heating, turbulent dissipation, whether mediated by numerical or physical viscosity, heating via adiabatic compression, and energy input from streaming cosmic rays through the excitation and damping of Alfv\'en waves. At larger cluster radii, the cooling times exceed several Gyrs \citep{Sanderson2006,Cavagnolo2009}, implying inefficient cooling. At the centre, the cooling time drops below a few Gyr, requiring a heating process for stabilisation, in particular for cool core clusters. Here, we examine whether turbulent dissipation via (anisotropic) viscosity can provide the necessary heating rate.

The ICM is a weakly-collisional medium, i.e., the Coulomb collision frequency is smaller than the gyro-frequencies of charged ions and electrons (see e.g. \citealt{Kunz2022}). This ordering of frequencies arises due to the very high temperatures and low densities found both in real clusters and in the PICO-Cluster simulations. This makes ideal MHD formally not valid, and calls for an extension of MHD known as Braginskii-MHD, which includes terms that model the resulting anisotropic transport of temperature and momentum along magnetic field lines \citep{Braginskii1965}. While the PICO-Cluster simulations presented here do not include these terms, \textsc{Arepo} features an implementation of anisotropic heat transport \citep{Talbot2025} and anisotropic viscosity, also known as Braginskii viscosity \citep{Berlok2020}.

The impact of Braginskii viscosity on the properties of ICM turbulence has not yet been fully quantified. A number of studies argued that Braginskii viscosity may help explain several observational features in the ICM, such as the lack of Kelvin-Helmholtz instabilities at cold fronts \citep{Sijacki2006,ZuHone2015a,ZuHone2016}, and the detection of small-scale velocity fluctuations below the nominal Kolmogorov viscous scale \citep{Zhuravleva2019,Heinrich2024}. These questions cannot properly be addressed without simulations that explicitly include anisotropic viscosity and heat conduction.

Anisotropic Braginskii viscosity has also been suggested as a viable heating mechanism in the ICM \citep{Kunz2011a,Squire2023}. The basic idea is the same as for ordinary isotropic viscosity (see \citealt{Reynolds2005}), i.e., motions are dissipated and the kinetic energy is converted into thermal energy. We have already seen in Figs.~\ref{fig:convergence_panels} and \ref{fig:energy_evolutions} that the amount of vorticity and kinetic energy depends on resolution. This is a consequence of the simulations currently relying on numerical viscosity for setting the dissipation scale.

Despite being ideal MHD simulations, we argue that we can still use our high-resolution PICO-Cluster simulations to put an upper bound on the amount of viscous heating one would find in \emph{resolved} simulations explicitly including Braginskii viscosity. Our reasoning is as follows: since our high-resolution Z24 simulation resolves the mean-free-path of ion collisions everywhere but the dense cluster core (i.e., the numerical dissipation scale is smaller than the dissipation scale one would find in a simulation with explicit isotropic Spitzer viscosity, see Fig.~\ref{fig:brag_6_panels}), we can compute the viscous heating due to Braginskii viscosity in post-processing without being affected by numerical dissipation. We acknowledge that this procedure is not entirely self-consistent, since we are neglecting the back-reaction of the anisotropic viscosity on the flow, which would act to reduce the heating by dampening the velocity shears aligned with the magnetic field direction \citep[in weakly-collisional plasmas this effect is known as ``magneto-immutability'', see, e.g.,][]{Squire2019,Squire2023,Majeski2024b}. Moreover, we do not capture the excitation of plasma microinstabilities that are excited in weakly-collisional, high-$\beta$ plasmas (e.g., mirror and firehose instability) that enhance particle scattering and reduce the Braginskii viscous heating \citep[][]{Kunz2011a,Kunz2014b}. Nevertheless, our approach allows us to place an upper bound on the Braginskii heating rate. 

The anisotropic viscosity tensor is given by (see e.g. \citealt{Berlok2020})
\be
    \mathbf{\Pi} = - \Delta P \left(\b\b - \frac{\mathbf{1}}{3}\right) ,
    \label{eq:Pi}
\en
where $\b$ is the local field direction and the pressure anisotropy, $\Delta P\equiv P_\perp-P_\parallel$ (which is the difference between the pressures perpendicular and parallel to $\b$, respectively), is given by
\be
    \Delta P =
    \eta_\para\left(3\b\b\vec{:}\del \vec{\varv}
     -\del\bcdot\vec{\varv}\right) ,
     \label{eq:p-aniso-no-dt}
\en
and $\vec{\varv}$ is the velocity. The viscosity parameter is given by $\eta_\para = 0.48 P \tau_\mathrm{ii}$ where $P$ is the thermal pressure and
\be
    \tau_\mathrm{ii} =
    \f{3\sqrt{\mH}}{4\sqrt{\uppi} \ln \Lambda_\rmn{Coul} e^4 } \f{\left(\kb T\right)^{3/2}}{n_\mathrm{i}}
\en
is the time scale for ion-ion Coulomb collisions \citep{Huba2013}. Here, $n_\mathrm{i}$ is the ion number density, $\ln \Lambda_\rmn{Coul}=30$ is the Coulomb logarithm and we note that we are for simplicity assuming a fully ionised hydrogen plasma.\footnote{The electron contribution to the viscosity is negligible. The contribution of higher mass ions, such as helium, is discussed in e.g., \citet{Berlok2015} and \citet{Berlok2016b}. We ignore these complications here.}

The mean-free-path of ion-ion particle collisions is given by $\lambda_{\mathrm{mfp}} = \sqrt{P/\rho}\,\tau_\mathrm{ii}/\sqrt{2}$. We can write this as
\be
    \lambda_{\mathrm{mfp}} =
     \sqrt{\f{\kb T}{\mH}}\,\tau_\mathrm{ii}
     =
     8  \times
     \left(\f{n_\mathrm{i}}{0.001 \;\si{cm^{-3}}}\right)^{-1}
     \left(\f{\kb T}{6 \;\si{keV}}\right)^{2} \si{kpc}\ ,
\en
which shows how the mean-free-path depends on ion number density and temperature.

The top left panel of Fig.~\ref{fig:brag_6_panels} shows $\lambda_\mathrm{mfp}$ in a slice through the Z24 simulation of halo 4. The very large values for $\lambda_\mathrm{mfp}$ are found in very dilute and hot regions (the median temperature and ion number densities in the slice are ${\sim} 6$ keV and $6\times 10^{-5}\;\mathrm{cm}^{-3}$) while the brightest cluster galaxy, which is dense and cold, has small values for $\lambda_\mathrm{mfp}$ and is in the collisional regime. The mean-free-path is of particular interest because it sets the scale on which turbulence is dissipated for plasmas in the collisional regime. 
In our high-resolution run Z24 we attempt to resolve $\lambda_\mathrm{mfp}$ in the bulk of the ICM, which poses a severe computational challenge (see also \citealt{Steinwandel2024}). When the scale is not resolved, the turbulent cascade is artificially truncated at the numerical dissipation scale (i.e. due to numerical viscosity). The top middle panel of Fig.~\ref{fig:brag_6_panels} shows the sizes of the Voronoi cells\footnote{Calculated as the diameter of the sphere that has the same volume as the Voronoi cell, which is regularised so that it does not deviate too much from a sphere.} and the top right panel shows their ratio. The simulation has cell sizes smaller than $\lambda_\mathrm{mfp}$, except for the central blue region, which is cold and dense. Quantitatively, the Z24 simulation resolves $\lambda_\mathrm{mfp}$ with at least 1 (10) cell(s) in 98 per cent
(45 per cent) of the image.

\begin{figure*}
        \includegraphics{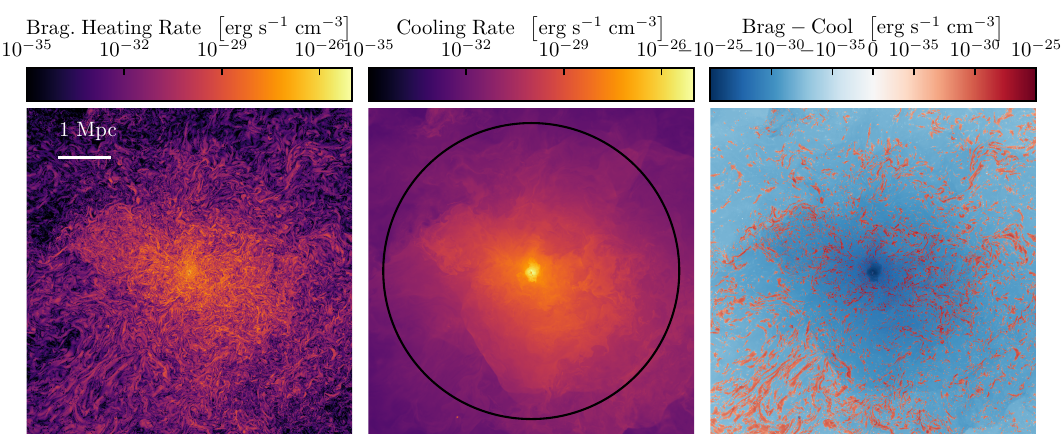}
        \caption{Left: the Braginskii viscous heating rate estimated using equation~\eqref{eq:heating-term-limited} in the same image slice as shown in Fig.~\ref{fig:brag_6_panels}. Middle: the radiative cooling rate from the simulations (computed as $n_\mathrm{H}^2 \Lambda_\mathrm{GFM}$ where $\Lambda_\mathrm{GFM}$ is the cooling rate output from the IllustrisTNG model). Right: the difference between Braginskii viscous heating and cooling. The cooling rate, which depends primarily on density and temperature, is almost spherically symmetric and varies only on large scales. The viscous heating, in contrast, occurs primarily in localised patches due to the spatial intermittency of the plasma-$\beta$. Although viscous heating arises from the dissipation of motions, $\beta$ controls its rate because the firehose/mirror instability limiters (equation~\ref{eq:fire_and_mirror}) are nearly always active. As a result, the viscous heating can exceed the cooling rate locally but not in a volume-filling fashion.}
        \label{fig:brag_heating_panels}
\end{figure*}

In weakly-collisional plasmas, fluid motions like shear and compression can drive a non-zero pressure anisotropy, $\Delta P$, and Braginskii viscosity relaxes $\Delta P$ while simultaneously heating the plasma. Concretely,
the heating rate that enters in the internal energy equation is (see e.g. \citealt{Berlok2020})
\be
        \mathcal{H}_\mathrm{Brag} =  -\mathbf{\Pi} \vec{:} \del \vec{\varv} = \f{\eta_\para}{3}
    \left(3\b\b\vec{:}\del \vec{\varv}
     -\del
     \bcdot
     \vec{\varv}
     \right)^2 =
    \f{(\Delta P)^2}{3\eta_\para} \ ,
    \label{eq:heating-term}
\en
and the heating rate thus depends on the amplitude of the driving, the viscosity coefficient and the direction of the local magnetic field.

This is not the full story, however, because micro-scale instabilities, such as the firehose and mirror instabilities, can be excited and modify $\Delta P$, driving it back towards the thresholds for marginal stability \citep{Kunz2014b,Riquelme2015a,Melville2016}. These are kinetic instabilities driven by anisotropies in velocity-space, and our simulations do not capture this regularization of the pressure anisotropy. The criterion for stability to firehose and mirror instability can be written as \citep{Kunz2012}
\be
    \label{eq:fire_and_mirror}
    -\f{2}{\beta} < \f{\Delta P}{P} < \f{1}{\beta} \ ,
\en
where $\beta$ is the plasma-$\beta$. Braginskii-MHD simulations conventionally use equation~\eqref{eq:fire_and_mirror} to employ hard-wall limiters on $\Delta P$ when evaluating the viscosity tensor, $\mathbf{\Pi}$, and we will use the same approach when estimating the heating rate below. This approach is motivated by kinetic plasma simulations \citep{Schekochihin2008,Rosin2011,Kunz2014a} and solar wind observations \citep{Bale2009,Chen2016}, and is discussed in more detail in e.g. \citet{Kunz2012}.

The lower left panel of Fig.~\ref{fig:brag_6_panels} shows the unlimited values of $\Delta P / P$, the lower middle panel shows $\beta^{-1}$, and the lower right panel shows $\beta \Delta P/P$. In our simulations, fluid motions drive $\Delta P$ with its sign fluctuating on small scales and a root-mean-square value of $|\Delta P/P| \sim 0.02$. The plasma-$\beta$, which was found to be $\beta\sim 100$ when volume-averaged inside $R_{200}$ or $R_{500}$ (see Fig.~\ref{fig:evolution_of_energy_ratios}), similarly fluctuates. The saturated colours in $\beta \Delta P / P$ show that most of the region is unstable to either firehose (${\approx} 20$ per cent) or mirror instability (${\approx} 50$ per cent), which means that the limiters will be active in most of the region.\footnote{A comparison between the Z12 and Z24 simulations of halo 4 reveals that these numbers are converged. Across images produced for all Z12 simulations, we find that the mean and standard deviations of these percentages are $18.9\pm1.7$ for the firehose instability and $52.5\pm3.2$ for the  mirror instability.}

With the limiters given in equation~\eqref{eq:fire_and_mirror} employed on $\Delta P$, the Braginskii heating rate becomes
\be
   \mathcal{H}_\mathrm{Brag} =
    \f{(\Delta P)_\mathrm{lim}^2}{3\eta_\para}
    \ .
    \label{eq:heating-term-limited}
\en
Given that the turbulent motions drive $\Delta P$ to exceed the firehose/mirror thresholds almost everywhere (see bottom right panel in Fig.~\ref{fig:brag_6_panels}), $(\Delta P)_\mathrm{lim}$ is essentially the same as $P \xi \beta^{-1}$ where $\xi$ is either $-2$ or $1$. The heating rate found using equation~\eqref{eq:heating-term-limited} is thus approximately $(P \xi / \beta)^2 / (3 \eta_\para)$, which is equation~(14) in \citet{Kunz2011a}.
We show the heating rate using equation~\eqref{eq:heating-term-limited} in the left panel of Fig.~\ref{fig:brag_heating_panels}. One striking feature of $\mathcal{H}_\mathrm{Brag}$ is its spatial intermittency, which is a consequence of the intermittency of the plasma $\beta$ together with the hard-wall limiters, since $P$ and $\eta_\para$ are spatially smooth fields ($\eta_\para$ depends only on $T$, which varies slowly). As a result, the heating rate visually appears as a radial modulation of $\beta^{-1}$ (shown in the bottom middle panel of Fig.~\ref{fig:brag_6_panels}). This is in contrast to standard isotropic viscous dissipation in turbulent flows at high-Reynolds number, where spatial intermittency of the heating rate is due to the spatial intermittency of the velocity gradients \citep[see, e.g.,][]{Buaria2019}.

\begin{figure*}
        \includegraphics[trim = 0 15 0 0]{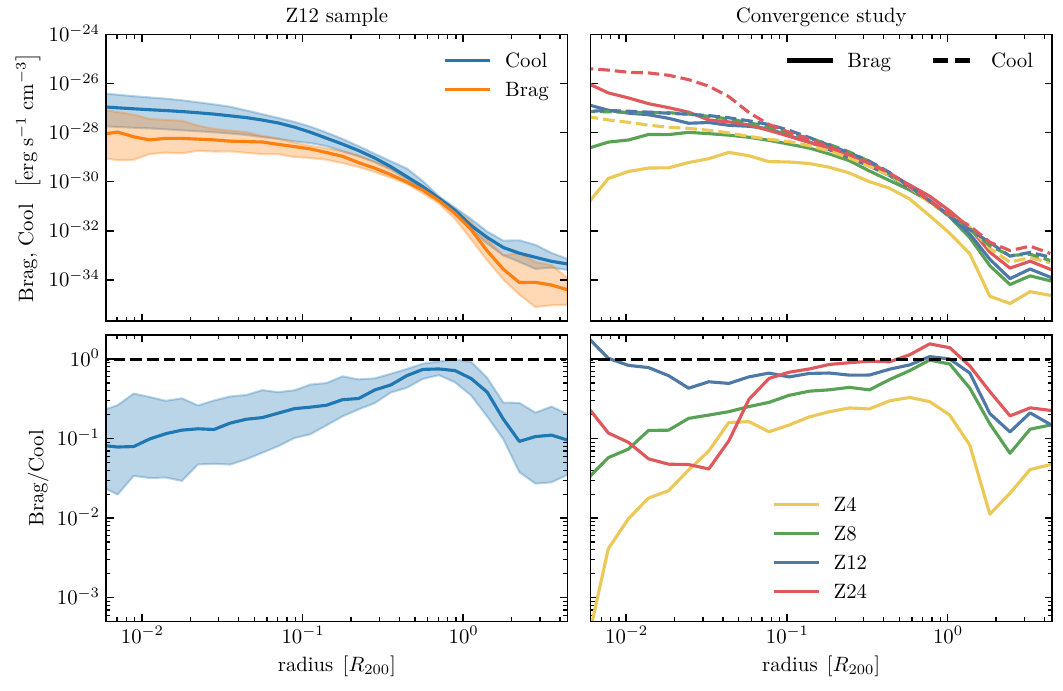}
        \caption{A comparison of the viscous heating rate and the cooling rate as a function of radius in the Z12 cluster sample (left column) and for different zoom factors for halo 4 (right column). The median heating rate of the cluster sample is systematically below the cooling rate (top left), but at distances approaching the virial radius, Braginskii heating can exceed 50 per cent of the cooling (bottom left). Increasing the resolution, Braginskii heating and cooling are converged at large radii already at Z12 (top right), while at small radii they both increase with resolution.}
        \label{fig:brag_radial}
\end{figure*}

The middle panel of Fig.~\ref{fig:brag_heating_panels} shows the cooling rate, which in comparison appears smooth and approximately spherically symmetric. The right panel shows the difference between the Braginskii heating rate and the cooling rate with red (blue) colours corresponding to net heating (cooling). The spatial intermittency of the viscous heating rate, combined with the smooth distribution of the volumetric cooling rate, gives rise to localized patches of net heating. In order to quantify this non-volume-filling heating, we show in Fig.~\ref{fig:brag_radial} the radial profiles of the viscous heating rate and the cooling rate.\footnote{Since we are only interested in the heating/cooling balance of the ICM, we have here excluded star-forming cells, cells with a temperature below $10^5$ K, and cells for which the galaxy formation model gives net heating (see \citealt{Pop2022}).}

In the top-left panel of Fig.~\ref{fig:brag_radial}, we observe over the entire Z12 cluster sample that the median heating rate due to Braginskii viscosity is systematically below the cooling rate. The relative importance of heating and cooling however is not uniform throughout the cluster: looking at the ratio of the two terms (bottom left), we see that at radii approaching $R_{200}$, where the cooling rate drops off due to the $\rho^2$ dependence (see equation~\ref{eq:x_ray_emissivity}), Braginskii heating exceeds 50 per cent of the cooling. This result suggests that in large portions of the cluster outside of the core, heating due to Braginskii viscosity may play a sizeable role in offsetting cooling losses. Repeating the same analysis at different resolutions for halo 4 (top right), we see that both Braginskii heating and cooling are converged at Z12 resolution for large radii. In contrast, both heating and cooling increase with resolution in the central region, which is a consequence of the slow convergence of the simulations on e.g. density, pressure and magnetic field strengths in this region (see Fig.~\ref{fig:convergence_radial_profiles} and discussion in Section~\ref{sec: convergence}).

In conclusion, we have used post-processing to analyse the PICO-Cluster simulations to place an upper bound on the expected amount of viscous heating that would be found in simulations explicitly including Braginskii viscosity. In such simulations, the numerical dissipation present in our MHD simulations would be replaced by an actual physical modelling of the dissipation processes. By volume-integrating the radial heating and cooling profiles for the Z12 simulations shown in Fig.~\ref{fig:brag_radial}, we find that up to $(17{-}69)$ per cent of the galaxy cluster cooling inside $R_{200}$ can be offset by viscous Braginskii heating. The mean and standard deviation of this value is  $(41\pm16)$ per cent. While this is not enough to maintain the cluster temperature, and other heating mechanisms such as shocks due to AGN feedback are required, our analysis suggests that the viscous Braginskii heating may nevertheless play an important role. In \citet{Kunz2011a}, the heating rate as a function of radius was calculated by assuming that the turbulent driving is sufficient to drive the pressure anisotropy to the limits set by the thresholds for firehose/mirror instabilities. The PICO-Cluster simulations confirm this prediction, albeit under the assumptions of ideal MHD and post-processed Braginskii heating. However, the plasma-$\beta$ is highly intermittent in space, and the regions with very weak fields (large $\beta$) contribute little heating, so that viscous Braginskii heating is unable to offset cooling on its own. 

Further work is clearly necessary in order to characterize the thermodynamics of the ICM, both on the theoretical side \citep[e.g., with the inclusion of other plasma effects that might contribute to heating;][]{Ley2023,Squire2023}, and on the simulation side with high-resolution simulations explicitly including Braginskii viscosity (and the associated reduced heating due to ``magneto-immutability'', see, e.g., \citealt{Squire2019,Squire2023,Majeski2024b}), together with more advanced diagnostics to accurately determine the relative contribution of the various heating terms \citep{Meenakshi2026}.

\section{Summary and Conclusions}
\label{sec:conclusions}

We have presented the baseline simulations of the PICO-Cluster project, a new suite of zoom-in simulations of very massive galaxy clusters performed using the \textsc{Arepo} code and the IllustrisTNG galaxy formation model. We have carried out a new parent simulation with a box size of $1~h^{-1}\mathrm{Gpc}$ (Fig.~\ref{fig:parent_box})\, followed by re-simulations of 24 clusters at a resolution comparable to TNG300 (Fig.~\ref{fig:B_panels}), as well as a single exceptionally massive cluster at varying resolution with up to eight times better mass resolution, which is comparable to TNG100 (Fig.~\ref{fig:zoom24_halo4}). We summarize our findings as follows:

\begin{description}
	\item[\textbf{Initial conditions:}] We used the new code \textsc{CosmoZoomIC} to generate zoom-in initial conditions, ensuring that all simulations are free from contamination of heavy (low-resolution) DM particles inside $r/\Rvir<2.74$ across all redshifts (Section~\ref{sec:particle_load} and Fig.~\ref{fig:contamination_radius}).
	\item[\textbf{Verification:}] We verified that the PICO-Clusters broadly agree with the previous MTNG simulation \citep{Pakmor2023} on global properties such as the stellar mass-halo relation and the gas fraction in the cluster (Fig.~\ref{fig:validation_MTNG}).  We also find that PICO-Clusters are consistent with cluster mass--observable scaling relations derived from the MTNG simulations as well as from observations. This includes scaling relations for cluster richness, integrated Compton-$y$ parameter, and core-excised X-ray luminosities. However, we find that the inclusion of magnetic fields in PICO-Clusters increases the masses of the SMBHs, which increases the feedback efficiency and reduces the stellar mass content (Fig.~\ref{fig:validation_MTNG}). The decreased star formation efficiency causes the PICO-Cluster simulations to have systematically lower K-band luminosities (Fig.~\ref{fig:cluster_scaling_relations}). 
	\item [\textbf{Validation:}] We compared emission-weighted thermodynamic profiles obtained with PICO-Clusters with the empirical X-COP profiles \citep{Eckert2019,Ettori2019,Ghirardini2019} and found good agreement at larger radii and a discrepancy at the centre: while our profiles match non-cool-core cluster data, none of our clusters shows a cool core, thereby posing a serious problem to the theoretical modelling of these systems (Fig.~\ref{fig:radial_profiles_obs}). The discrepancy at the centre is due to the inability of the kinetic AGN feedback to self-regulate cooling while maintaining a low entropy consistent with the values found in cool-core clusters, an effect which is exacerbated by the over-massive SMBHs. This motivates improvements on the models for SMBH growth and AGN jet feedback \citep[e.g.,][]{Weinberger2025,Weinberger2026}. The PICO-Cluster gas metallicity profiles are on the lower side of the observed non-cool-core cluster profiles \citep{Ghizzardi2021}, which could be due to an observational bias towards high-metallicity plasmas or due to insufficient star formation in PICO-Clusters. Similarly, the PICO-Cluster cumulative stellar mass profiles fall somewhat short of the observed profiles either due to a combination of incomplete numerical convergence and overabundance of the intracluster light in simulations or due to the intrinsic uncertainty of converting observed stellar light to mass.
	\item [\textbf{Numerical convergence:}] We performed a series of simulations with successively increasing resolution to test the convergence of the IllustrisTNG galaxy formation model and the quality of our initial conditions. Focusing on a single galaxy cluster, halo 4, we conducted four simulations at Z4, Z8, Z12, and Z24 resolution, the latter of which corresponds to a baryonic mass resolution of $1.4\times10^{6}\,\mathrm{M}_\odot$. The combination of initial conditions and computational code to execute the simulations performs extremely well, with higher resolution simulations able to capture more substructure but keeping the location and masses of collapsed objects that were already resolved at lower resolution (Fig.~\ref{fig:convergence_panels}). We find that the magnetic field and the gas density, in particular, converge rapidly for radii $>100$~kpc, whereas vorticity continues to increase in magnitude with increasing resolution. This is a consequence of the PICO-Cluster simulations presented here relying on numerical dissipation. Nevertheless, volume-integrated energies inside $R_{200}$ have converged to a satisfactory degree already at Z12 resolution (with a difference between Z12 and Z24 on the values of $E_\mathrm{th}$, $E_\mathrm{kin}$ and $E_\mathrm{mag}$ of 6--7 per cent at $z=0$, see Fig.~\ref{fig:energy_evolutions}). Time-averaged radial profiles for gas density, temperature, pressure, and magnetic field in the bulk of the ICM agree between Z12 and Z24 (Fig.~\ref{fig:convergence_radial_profiles}). By contrast, metallicity profiles as well as the central thermodynamic profiles including that of the magnetic field strength keep increasing with resolution. These discrepancies are interpreted to be due to the slow convergence of the stellar mass content (lower panels of Fig.~\ref{fig:SFR}).
	\item [\textbf{Magnetic fields:}] Following up on \citet{Tevlin2025}, which analysed magnetic field amplification in four PICO-Clusters and found that galaxy formation plays an essential role, we analysed the magnetic field in all 24 PICO-Clusters (Fig.~\ref{fig:B_panels}). We investigated the correlation between magnetic field strength and gas density (Fig.~\ref{fig:B_correlations}), and found that there is an overall $B\propto \rho^{2/3}$ dependence with what appears to be a $B\propto \rho^{1/2}$ ($B\propto \rho^{5/6}$) at high (low) densities, corresponding to the central regions (and outskirts) of the clusters. The high cadence of PICO-Cluster simulation output allows a detailed analysis of the generation of vorticity and the amplification of magnetic fields (Fig.~\ref{fig:x_t_plots}). We find that the magnetic dynamo in the PICO-Cluster simulations saturates already at $z=4$ with a median magnetic field strength inside $R_{200}$ of order $1 \muG$ (Fig.~\ref{fig:energy_evolutions}). The ratio of integrated magnetic-to-thermal energies remains essentially constant from $t = 5\;\mathrm{Gyr}$ (corresponding to $z=1.2$) until today (Fig.~\ref{fig:evolution_of_energy_ratios}). This translates to a volume-averaged plasma-$\beta$ of order 100, although the spatial scatter can be very high. The PICO-Cluster prediction that $E_\mathrm{mag}/E_\mathrm{th} \sim \mathrm{const.}$ implies that low redshift clusters should have the same scaling relation for magnetic energy as for the thermal energy, i.e., we deduce from $E_\mathrm{th}\propto M_{200}^{5/3}$ the scalings $E_\mathrm{mag}\propto M_{200}^{5/3}$ and $B\propto M_{200}^{1/3}$ (Fig.~\ref{fig:cluster-energy-scalings}).
	\item[\textbf{Faraday rotation measures:}] We computed a high-resolution Faraday rotation map of the Z24 simulation (Fig.~\ref{fig:FRM}) and analysed the relative contributions to the RM signal from the ICM and the cluster galaxies. Our simulations predict that the galaxy contribution becomes more important at larger radii. The galaxy signal appears as strong sources on small angular scales, so that the mean absolute value of RM declines faster than the root-mean-squared value. Histograms of RM at different cylindrical radii revealed a log-normally distributed contribution from the ICM  and a high RM tail contribution from the galaxies (Fig.~\ref{fig:FRM_histograms}). Although the simulated $\sigma_\mathrm{RM}$ within $R_{500}$ exceeds the observed value by a factor of two, excising the innermost region ($r < 0.1\,R_{500}$) brings the two into agreement. Resolving this discrepancy will require future work that carefully accounts for observational biases.
	\item[\textbf{Braginskii viscosity:}] We post-processed our ideal MHD simulations to place an upper bound on the amount of viscous heating one would find in a very high resolution simulation performed with explicit Braginskii viscosity. The high-resolution PICO-Cluster simulations resolve spatial scales on the order of the mean-free-path for ion-ion Coulomb collisions (Fig.~\ref{fig:brag_6_panels}). As proposed by \citet{Kunz2011a}, motions driven by mergers and AGN activity are sufficient to drive the pressure anisotropy above the thresholds for micro-scale plasma instabilities (Fig.~\ref{fig:brag_6_panels}). With the hard wall limiters for pressure anisotropy employed (equation~\ref{eq:fire_and_mirror}), a spatially intermittent viscous heating rate thus arises indirectly via the turbulent magnetic field rather than directly via the turbulent velocity field. We found that this spatially intermittent  viscous heating does not offset cooling in a volume filling fashion (Fig.~\ref{fig:brag_heating_panels}). However, the construction of radial profiles of heating/cooling reveals that viscous heating, while systematically lower than cooling, could nevertheless contribute at a non-negligible 40 per cent level inside $R_{200}$ (Fig.~\ref{fig:brag_radial}). Additional sources of heating are thus required to prevent excessive cooling flows, such as shocks driven by the AGN feedback and mixing of hot gas from larger radii into the core region.
\end{description}
Thus, with this initial PICO-Cluster paper, we have made progress towards elucidating two of the eleven outstanding cluster problems identified in Section~\ref{sec:introduction}; namely quantifying the thermodynamic effects of the weakly collisional ICM (problem iv) and understanding the origin and scaling properties of cluster magnetic fields \citep[problem (v), complementing][]{Tevlin2025}. 

In future PICO-Cluster papers, we plan to make substantial progress towards answering all of the identified problems. In most cases, this will require new simulations with dedicated additional cluster physics to be incorporated and compared against the baseline simulations presented here, which adopt the IllustrisTNG galaxy formation model. Examples include improved AGN feedback simulations that explicitly model jet launching and propagation from SMBHs \citep{Weinberger2026b} to address the bimodal distribution of central cooling times and the cooling flow problem (problems i and ii), and including cosmic rays from AGNs and structure formation shocks (Talbot et al., in prep.), thereby improving our understanding of effective re-acceleration models underlying radio (mini-)halos and the missing gamma-ray emission from clusters (problems vi and viii). Clearly, a full understanding of diffuse cluster radio emission requires kinetic plasma simulations with boundary conditions informed by cosmological simulations to develop a small-scale effective cosmic ray re-acceleration model that can then be ported to and applied in cosmological simulations. Post-processing the PICO-Cluster simulations with advanced filtering algorithms \citep{Perrone2026a} will elucidate the turbulent pressure support (problem iii) and further shed light on magneto-genesis in clusters (problem v, Perrone et al., in prep.).

However, addressing some of these questions, such as the physics underlying jellyfish galaxies (problem ix), will clearly require going beyond cosmological MHD simulations, calling for high-resolution, idealized meso-scale simulations. These so-called wind-tunnel simulations would draw on boundary conditions informed by our cosmological PICO-Cluster simulations (Dusch et al., in prep.), while extending the ISM modelling to a multiphase structure with two-moment cosmic ray transport  \citep[][which would currently not be possible in the full cosmological cluster setting]{Thomas2025}, thus enabling a set-up in which we can isolate individual physical effects by running several simulations in a cost-efficient manner. Another example of such hybrid high-resolution MHD simulations concerns the physics underlying radio relics (problem vii), which demanded fresh insight from cosmological simulations, augmented by additional numerical resolution to capture the small-scale density fluctuations in the cluster outskirts that appears to be critical to resolve the hydrodynamic instabilities that amplify cluster magnetic fields in the post-shock regions \citep{Whittingham2026a,Whittingham2026b}. We believe the PICO-Cluster project holds great promise for advancing our understanding of multi-scale cluster physics, spanning from cosmological scales to high-resolution meso-scale physics with (partially extended) MHD models down to kinetic plasma simulations of setups informed by cosmological initial conditions to build a grand picture of plasma effects in cosmological galaxy clusters.

\section*{Acknowledgements}
TB gratefully acknowledges funding from the European Union’s Horizon Europe research and innovation programme under the Marie Skłodowska-Curie grant agreement No 101106080 and financial support by the Carlsberg Foundation via grant CF23-0417. CP, LP, JW, and LT acknowledge support from the European Research Council via the ERC Advanced Grant ``PICOGAL'' (project ID 101019746). CP and ND acknowledge support by the Deutsche Forschungsgemeinschaft (German Research Foundation) for the Research Grant (project ID 567737545). RW acknowledges funding of a Leibniz Junior Research Group (project number J131/2022). The authors gratefully acknowledge the computing time granted by the Resource Allocation Board and provided on SuperMUC-NG through the project ``The plasma physics of galaxy clusters in a cosmological context'' (ID: pn68cu). The Tycho supercomputer hosted at the SCIENCE HPC centre at the University of Copenhagen was used for supporting this work.

\emph{Software:} Analysis and visualization was performed using \textsc{paicos} \citep{Berlok2024} with \textsc{matplotlib} \citep{Hunter2007} for plotting.

\section*{Data availability}
After an initial proprietary period, the data underlying this article will be made available in a repository hosted at the Leibniz Institute for Astrophysics Potsdam, and can be accessed via the \href{https://pico-cluster.aip.de}{PICO-Cluster website}.




\bibliographystyle{mnras}
\bibliography{references} 


\appendix

\section{Band-integrated X-ray emissivity}
\label{sec:x_ray_emissivity}
The frequency-dependent thermal Bremsstrahlung emissivity is given by
(\citealt{Rybicki1979}, equation 5.14a)
\begin{align}
    j(\nu) =
    \f{2^5\uppi e^6}{3 m_\mathrm{e} c^3} \left(\f{2\uppi}{3 k_\mathrm{B} m_\mathrm{e}}\right)^{1/2} T^{-1/2} Z^2 n_\mathrm{e}
    n_\mathrm{i}
    \mathrm{e}^{-h\nu/k_\mathrm{B}T} \bar{g}_\mathrm{ff} \left(\nu,\,T\right)
    \nonumber \\
    =
    A Z^2 n_\mathrm{e}
    n_\mathrm{i}
    T^{-1/2} \mathrm{e}^{-h\nu/k_\mathrm{B}T} \bar{g}_\mathrm{ff}\left(\nu,\,T\right)\ ,
\end{align}
where $\bar{g}_\mathrm{ff}$ is the velocity-averaged free-free Gaunt factor and we have implicitly defined the constant $A$ in the second step.
This can be generalized to a multi-species plasma by introducing a sum over ion-species, $s$:
\be
    j(\nu) = A n_\mathrm{e}
    T^{-1/2} \mathrm{e}^{-h\nu/k_\mathrm{B}T} \sum_s Z_s^2 n_\mathrm{s} \bar{g}_\mathrm{ff,\,s}\left(\nu,\,T\right) .
\en
%
We find the band-integrated volume emissivity by integrating from photon frequencies $\nu_\mathrm{low} = E_\mathrm{low}/h $ to $\nu_\mathrm{high} = E_\mathrm{high}/h$. We assume for simplicity that $\bar{g}_\mathrm{ff}$ is constant and the same for both ion species, i.e., we introduce the $g_\mathrm{B}$ notation for a frequency-averaged Gaunt factor also used in \citet{Rybicki1979}.
The integral yields
\begin{align}
    \varepsilon
    &= \int_{\nu_\mathrm{low}}^{\nu_\mathrm{high}} j(\nu) \,\mathrm{d}\nu \nonumber \\
    &=
    n_\mathrm{e} n_\mathrm{H} A \f{k_\mathrm{B}T^{1/2}}{h} \left(\mathrm{e}^{-E_\mathrm{low}/k_\mathrm{B}T}-\mathrm{e}^{-E_\mathrm{high}/k_\mathrm{B}T}\right)\sum_s \bar{g}_{\mathrm{B},\,s} Z_s^2 n_\mathrm{s}/n_\mathrm{H} \ .
    \label{eq:epsilon}
\end{align}
Setting $\varepsilon \equiv n_\mathrm{e} n_\mathrm{H} \Lambda$ thus defines $\Lambda$ as in equation~\eqref{eq:Lambda_xray}.






\bsp  
\label{lastpage}
\end{document}